\documentclass[11pt]{article}
\usepackage[english]{babel}
\usepackage{amsmath}
\usepackage{amsfonts}
\usepackage{graphics,graphicx}
\usepackage{cool}
\usepackage{rotating}
\usepackage[dvipsnames]{xcolor}
\usepackage{array}

\DeclareMathOperator{\arctanh}{arctanh}

\begin{document}
\title{\bf Particle velocity based hydrofracturing algorithm for a penny-shaped crack}
\author{D.~Peck, M.~Wrobel, M.~Perkowska, G.~Mishuris} 
\date{}
\maketitle

\section{Introduction}

Hydraulic fracture (HF) is the phenomenon of a fluid driven crack propagating in a solid material. It can be encountered in various natural processes, such as  subglacial drainage of water or during the extension of magmatic intrusions in the earth's crust. Simultaneously the underlying physical mechanism is very important in numerous man-made activities. Hydrofracturing can appear as an unwanted and detrimental factor in underground CO$_2$ or waste repositories \cite{wasted}. On the other hand, intentionally induced hydraulic fractures constitute the essence of fracking technology - a method used when stimulating unconventional hydrocarbon reservoirs \cite{Clark2013} or for geothermal energy exploitation \cite{geothermal}. All of these applications create demand for a proper understanding and prediction of the process of hydraulic fracture.

As a result of the multiphysical nature of the underlying physical phenomenon and complex interactions between the component physical fields, the mathematical modeling of hydraulic fractures represents a significant challenge. The main difficulties arise due to: i) strong non-linearities resulting from interaction between the solid and fluid phases, ii) singularities in the physical fields, iii) moving boundaries, iv) degeneration of the governing equations at the crack tip, v) leak-off to the rock formation, vi) pronounced multiscaling effects, vii) complex geometry.

The first theoretical models of hydraulic fracture were created in 1950s (see for example \cite{Harrison} and \cite{Hubbert}). Subsequent research led to the formulation of the so-called classic 1D models: PKN \cite{Perkins1961,Nordgren1972}, KGD (plane strain) \cite{Khristianovic1955,Geertsma1969} and penny-shaped/radial \cite{Sneddon1946}. Up to the 1980s these very simplified models were used to design and optimize the treatments used in HF. The increasing number and size of fracking installations, alongside the simultaneous advance in computational techniques, resulted in the formulation of more sophisticated and realistic models of HFs. A comprehensive review of the topic can be found in \cite{Adachi2007}.

Though superseded in most practical applications, the classic 1D models remain a significant avenue of research into the fundamentals of HF. They enable one to investigate some inherent features of the underlying physical process, the mathematical structure of the solution, and finally to construct and validate computational algorithms. Substantial advances have been achieved in this area throughout the last thirty years by way of a cyclical revision of these classic formulations. It was not until 1985 \cite{Spence} that the importance of the solution tip asymptotics was first noticed, specifically for the KGD and penny shaped cracks. The explicit form of the tip asymptote for the PKN model was given in 1990 by \emph{Kemp} \cite{Kemp}. Moreover, in this publication the author remarked, for the first time, that when properly posed the Nordgren's model constitutes a Stefan-type problem and as such needs an additional boundary condition which equates the crack propagation speed with the velocity of the fluid front. However, this important idea was abandoned for the next twenty years until rediscovered by \emph{Linkov} \cite{Linkov2011} in 2011. The author proved that the general HF problem is ill-posed and proposed a regularization technique based on application of the aforementioned Stefan condition - called there the speed equation. The numerous investigations carried out since the beginning of the present century for the KGD \cite{Garagash_small_toughness,garagash_large_toughenss,Garagash_shut_in,Garagash2011} and penny-shaped models \cite{Savitski2002,Bunger2005,Bunger2005b} have led to the importance of the problem's multiscale character being recognized. It is now well understood that the global response of the fluid driven fracture is critically dependent on the interaction between competing physical processes at various temporal and spatial scales. Depending on the intensity of various energy dissipation mechanisms, as well as the fracturing fluid and solid material properties, the hydraulic fracture evolves in the parametric space encompassed by the limiting regimes: i) viscosity dominated, ii) toughness dominated, iii) storage dominated, iv) leak-off dominated.

Bearing in mind the whole complexity of the problem, it still remains an extremely challenging task to deliver credible solutions which reflect all of the desired features. The relative simplicity of the classic 1D models means that they are well suited to the task of creating benchmarks, 
used when developing and verifying more advanced solutions and algorithms. For the KGD and PKN models one can find in the literature a number of credible results, including recently developed simple and accurate approximate solutions, that can be used for the aforementioned purposes \cite{Linkov_3,Mishuris2012,Wrobel2015,Perkowska2015}.

Unfortunately there is not a substantial body of suitable benchmarks available for the radial model. One can mention here the work by \emph{Advani et al} \cite{Advani1987}, where the approximate time-dependent solution for both Newtonian and non-Newtonian fluids is given. However, its accuracy has not been convincingly proved. An early simulator of penny-shaped fracture was presented in \cite{Meyer}, where comparison with previous results was also provided. However again, the error level of the final results is unknown. In \cite{Savitski2002} the asymptotic solutions for zero and large toughness regimes were delivered for a Newtonian fluid. An additional asymptotic solution for the toughness dominated regime, for a Newtonian fluid, over small and large time scales was presented by \emph{Bunger et al} \cite{Bunger2005}. These asymptotic solutions were later shown to correspond reasonably well to experimental results \cite{Bunger2008}.

The field has become more active in the past year however. There is a work of \emph{Kanaun} \cite{Kanaun}, which provides a discretized approach to the time-dependent form of the problem. Unfortunately the model only provides an approximate solution for Newtonian fluids in the toughness dominated regime without fluid leak-off. There has also been an experimental paper by \emph{Lai et al} \cite{Lai2016}, which examined the growth of a penny-shaped fracture in a gelatin matrix. This study was able to demonstrate the effect of varying experiment parameters for small values of the fracture toughness, and suggests that such fractures behave according to the scaling arguments of \emph{Spence \& Sharp} \cite{Spence} over long times. Finally there is a recent numerical solution provided by \emph{Linkov} \cite{Linkov2015,Linkov2016}, for the class of Newtonian and shear-thinning fluids, but only in the viscosity dominated case. Unfortunately, the accuracy of the aforementioned penny-shape benchmarks is still to be confirmed. Additionally, neither of the recalled solutions takes the convenient form of a simple formula (such as those for the KGD model from \cite{Wrobel2015,Perkowska2015}) that can be easily used for comparison.

The aim of this paper is to meet the demand for benchmark solutions to the radial HF model and: i) by means of a dedicated computational scheme deliver a highly accurate numerical solution,  ii) provide simple solution approximations, which maintain a reasonable level of accuracy, for the zero leak-off case, iii) verify the accuracy of existing benchmarks, iv) introduce purely analytical solutions to the problem obtained for a predefined non-zero leak-off function.

To this end the self-similar formulation of the penny-shaped model will be analyzed. The numerical computations will be performed using a modified form of the universal algorithm introduced in \cite{Wrobel2015,Perkowska2015}. It employs a mechanism of fracture front tracing, based on the speed equation approach \cite{Mishuris2012}, coupled with an extensive use of  information on the crack tip asymptotics and regularization of the Tikhonov type (the technical details of both concepts can be found in  \cite{Kusmierczyk,Wrobel2013}). The modular architecture of the computational scheme facilitates its adaptation to the problem of radial HF.

The paper is organized as follows. The basic system of equations describing the problem is given in Sect.~\ref{Sect:ProbForm}. Next, normalization to the dimensionless form is carried out. In Sect.~\ref{TipVar},
comprehensive information about the solution asymptotics is presented, which is heavily utilized in the subsequent numerical implementation. New computational variables, the reduced particle velocity and modified pressure derivative, are introduced. The advantages of both are outlined, and the problem is reformulated in terms of the new variables. In Sect.~\ref{Sect:SelfSim} the governing system of equations is reduced to the time independent self-similar form. This formulation is used in  Sect.~\ref{Sect:NumBench} to construct the computational algorithm. The accuracy and efficiency of computations are examined against newly introduced analytical benchmark examples. Alternative error measures are proposed for the cases where no closed-form analytical solution is available. Then, numerical reference solutions are proposed for the variant of an impermeable solid. Simple and accurate solution approximations are given for various fixed values of the material toughness, over the whole range of the fluid behaviour index. Next, the computational algorithm is used to verify other solutions available in the literature. Sect.~\ref{Sect:Conclusions} contains the final discussion and conclusions. Some additional information concerning the limiting cases of Newtonian and perfectly plastic fluids, together respective models of the approximation, is collected in the appendices.


\section{Problem formulation} \label{Sect:ProbForm}

Let us consider a 3D penny-shaped crack, defined in polar coordinates by the system $\{r,\theta,z\}$, with associated crack dimensions $\{l(t),w(t)\}$ as the fracture radius and aperture respectively, noting that both are a function of time. The crack is driven by a point source of power-law fluid located at the origin, and has a known pumping rate: $Q_0(t)$. The fluid's rheological properties are described by a power-law \cite{Cameron1989}. We have that, as the flow is axisymmetric, all variables will be independent of the angle $\theta$.

The fluid mass balance equation is as follows:
\begin{equation}
\frac{\partial w}{\partial t} + \frac{1}{r} \frac{\partial}{\partial r}\left( r q \right) + q_l = 0 , \quad 0<r<l(t),
\label{fluidmass}
\end{equation}
where $q_l (r,t)$ is the fluid leak-off function, representing the volumetric fluid loss to the rock formation in the direction perpendicular to the crack surface per unit length of the fracture. Throughout this paper we will assume it to be predefined and bounded at the fracture tip.

Meanwhile, $q(r,t)$ is the fluid flow rate inside the crack, given by the Poiseuille law:
\begin{equation}
q^n =-\frac{w^{2n+1}}{M^\prime}\frac{\partial p}{\partial r} ,
\label{Poiseville}
\end{equation}
with $p(r,t)$ being the net fluid pressure on the fracture walls (i.e. $p=p_f - \sigma_0$, $\sigma_0$ is the confining stress), while the constant $M^\prime$ is a modified fluid consistency index $M^\prime = 2^{n+1} (2n+1)^n / n^n M$, where $0\leq n \leq 1$ is the fluid behaviour index.

The non-local relationships between the fracture aperture and the pressure (elasticity equations) are as follows:
\begin{equation}
p(r,t) =  \frac{E^\prime}{l(t)} {\cal A}[w](r,t) , \quad w(r,t) = \frac{l(t)}{E^\prime} {\cal A}^{-1}[p](r,t) ,
\label{pressure1}
\end{equation}
where $E^\prime=Y/(1-\nu^2)$, with $Y$ being the Young's modulus and $\nu$ the Poisson ratio. The operator ${\cal A}$ and its inverse take the form:
\begin{equation}
{\cal A}[w] = - \int_0^1 \frac{\partial w(\eta l(t), t)}{\partial \eta} M\left[ \frac{r}{l(t)},\eta\right] \, d\eta ,
\label{aperture1}
\end{equation}
\begin{equation}
 \begin{aligned}
{\cal A}^{-1} [p] &=\frac{8}{\pi} \int_{r/l(t)}^1 \frac{\xi}{\sqrt{\xi^2 - (r/l(t))^2}} \int_0^1 \frac{\eta p(\eta \xi l(t),t)}{\sqrt{1-\eta^2}} \, d\eta \, d\xi &\\
&\equiv \frac{8}{\pi} \int_0^1 \eta p(\eta l(t), t) G\left[\frac{r}{l(t)} , \eta \right] \, d\eta \, , &
\label{aperture2}
 \end{aligned}
\end{equation}

for the pertinent kernels:
\begin{equation}
M\left[\xi , s\right] = \frac{1}{2\pi} \begin{cases} \frac{1}{\xi} \EllipticK{\frac{s^2}{\xi^2}} + \frac{\xi}{s^2 - \xi^2} \EllipticE{ \frac{s^2}{\xi^2}} , & \xi>s \\  \frac{s}{s^2 - \xi^2} \EllipticE{ \frac{\xi^2}{s^2}} , & s>\xi \end{cases}
\label{M1}
\end{equation}
\begin{equation}
G(\xi , s) = \begin{cases} \frac{1}{\xi} \IncEllipticF{\arcsin \left(\sqrt{\frac{1-\xi^2}{1-s^2}}\right) }{ \frac{s^2}{\xi^2}}   , & \xi>s \\ \frac{1}{s}  \IncEllipticF{\arcsin \left(\sqrt{\frac{1-s^2}{1-\xi^2}}\right) }{ \frac{\xi^2}{s^2} } , & s>\xi \end{cases}
\label{G1}
\end{equation}
$K$, $E$ are the complete elliptic integrals of the first and second kinds respectively, and $F$ the incomplete elliptic integral of the first kind, given in \cite{Abramowitz1972}.

These equations are supplemented by the boundary condition at $r=0$, which defines the intensity of the fluid source, $Q_0$:
\begin{equation}
\lim_{r \rightarrow 0} r q(r,t) = \frac{Q_0(t)}{2\pi} ,
\label{source1}
\end{equation}
the tip boundary conditions:
\begin{equation}
w(l(t),t)=0 , \quad q(l(t),t)=0 ,
\label{BC1}
\end{equation}
and appropriate initial conditions describing the starting crack opening and length:
\begin{equation}
w(r,0) = w_* (r) , \quad l(0) = l_0 .
\label{IC1}
\end{equation}

Additionally, it is assumed that the crack is in continuous mobile equilibrium, and as such the classical crack propagation criterion of linear elastic fracture mechanics is imposed:
\begin{equation}
K_I  = K_{I c} ,
\label{fracCrit}
\end{equation}
where $K_{I c}$ is the material toughness while $K_I$ is the stress intensity factor. The latter is computed according to the following formula \cite{Rice1968}: 
\begin{equation}
K_I (t) = \frac{2}{\sqrt{\pi l(t)}} \int_0^{l(t)} \frac{r p(r,t)}{\sqrt{l^2(t) - r^2}} \, dr .
\label{criterion1}
\end{equation}
Throughout this paper we accept the convention that when $K_{Ic}=0$ the hydraulic fracture propagates in the viscosity dominated regime. Otherwise the crack evolves in the toughness dominated mode. Each of these two regimes is associated with qualitatively different tip asymptotics, which constitutes a singular perturbation problem as $K_{Ic} \to 0$, and leads to serious computational difficulties in the small toughness range.

Finally, noting (\ref{fluidmass}) and (\ref{source1}), the global fluid balance equation is given by:
\begin{equation}
\int_0^{l(t)} r\left[ w(r,t) - w_0(r) \right] \, dr \, + \int_0^t \int_0^{l(t)} r q_l (r,\tau) \, dr \, d\tau = \frac{1}{2\pi} \int_0^t Q_0 (\tau) \, d\tau .
\label{fluidbalance1}
\end{equation}

The above set of equations and conditions represent the typically considered formulation for a penny-shaped hydraulic fracture \cite{Savitski2002}.

In order to facilitate the analysis we shall utilize an additional dependent variable, $v$, which describes the average speed of fluid flow through the fracture cross-section \cite{Mishuris2012}. It will be referenced to in the text as the particle velocity,
and is defined as:
\begin{equation}
v(r,t) = \frac{q(r,t)}{w(r,t)}  , \quad v^n (r,t) =  -\frac{1}{M^\prime} w^{n+1} \frac{\partial p}{\partial r}  .
\label{particlev1}
\end{equation}
Provided the fluid leak-off $q_l$ is finite at the crack tip, $v$ has the following property:
\begin{equation}
\lim_{r\rightarrow l(t)} v(r,t)  = v_0 (t) <\infty .
\label{v0infty}
\end{equation}
Additionally, given that the fracture apex coincides with the fluid front (no lag), and that the fluid leak-off at the fracture tip is weaker than the Carter law variant, the so-called \emph{speed equation} \cite{Linkov2011} holds:
\begin{equation}
\frac{d l}{dt} =v_0(t) .
\label{particlev2}
\end{equation}
This Stefan-type boundary condition constitutes an explicit level set method, as opposed to an implicit method \cite{Peirce2008}, and can be effectively used to construct a mechanism of fracture front tracing. The advantages of implementing such a condition have been shown in \cite{Wrobel2015,Perkowska2015,Linkov2016}.

\subsection{Problem normalization}

For the main body of the text, in order to make the presentation clearer, we will assume during derivations that $0<n<1$, however all results shown will be calculated according to their respective models. Any modification to the governing equations and numerical scheme in the limiting cases $n=0$ and $n=1$ are detailed in Appendix~\ref{Append:Cases}.

We normalize the problem by introducing the following dimensionless variables:
\begin{equation}
 \begin{aligned}
\tilde{r} = \frac{r}{l(t)} &, \quad \tilde{t}=\frac{t}{t_n^{1/n}}, \quad \tilde{w}(\tilde{r},\tilde{t})=\frac{w(r,t)}{l_*}  , \quad L(\tilde{t})=\frac{l(t)}{l_*} , \quad \tilde{q}_l (\tilde{r}, \tilde{t}) = \frac{t_n^{1/n}}{l_*} q_l (r,t) ,& \\
&\tilde{q}(\tilde{r},\tilde{t}) = \frac{t_n^{1/n}}{l_*^2} q(r,t) , \quad \tilde{Q}_0(\tilde{t})=\frac{t_n^{1/n}}{l_*^2 l(t)}Q_0(t), \quad \tilde{v}(\tilde{r},\tilde{t})=\frac{t_n^{1/n}}{l_*}v(r,t), & \\
& \tilde{p}(\tilde{r},\tilde{t}) = \frac{t_n}{M^\prime} p(r,t) , \quad \tilde{K}_{Ic} = \frac{1}{E^\prime \sqrt{l_*}}K_{Ic} , \quad t_n=\frac{M^{\prime}}{E^\prime} , &
\end{aligned}
\label{Normalizations1}
\end{equation}
where $\tilde{r}\in \left[0,1\right]$ and $l_*$ is chosen for convenience.

We note that such a normalization scheme has previously been used in \cite{Wrobel2015,Perkowska2015,Linkov2016}, and that it is not attributed to any particular influx regime or asymptotic behaviour of the solution.\\

Under normalization scheme \eqref{Normalizations1}, the continuity equation \eqref{fluidmass} can be rewritten in terms of the particle velocity \eqref{particlev1} to obtain:
\begin{equation}
\frac{\partial \tilde{w}}{\partial \tilde{t}} - \frac{L^\prime (\tilde{t})}{L(\tilde{t})} \tilde{r} \frac{\partial \tilde{w}}{\partial \tilde{r}} + \frac{1}{L(\tilde{t})\tilde{r}} \frac{\partial}{\partial \tilde{r}}\left( \tilde{r}\tilde{w}\tilde{v}\right) + \tilde{q}_l = 0 .
\label{fluidmassN2}
\end{equation}
The particle velocity \eqref{Poiseville} is expressed as:
\begin{equation}
\tilde{v}= \left[ -\frac{\tilde{w}^{n+1}}{L(\tilde{t})}\frac{\partial \tilde{p}}{\partial \tilde{r}}  \right]^{\frac{1}{n}} ,
\label{particlevN1}
\end{equation}
while the speed equation is now given by combining \eqref{particlev1}-\eqref{particlev2}:
\begin{equation}
\tilde{v}_0 (\tilde{t}) = L'(\tilde{t}) = \left[ - \frac{\tilde{w}^{n+1}}{L(\tilde{t})}\frac{\partial \tilde{p}}{\partial \tilde{r}} \right]^{\frac{1}{n}}_{\tilde{r}=1} < \infty .
\label{particlevN2}
\end{equation}
The global fluid balance equation (\ref{fluidbalance1}) is transformed to:
\begin{equation}
\begin{aligned}
&\int_0^1 \tilde{r} \left[ L^2(\tilde{t})\tilde{w}(\tilde{r},\tilde{t}) - L^2(0)\tilde{w}_0(\tilde{r}) \right] \, d\tilde{r} + \int_0^{\tilde{t}} \int_0^1 \tilde{r} L^2 (\tau) \tilde{q}_l (\tilde{r},\tau) \, d\tilde{r} \, d\tau &\\
&\quad = \frac{1}{2\pi} \int_0^{\tilde{t}} L(\tau) \tilde{Q}_0 (\tau) \, d\tau . &
 \end{aligned}
\label{fluidbalanceN1}
\end{equation}
The notation for the elasticity equations  \eqref{pressure1}-\eqref{aperture2} takes the form:
\begin{equation}
\tilde{p}(\tilde{r},\tilde{t}) = \frac{1}{L(\tilde{t})} {\cal A} [\tilde{w}](\tilde r,\tilde t),\quad
\tilde{w}(\tilde{r},\tilde{t}) = L(\tilde{t}) {\cal A}^{-1} [\tilde{p}](\tilde r,\tilde t),
\label{pressureN1}
\end{equation}
where the operators denote:
\begin{equation}
{\cal A}[\tilde{w}](\tilde{r},\tilde{t})= -\int_0^1 \frac{\partial \tilde{w}(\eta,\tilde{t})}{\partial \eta} M\left[ \tilde{r},\eta\right] \, d\eta,\quad
\label{apertureN1}
\end{equation}
\begin{equation}
{\cal A}^{-1}[\tilde p](\tilde{r},\tilde{t})= \frac{8}{\pi}\int_{\tilde{r}}^1 \frac{\xi}{\sqrt{\xi^2 - \tilde{r}^2}} \int_0^1 \frac{\eta \tilde{p}(\eta \xi,\tilde{t})}{\sqrt{1-\eta^2}} \, d\eta \, d\xi .
\label{apertureN2}
\end{equation}

From definition \eqref{criterion1} and the fracture propagation condition \eqref{fracCrit} we have that:
\begin{equation}
\tilde{K}_I = \tilde{K}_{Ic} =  \, \frac{2}{\sqrt{\pi}} \sqrt{L(\tilde{t})} \int_0^1 \frac{\tilde{r} \tilde{p}(\tilde{r},\tilde{t})}{\sqrt{1-\tilde{r}^2}} \, d\tilde{r}  .
\label{criterionN1}
\end{equation}
Note that through proper manipulation of \eqref{apertureN2} and the use of \eqref{criterionN1}, \eqref{pressureN1}$_2$  can be expressed in the following form:
\begin{equation}
\tilde{w}(\tilde{r},\tilde{t}) = \frac{8}{\pi}L(\tilde{t}) \int_0^1 \frac{\partial \tilde{p}}{\partial y}(y,\tilde{t}) {\cal K}(y,\tilde{r}) \, dy  + \frac{4}{\sqrt{\pi}}\sqrt{L(\tilde{t})}\tilde{K}_I \sqrt{1-\tilde{r}^2}  ,
\label{apertureN3}
\end{equation}
for the kernel function ${\cal K}$ given by:
\begin{equation}
\mathcal{K}(y,\tilde{r}) = y\left[  \IncEllipticE{\arcsin\left(y\right)}{\frac{\tilde{r}^2}{y^2}} -  \IncEllipticE{\arcsin\left(\chi \right)} {\frac{\tilde{r}^2}{y^2}} \right] ,
\label{Kernel1}
\end{equation}
where:
\begin{equation}
\chi = \min\left(1,\frac{y}{\tilde{r}}\right) ,
\label{chi1}
\end{equation}
with the function $\IncEllipticE{\phi}{m}$ denoting the incomplete elliptic integral of the second kind \cite{Abramowitz1972}.\\

While this form of the elasticity operator has not previously been used in the case of a penny-shaped fracture,
an analogous form of the elasticity equation for the KGD model has been utilized in \cite{Wrobel2015,Perkowska2015}, where its advantages in numerical computations have been demonstrated. Notably, the kernel function ${\cal K}$ exhibits better behaviour than the weakly singular kernel $G$ \eqref{G1}, having no singularities for any combination of $\left\{\tilde{r},y\right\}$. Additionally, equation \eqref{particlevN1} can be easily transformed to obtain $p^\prime$ and then substituted into \eqref{apertureN3}, meaning that the latter can be utilized without the additional step of deriving the pressure function needed for the classic form of the operator.


Next the boundary conditions \eqref{BC1}, in view of \eqref{v0infty}, transform to a single condition:
\begin{equation}
\tilde{w}(1,\tilde{t})=0, 
\label{BCN1}
\end{equation}
alongside the initial conditions \eqref{IC1}:
\begin{equation}
\tilde{w}(\tilde{r},0) = \frac{w_*(r)}{l_*} , \quad L_0=\frac{l_0}{l_*} .
\label{ICN1}
\end{equation}
The source strength (\ref{source1}) is now defined as:
\begin{equation}
\frac{\tilde{Q}_0(\tilde{t})}{2\pi}=\lim_{\tilde{r}\rightarrow 0}\tilde{r}\tilde{w}(\tilde{r},\tilde{t})\tilde{v}(\tilde{r},\tilde{t}) .
\label{sourceN1}
\end{equation}
While combining the above with \eqref{particlevN1} we obtain the following relationship:
\begin{equation}
\lim_{\tilde{r}\rightarrow 0}\tilde{r}^n \frac{\partial \tilde{p}}{\partial \tilde{r}} =- \left(\frac{\tilde{Q}_0 (\tilde{t})}{2\pi} \right)^{n} \frac{L (\tilde{t})}{\tilde w^{2n+1} (0,\tilde{t})} ,
\label{sourceN2}
\end{equation}
which provides a valuable insight into how the behaviour of the fluid pressure function near to the source varies for differing values of $n$. The resulting pressure asymptotics at the injection point, with corresponding aperture,  are detailed below:
 \begin{equation}
 \tilde{p}(\tilde{r},\tilde{t}) = \tilde{p}_0^o (\tilde{t}) + \tilde{p}_1^o (\tilde{t}) \tilde{r}^{1-n} +  O\left(\tilde{r}^{2-n} \right) , \quad \tilde{r}\to 0 ,
 \label{presAsym1}
 \end{equation}
  \begin{equation}
 \tilde{w}(\tilde{r},\tilde{t}) = \tilde{w}_0^o (\tilde{t}) +\tilde{w}_1^o (\tilde{t}) \tilde{r}^{2-n} + O\left(\tilde{r}^2 \log(\tilde{r})\right) , \quad \tilde{r}\to 0 .
 \label{wAsym02}
\end{equation}

It is worth restating that there are minor differences to both the asymptotics and fundamental equations in the limiting cases $n=0$ and $n=1$. These are explained in further detail in Appendix~\ref{Append:Cases}.

\section{Crack tip asymptotics, the speed equation and proper variables} \label{TipVar}

A universal algorithm for numerically simulating hydraulic fractures has recently been introduced in \cite{Wrobel2015,Perkowska2015} and tested against the PKN and KGD (plane strain) models for Newtonian and shear-thinning fluids. It proved to be extremely efficient and accurate. Its modular architecture enables one to adapt it to other HF models by simple replacement or adjustment of the basic blocks. In the following we will construct a computational scheme for the radial fracture based on the universal algorithm. To this end we need to introduce appropriate computational variables, and to define the basic asymptotic interrelations between them. For the sake of completeness detailed information on the solutions tip asymptotic behaviour, for different regimes of crack propagation, are presented below.

\subsection{Crack tip asymptotics}

\subsubsection{Viscosity dominated regime ($\tilde{K}_{Ic} = 0$)}

In the viscosity dominated regime the crack tip asymptotics of the aperture and pressure derivative can be expressed as follows:
\begin{equation}
 \begin{aligned}
\tilde w(\tilde r,\tilde t) &=\tilde w_0(\tilde t)  \left(1-\tilde{r}^2 \right)^{\alpha_0} + \tilde w_1(\tilde t)\left(1-\tilde r^2 \right)^{\alpha_1}+ \tilde w_2(\tilde t)\left(1-\tilde r^2 \right)^{\alpha_2}&\\
& \quad + O\left(\left(1-\tilde{r}^2 \right)^{\alpha_2 + \delta}\right) , \quad \tilde r \to 1, &
 \end{aligned}
\label{apertureasymp1_otherb}
\end{equation}
\begin{equation}
\frac{\partial \tilde{p}}{\partial \tilde{r}} (\tilde{r}, \tilde{t}) = \tilde{p}_0 (\tilde{t}) \left(1-\tilde{r}^2\right)^{\alpha_0-2} + \tilde{p}_1 (\tilde{t}) \left(1-\tilde{r}^2\right)^{\alpha_0-1} + O\left(1 \right)  , \quad \tilde{r}\to 1 .
\label{dp_asym_1}
\end{equation}
The crack tip asymptotics of the pressure function can be derived from the above, however this form is given due to its use in computations (this will be explained in further detail later).

As a consequence the particle velocity behaves as:
\begin{equation}
\tilde v (\tilde r , \tilde t ) = \tilde v_0 (\tilde t )  + \tilde v_1 \left(\tilde t \right) \left(1-\tilde{r}^2 \right)^{\beta_1} + O\left(\left(1-\tilde{r}^2 \right)^{\beta_2}\right) , \quad \tilde r \to 1 . 
\label{v_1_asymp}
\end{equation}
Note that we require $\tilde v_0 (\tilde t )>0$ to ensure the fracture is moving forward. The values of constants $\alpha_i$, $\beta_i$ are given in Table~\ref{table:albe}. The general formulae for the limiting cases $n=0$ and $n=1$ remain the same as \eqref{apertureasymp1_otherb}-\eqref{v_1_asymp}, with the respective powers $\alpha_i$, $\beta_i$ again being determined according to Table~\ref{table:albe}.


Now, let us adopt the following notation for the crack propagation speed, based on the \emph{speed equation} \eqref{particlevN2} and the tip asymptotics \eqref{v_1_asymp}:
\begin{equation}
\tilde v_0 (\tilde t) = L^\prime (\tilde t) = \left[ \frac{{\cal C} {\cal L}(\tilde w )}{L^2 ( \tilde t )} \right]^{\frac{1}{n}} .
\label{particlevN3}
\end{equation}
Here ${\cal{L}}(\tilde{w})>0$ is a known functional and ${\cal{C}}$ is a positive constant. In the viscosity dominated regime we have that: 
 \begin{equation}
 C = \frac{2n}{(n+2)^2} \cot \left( \frac{n\pi}{n+2}\right) , \quad {\cal{L}} (\tilde{w}) = \tilde{w}_0^{n+2} .
 \label{w0v0_1}
 \end{equation}
Additionally, we can directly integrate \eqref{particlevN3} in order to obtain an expression for the fracture length:
\begin{equation}
L(\tilde{t}) = \left[ L^{1+\frac{2}{n}}(0) + \left(1+\frac{2}{n}\right)\mathcal{C}^{\frac{1}{n}} \int_0^{\tilde{t}} \mathcal{L}^{\frac{1}{n}} (\tilde{w}) \, d\tau \right]^{\frac{n}{n+2}} .
\label{fraclengthN2}
\end{equation}

\begin{table}[t]
 \centering
 \begin{tabular}{|c|c|c|c|c|c|}
 \hline
 Crack propagation regime & $\alpha_0$ & $\alpha_1$ & $\alpha_2$ & $\beta_1$ & $\beta_2$ \\[0.5ex]
\hline \hline
&&&&&\\
Viscosity dominated & $\dfrac{2}{n+2}$ & $\dfrac{n+4}{n+2}$ & $\dfrac{2n+6}{n+2}$ & $1$ & $\dfrac{2n+2}{n+2}$ \\[9pt]
\hline
&&&&&\\
Toughness dominated & $\dfrac{1}{2}$ & $\dfrac{3-n}{2}$ & $\dfrac{5-2n}{2}$ & $\dfrac{2-n}{2}$ & $1$ \\[9pt]
\hline
\end{tabular}
 \caption{Values of the basic constants used in the asymptotic expansions for $\tilde{w}$ and $\tilde{v}$ for $0<n<1$.}
\label{table:albe}
\end{table}

\subsubsection{Toughness dominated regime ($\tilde{K}_{Ic} > 0$)}

Near the fracture front the form of the aperture and particle velocity asymptotics remains the same as in the viscosity dominated regime \eqref{apertureasymp1_otherb}, \eqref{v_1_asymp}. Meanwhile the pressure derivative asymptotics yields:
\begin{equation}
\frac{\partial \tilde{p}}{\partial \tilde{r}} (\tilde{r}, \tilde{t}) = \tilde{p}_0 \left(1-\tilde{r}^2\right)^{\alpha_1-2} + \tilde{p}_1 \left(1-\tilde{r}^2\right)^{\alpha_2-2} + O\left(1 \right)  , \quad \tilde{r}\to 1 .
\label{dp_asym_2}
\end{equation}
The values of $\alpha_i$, $\beta_i$ for this regime are provided in Table~\ref{table:albe}. The asymptotics in the limiting cases $n=0$ and $n=1$ is given in Appendix~\ref{Append:Cases} (equations \eqref{apertureasymp1_Tn0} and \eqref{apertureasymp1_Tn1} respectively).

We again use notation \eqref{particlevN3} for the crack propagation speed, however the values of the functional ${\cal L}$ and the $C$ will in this case be:
 \begin{equation}
C= \frac{(3-n)(1-n)}{4} \tan \left(\frac{n\pi}{2}\right)  , \quad {\cal {L}}(\tilde{w}) = \tilde{w}_0^{n+1} \tilde{w}_1 ,
\label{w1v0_1}
 \end{equation}
while the fracture length will be given by \eqref{fraclengthN2}.

\subsection{Reformulation in terms of computational variables}


It is readily apparent that the choice of computational variables plays a decisive role in ensuring the accuracy and efficiency of the computational algorithm \cite{Mishuris2012,Wrobel2015,Kusmierczyk}. Let us introduce a new system of proper variables which are conducive to robust numerical computing.

\begin{itemize}
 \item The reduced particle velocity $\Phi(\tilde{r},\tilde{t})$:
\begin{equation}
\Phi(\tilde r,\tilde t)=\tilde r\tilde v(\tilde r,\tilde t)-\tilde r^{2} \tilde v_0(\tilde t) .
\label{new_V}
\end{equation}
It is a smooth, well behaved and non-singular variable that facilitates the numerical computations immensely. It is bounded at the crack tip and the fracture origin. The advantages of using an analogous variable in the PKN and  KGD models have previously been demonstrated in \cite{Wrobel2015,Perkowska2015}.
 \item The modified pressure derivative $\Omega(\tilde{r},\tilde{t})$:
\begin{equation}
\tilde r^n \Omega(\tilde r,\tilde t)=\tilde r^n \frac{\partial \tilde p}{\partial \tilde r}-\Omega_0(\tilde t),
\label{new_P}
\end{equation}
\begin{equation}
\Omega_0 (\tilde{t}) = - \left(\frac{\tilde{Q}_0 (\tilde{t})}{2\pi} \right)^{n} \frac{L (\tilde{t})}{\tilde w^{2n+1} (0,\tilde{t})}.
\label{Omega0}
\end{equation}
It reflects the singular tip behavior of $\tilde p'_{\tilde{r}}$, having the same tip asymptotics as the pressure derivative, however it is bounded at the fracture origin.
From \eqref{new_P} the pressure can be immediately reconstructed as:
\begin{equation}
\tilde p(\tilde r,\tilde t)=  \frac{\Omega_0(\tilde t)}{1-n} \tilde{r}^{1-n}+C_p(\tilde t)+\int_0^{\tilde r}\Omega(\xi , \tilde{t} )d\xi  ,
\label{new_P2}
\end{equation}
where the term $C_p$ follows from \eqref{criterionN1}:
\begin{equation}
C_p (\tilde{t}) = \frac{1}{2}\sqrt{\frac{\pi}{L(\tilde{t})}} \tilde{K}_I - \frac{\sqrt{\pi}\Gamma\left(\frac{3-n}{2}\right)}{2\left(1-n\right)\Gamma\left(2-\frac{n}{2}\right)} \Omega_0 (\tilde{t})  - \int_0^1 \Omega(y , \tilde{t})\sqrt{1-y^2} \, dy .
\label{Cpt}
\end{equation}
This auxiliary variable will primarily be used in numerical computation of the elasticity operator.
\end{itemize}

The following interrelationship exists between the newly introduced variables:
\begin{equation}
\Omega(\tilde{r},\tilde{t}) = \left(\frac{\tilde{Q}_0 (\tilde{t})}{2\pi \tilde{r}} \right)^{n} \frac{L(\tilde{t})}{\tilde{w}^{2n+1} (0,\tilde{t})} - \frac{L(\tilde{t})}{\tilde{w}^{n+1}(\tilde{r},\tilde{t})}\left[\frac{\Phi(\tilde{r},\tilde{t})}{\tilde{r}} + \tilde{r} \tilde{v}_0(\tilde{t}) \right]^{n} .
\label{relationPV}
\end{equation}

Since under this new scheme $\Phi$ is bounded at the fracture tip, the source strength \eqref{sourceN1} and the boundary condition \eqref{BCN1} can now be expressed as:
\begin{equation}
 \tilde w (0,\tilde t) \Phi (0 , \tilde t ) = \frac{\tilde Q_0 (\tilde t)}{2\pi} , \quad \tilde w (1, \tilde t ) = 0 .
 \label{sourceK1}
\end{equation}
By utilizing the boundary condition \eqref{sourceK1}$_1$, the relationship between the new variables \eqref{relationPV} can be represented in the form:
\begin{equation}
\Omega\left(\tilde{r},\tilde{t}\right) = \frac{1}{\tilde{r}^n} \left[ \frac{ \Phi^n (0,\tilde{t})}{\tilde{w}^{n+1} (0,\tilde{t})} - \frac{\left(\Phi (\tilde{r},\tilde{t})+\tilde{r}^2 \tilde{v}_0  (\tilde{t})  \right)^n }{\tilde{w}^{n+1} (\tilde{r},\tilde{t})} \right] .
\label{relationPV2}
\end{equation}
Note that this is not only a more concise representation of \eqref{relationPV} but also does not depend on $L(\tilde{t})$, which will be beneficial when computing the self-similar formulation. In this way the computational scheme will be based on: the crack opening, $\tilde w$, the reduced particle velocity, $\Phi$, and an auxiliary variable, the modified fluid pressure, $\Omega$.

By substituting the new variable $\Phi$ from \eqref{new_V} into the continuity equation \eqref{fluidmassN2}, we obtain the modified governing equation:
\begin{equation}
\frac{\partial \tilde{w}}{\partial \tilde{t}} + \frac{1}{L(\tilde{t})\tilde{r}} \frac{\partial}{\partial \tilde{r}}\left(\tilde{w} \Phi\right) + \frac{2 \tilde{v}_0}{L(\tilde{t})} \tilde{w} + \tilde{q}_l  = 0 , \quad 0<\tilde{r}<1 .
\label{fluidmassK1}
\end{equation}

Additionally, the elasticity equation \eqref{apertureN3} can be now expressed as follows:
\begin{equation}
\tilde{w}(\tilde{r},\tilde{t}) = \frac{8}{\pi}L(\tilde{t})
\int_0^1 \Omega(y,\tilde{t}) \mathcal{K}(y,\tilde{r}) \, dy +\frac{4}{\sqrt{\pi}}\sqrt{L(\tilde{t})} \tilde{K}_I \sqrt{1-\tilde{r}^2} + \frac{8}{\pi}L(\tilde{t}) \Omega_0 (\tilde{t}) \mathcal{G}_n (\tilde{r}) ,
\label{newaperture5}
\end{equation}
where ${\cal K}$ is given in \eqref{Kernel1}, while 
$\mathcal{G}_n$ is defined by:
\begin{equation}
{\cal G}_n (\tilde{r}) = \frac{\sqrt{\pi} \Gamma\left(\frac{3-n}{2}\right)}{2\left(n-1\right)\Gamma\left(2-\frac{n} {2}\right)}\left[ \sqrt{1-\tilde{r}^2} + \frac{{_2F_1}\left(\frac{1}{2},\frac{n-2}{2};\frac{n}{2};\tilde{r}^2\right)}{n-2} - \frac{\sqrt{\pi}\tilde{r}^{2-n} \Gamma\left(\frac{n}{2}-1\right)}{2\Gamma\left(\frac{n-1}{2}\right)}\right] .
\label{aux1}
\end{equation}
It can be easily shown that this function is well behaved in the limits.

\section{Self-similar formulation} \label{Sect:SelfSim}

In this section we will reduce the problem to its time-independent self-similar version. This formulation will be used to define the computational scheme used later on in the numerical analysis.

We begin by assuming that a solution to the problem can be expressed in the following manner:
\begin{equation}
\tilde{w}(\tilde{r},\tilde{t})=\psi(\tilde{t})\hat{w}(\tilde{r}) , \quad \tilde{p}(\tilde{r},\tilde{t})=\frac{\psi(\tilde{t})}{L(\tilde{t})}\hat{p}(\tilde{r}), \quad \tilde{q}(\tilde{r},\tilde{t})=\frac{\psi^{2+\frac{2}{n}}(\tilde{t})}{L^{\frac{2}{n}}(\tilde{t})}\hat{q}(\tilde{r}),
\notag
\end{equation}
\begin{equation}
 \tilde{Q}_0 (\tilde{t}) = \frac{\psi^{2+\frac{2}{n}} (\tilde{t})}{L^{\frac{2}{n}}(\tilde{t})} \hat{Q}_0, \quad \tilde{v}(\tilde{r},\tilde{t})=\frac{\psi^{1+\frac{2}{n}}(\tilde{t})}{L^{\frac{2}{n}}(\tilde{t})}\hat{v}(\tilde{r}) , \quad \Phi(\tilde{r},\tilde{t})=\frac{\psi^{1+\frac{2}{n}}(\tilde{t})}{L^{\frac{2}{n}}}\hat{\Phi}(\tilde{r}) ,
\notag
\end{equation}
\begin{equation}
\begin{aligned}
& \tilde{K}_I(\tilde{t}) = \frac{\psi(\tilde{t})}{\sqrt{L(\tilde{t})}} \hat{K}_I, \quad &\Omega(\tilde{r},\tilde{t}) = \frac{\Psi(\tilde{t})}{L(\tilde{t})} \hat{\Omega}(\tilde{r}) ,\\
& \Omega_0 (\tilde{t}) = \frac{\Psi(\tilde{t})}{L(\tilde{t})} \hat{\Omega}_0 , \quad &C_p (\tilde{t}) = \frac{\Psi(\tilde{t})}{L(\tilde{t})} \hat{C}_p , &
\label{SelfSimilar1}
 \end{aligned}
\end{equation}
where $\Psi(t)$ is a smooth continuous function. By separating the variables in this manner it becomes possible to reduce the problem to a time-independent formulation when $\Psi$ is described by an exponential or a power-law type function. From here on the spatial components will be marked by a 'hat'-symbol, and will describe the self-similar quantities. It is worth noting that the separation of spatial and temporal components given in \eqref{SelfSimilar1} ensures that the qualitative bahaviour of the solution tip asymptotics remains the same as in the time-dependent variant.

\subsection{The self-similar representation of the problem}

We wish to examine two variants of the time dependent function:
\begin{equation}
 \Psi_1 (\tilde t) = e^{\gamma \tilde t} , \quad \Psi_2 (\tilde t ) = \left(a + \tilde t \right)^{\gamma} .
\label{casesS1}
\end{equation}
In both cases  the fluid leak-off function will be assumed to take the form:
\begin{equation}
\tilde q_l (\tilde r , \tilde t ) = \frac{1}{\gamma} \Psi^\prime (\tilde t ) \hat q_l (\tilde r ) .
\label{leakoffS1}
\end{equation}
The self-similar reduced particle velocity \eqref{new_V}, modified pressure derivative \eqref{new_P}, \eqref{Omega0} and pressure \eqref{new_P2} are defined by:
\begin{equation}
\hat{\Phi}(\tilde{r})= \tilde{r} \hat{v}(\tilde{r}) - \tilde{r}^2 \hat{v}_0 , \quad \tilde{r}\hat{\Omega}(\tilde{r}) = \tilde{r}\frac{d\hat{p}}{d\tilde{r}}-\hat{\Omega}_0 ,
\label{SSdef}
\end{equation}
\begin{equation}
\hat p(\tilde r)=  \frac{\hat{\Omega}_0}{1-n} \tilde{r}^{1-n}+\hat{C}_p+\int_0^{\tilde r} \hat{\Omega}(\xi )d\xi  ,
\end{equation}
with
\begin{equation}
\hat{\Omega}_0 = - \left(\frac{\hat{Q}_0}{2\pi} \right)^{n} \frac{1}{\hat w^{2n+1} (0)} ,
\label{SSOmega0}
\end{equation}
\begin{equation}
\hat{C}_p = \frac{\sqrt{\pi}}{2}\hat{K}_I - \frac{\sqrt{\pi}\Gamma\left(\frac{3-n}{2}\right)}{2\left(1-n\right)\Gamma\left(2-\frac{n}{2}\right)} \hat{\Omega}_0   - \int_0^1 \hat{\Omega} (y)\sqrt{1-y^2} \, dy .
\end{equation}
It is immediately apparent from (\ref{particlevN3}) and (\ref{SelfSimilar1}) that the self-similar crack propagation speed is given by:
\begin{equation}
\hat v_0 =  \lim_{\tilde{r}\rightarrow 1} \left[ - \hat{w}^{n+1} \frac{d \hat{p}}{d \tilde{r}}\right]^{\frac{1}{n}} = \left( \mathcal{C}\mathcal{L}(\hat{w}) \right)^{\frac{1}{n}} .
\label{particlevS1}
\end{equation}
Note again that the qualitative asymptotic behaviour of the aperture, pressure and particle velocity as $\tilde{r}\to 0$ and $\tilde{r}\to 1$ remains the same as in the time dependent version of the problem \eqref{apertureasymp1_otherb}, \eqref{dp_asym_1}, \eqref{v_1_asymp}, \eqref{dp_asym_2}. The respective asymptotic formulae hold provided that multipliers of the spatial terms are constant rather than being functions of time.

The self-similar counterparts of the elasticity equations \eqref{pressureN1} and \eqref{apertureN1} are now:
\begin{equation}
\hat p ( \tilde r ) = \hat {\cal A} [\hat w ] (\tilde r ) ,
\label{pressureS1}
\end{equation}
where:
\begin{equation}
\hat {\cal A} [\hat w ] ( \tilde r ) = - \int_0^1 \frac{d \hat w ( \eta )}{d \eta } M \left[ \tilde r , \eta \right] \, d\eta  ,
 \label{apertureS1}
\end{equation}
with its inverse being:
\begin{equation}
\hat{w}(\tilde{r}) = \frac{8}{\pi} \int_0^1 \hat{\Omega}(y) \mathcal{K}(y,\tilde{r}) \, dy + \frac{4}{\sqrt{\pi}} \hat{K}_I \sqrt{1-\tilde{r}^2} + \frac{8}{\pi} \hat{\Omega}_0 \mathcal{G}_n (\tilde{r}).
\label{newapertureS2}
\end{equation}
As the integral and function ${\cal G}_n (\tilde{r})$ both tend to zero faster than the square root term at the fracture tip, it immediately follows that, in the toughness dominated case ($\hat{K}_{Ic}>0$), the leading asymptotic term of the aperture \eqref{apertureasymp1_otherb} is given by:
\begin{equation}
\hat{w}_0 = \frac{4}{\sqrt{\pi}}\hat{K}_I . 
\label{w0asym1}
\end{equation}
The self-similar particle velocity  takes the form:
\begin{equation}
\hat{v}(\tilde{r}) =  \left[- \hat{w}^{n+1} (\tilde{r}) \frac{d \hat{p}(\tilde{r})}{d \tilde{r}} \right]^{\frac{1}{n}} .
\label{PoisevilleS1}
\end{equation}
The governing equation \eqref{fluidmassK1} becomes:
\begin{equation}
\frac{1}{\tilde r \hat{v}_0} \frac{d}{d\tilde r} \left( \hat w \hat \Phi \right) = -\left( 3 - \rho \right)\hat{w} - \left(1-\rho\right) \frac{\hat{q}_l}{\gamma},
\label{fluidmassS1}
\end{equation}
with the value of $\rho$ in each case, alongside the fracture length, provided in Table~\ref{table:S1}. Meanwhile the fluid balance condition \eqref{fluidbalanceN1} becomes:
\begin{equation}
\left(3-\rho\right)\int_0^1 \tilde r \hat w ( \tilde r ) \, d \tilde r + \frac{1-\rho}{\gamma} \int_0^1 \tilde r \hat q_l \ d\tilde r = \frac{\hat Q_0}{2\pi \hat{v}_0} .
\label{fluidbalanceS1}
\end{equation}

\begin{table}[t]
 \centering
\begin{tabular}{||c|c|c||}
\hline
Self-similar law & $\rho$ & $L(\tilde t)$  \\ [0.5ex]
\hline
&&\\
$\Psi (\tilde t ) = e^{\gamma \tilde t}$ & 0 & $\left[ \frac{\hat{v}_0}{\gamma} \right]^{\frac{n}{n+2}} e^{\gamma \tilde t }$ \\[2mm]
\hline
&&\\
$\Psi (\tilde t ) = \left(a + \tilde t \right)^{\gamma}$ & $\frac{n}{\gamma\left(n+2\right)+n}$ & $\left[\frac{\left(n+2\right)\hat{v}_0}{\gamma\left(n+2\right)+n}\right]^{\frac{n}{n+2}} \left(a+\tilde{t}\right)^{\gamma+\frac{n}{n+2}}$ \\[2mm]
\hline
\end{tabular}
 \caption{Table providing the fracture length $L(\tilde t)$, which is obtained using \eqref{fraclengthN2} and \eqref{particlevS1}, and the constant $\rho$, used in \eqref{fluidmassS1} and \eqref{fluidbalanceS1}, for different variants of the self-similar solution.}
\label{table:S1}
\end{table}

The self-similar stress intensity factor \eqref{criterionN1} is given by:
\begin{equation}
\hat{K}_I = \hat{K}_{Ic} =  \,  \frac{2}{\sqrt{\pi}} \int_0^1 \frac{\tilde{r}\hat{p}(\tilde{r})}{\sqrt{1-\tilde{r}^2}} \, d\tilde{r} .
\label{criterionS1}
\end{equation}

Finally, the system's boundary conditions \eqref{sourceK1} transform to:
\begin{equation}
 \hat w (0) \hat \Phi (0 ) = \frac{\hat Q_0 }{2\pi} , \quad \hat w (1 ) = 0 .
\label{BCS1}
\end{equation}

In the general case with $0<n<1$ these equations represent the full self-similar problem. Some modifications are necessary in the special cases when $n=0$ and $n=1$. These differences are outlined in Appendix~\ref{Append:Cases}.

\section{Numerical results} \label{Sect:NumBench}


In this section we will construct an iterative computational scheme for numerically simulating hydraulic fracture. The approach is an extension of the universal algorithm introduced in \cite{Wrobel2015,Perkowska2015}. The computations are divided between two basic blocks, the first of which utilizes the continuity equation and the latter using the elasticity operator. The previously introduced computational variables, alongside the known information about the solution tip asymptotics, are employed extensively. The accuracy and efficiency of the computations are verified against the newly introduced analytical benchmark examples. Then the numerical benchmark solutions are given. Finally, a comparative analysis with other data available in the literature is delivered.

\subsection{Computational scheme}

The algorithm is constructed using the approach framework introduced for the PKN and KGD models in \cite{Wrobel2015,Perkowska2015}. The numerical scheme is realized as follows:
\begin{enumerate}
 \item An initial approximation of the aperture $\hat{w}=\hat{w}^{j-1}$ is taken, such that it has the correct asymptotic behaviour and satisfies the boundary conditions.
 \item The fluid balance equation \eqref{fluidbalanceS1} is utilized to obtain the asymptotic term(s) $\hat{w}_{0,1}^{j}$  needed to compute the particle velocity $\hat{v}_0^{j}$ using \eqref{particlevS1}.
  \item Having the above values the reduced particle velocity $\hat{\Phi}^{j}$ is reconstructed by direct integration of \eqref{fluidmassS1}. Tikhonov type regularization is employed at this stage.
 \item Equation \eqref{PoisevilleS1} is then used to obtain an approximation of the modified pressure derivative $\hat{\Omega}$, and the elasticity equation 
\eqref{newapertureS2} serves to compute the next approximation of the fracture aperture $\hat{w}^j$.
 \item The system is iterated until all variables $\hat{\Phi}$, $\hat{w}$ and $\hat{v}_{0}$ converge to within prescribed tolerances.
\end{enumerate}
We will demonstrate in this section that this scheme, combined with an appropriate meshing strategy, yields a highly accurate algorithm. A more detailed description of the algorithm's construction has been outlined in \cite{Wrobel2015,Perkowska2015}.

It is worth noting that, due to the degeneration of the Poiseuille equation when $n=0$, it can no longer be used to compute the fluid flow rate or the particle velocity. However, thanks to the modular structure of the proposed algorithm, one can easily adapt it to this variant of the problem. In this case a special form of the elasticity equation \eqref{AperturePlastic} is utilized to obtain the aperture, with the particle velocity being reconstructed using relations \eqref{v0Plastic} and \eqref{PVPlastic}.

\subsection{Accuracy of computations} \label{Sect:Acc1}


In this subsection we will investigate the accuracy of computations delivered by the proposed numerical scheme. To this end a newly introduced set of analytical benchmark solutions with a non-zero fluid leak-off function will be used. Alternative measures for testing the numerical accuracy in the absence of exact solutions will then be proposed and analysed. Next, the problem of a penny-shaped hydraulic fracture propagating in an impermeable material will be considered. The accuracy of numerical solutions will be verified by the aforementioned alternative measures. Simple, semi-analytical approximations, which mimic the obtained numerical data to a prescribed level of accuracy, will be provided. Finally, a comparative analysis with other solutions available in the literature will be performed.

\subsubsection{Analysis of computational errors against analytical benchmarks} \label{Sect:Acc2}

The first method of testing the computational accuracy is by comparison with analytical benchmark solutions. Respective closed-form benchmarks with predefined, non-zero, leak-off functions are outlined in Appendix.~\ref{App:AnalBench}. They have been constructed for both the viscosity and toughness dominated regimes, for a class of shear-thinning and Newtonian fluids.
All of the analytical benchmarks used for comparison are designed to ensure physically realistic behaviour of the solution while maintaining the proper asymptotic behaviour. In all numerical simulations the power-law variant of the time dependent function $\Psi_2$ \eqref{casesS1}$_2$ is used.\\

\begin{figure}[h!]
 \centering
 \includegraphics[scale=0.5]{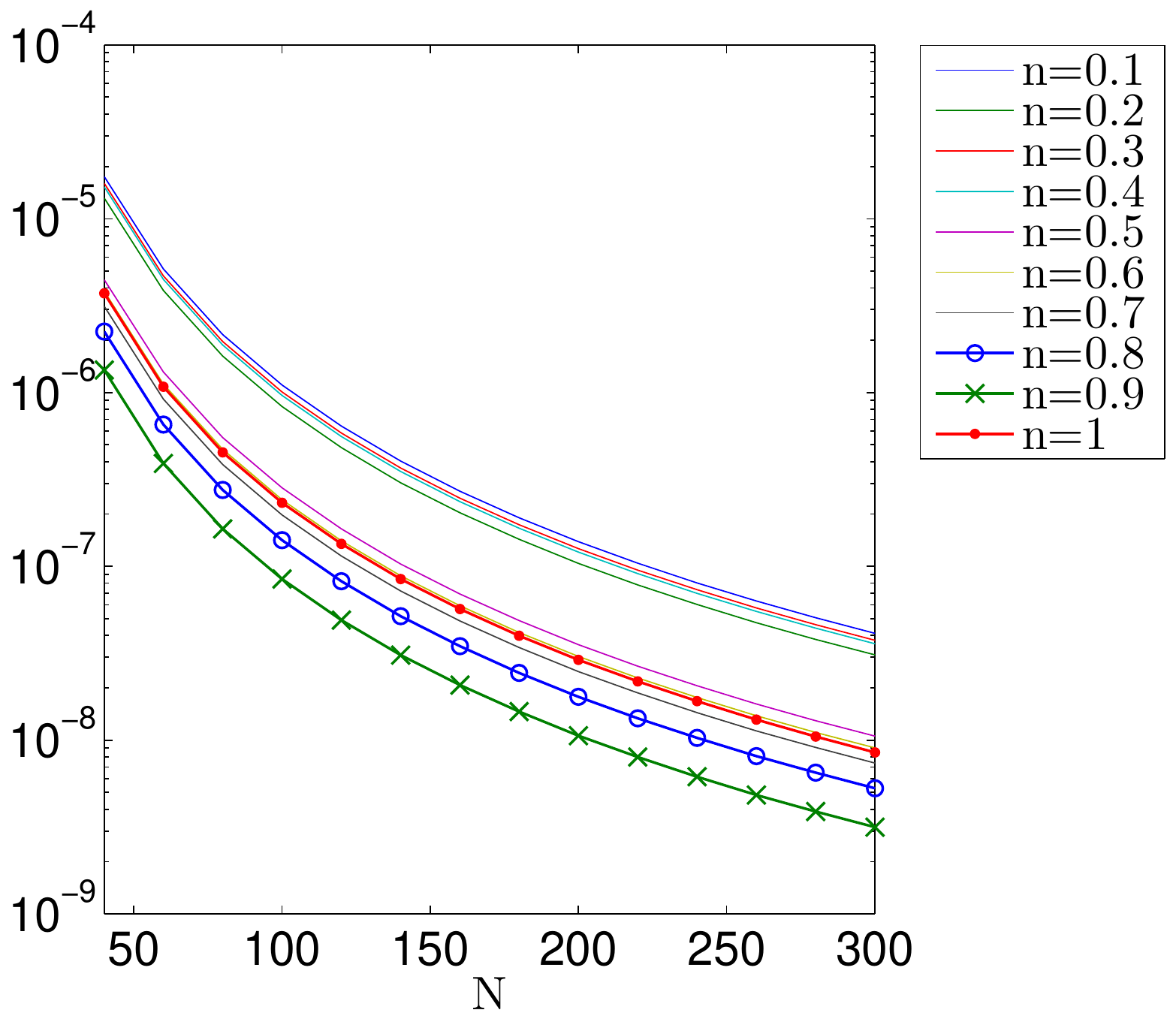}
 \put(-238,200) {{\bf (a)}}
 \put(-238,105) {\rotatebox{90}{{\bf $\delta w$}}}
 \includegraphics[scale=0.5]{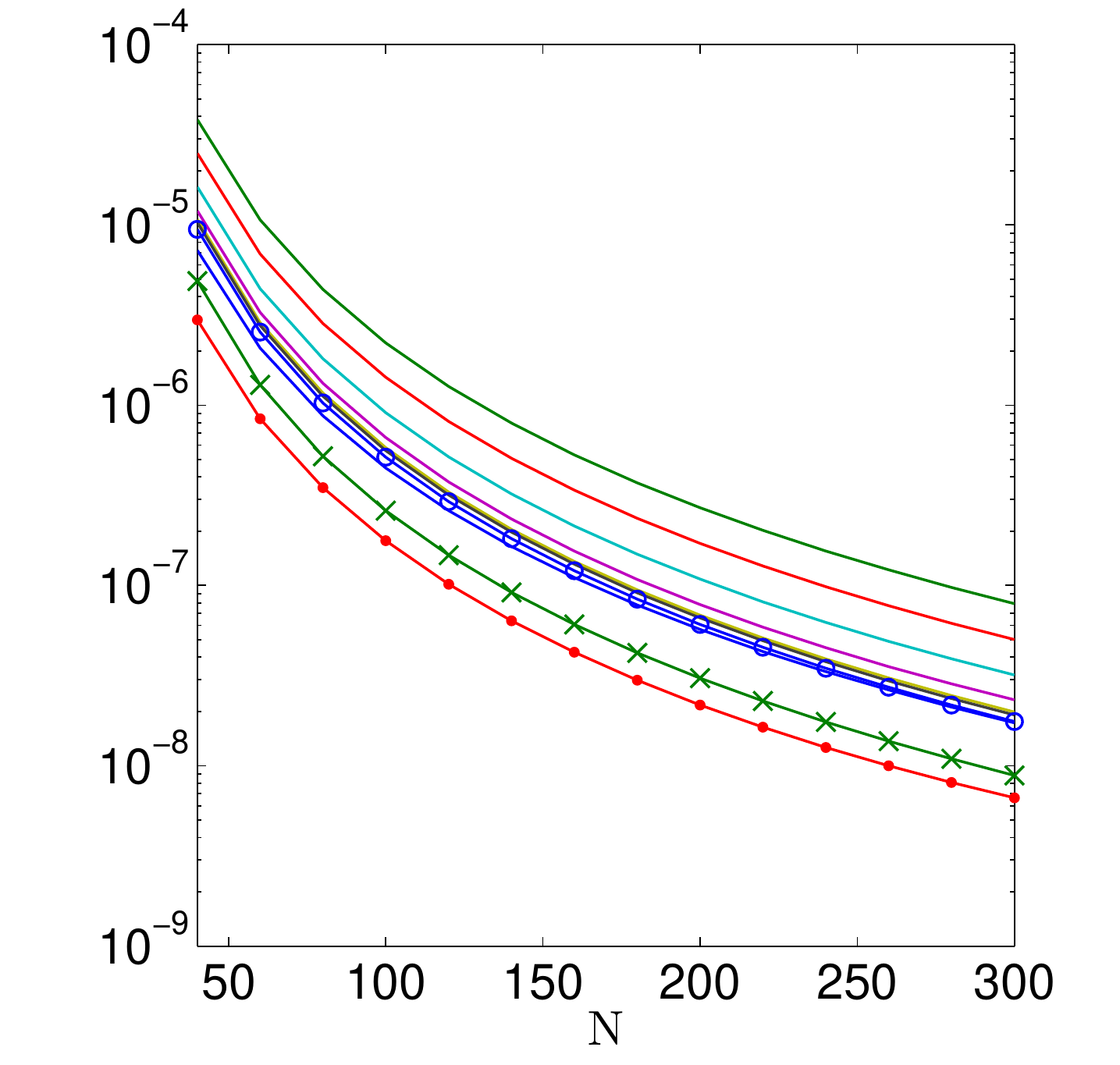}
 \put(-208,200) {{\bf (b)}}
 \caption{Relative average error of the crack aperture \eqref{BenchAcc}$_1$ obtained against the analytical benchmark over $N$ for the (a) viscosity dominated regime, (b) toughness dominated regime.}
 \label{Acc_Visc_01}
 \end{figure}

\begin{figure}[h!]
 \centering
 \includegraphics[scale=0.5]{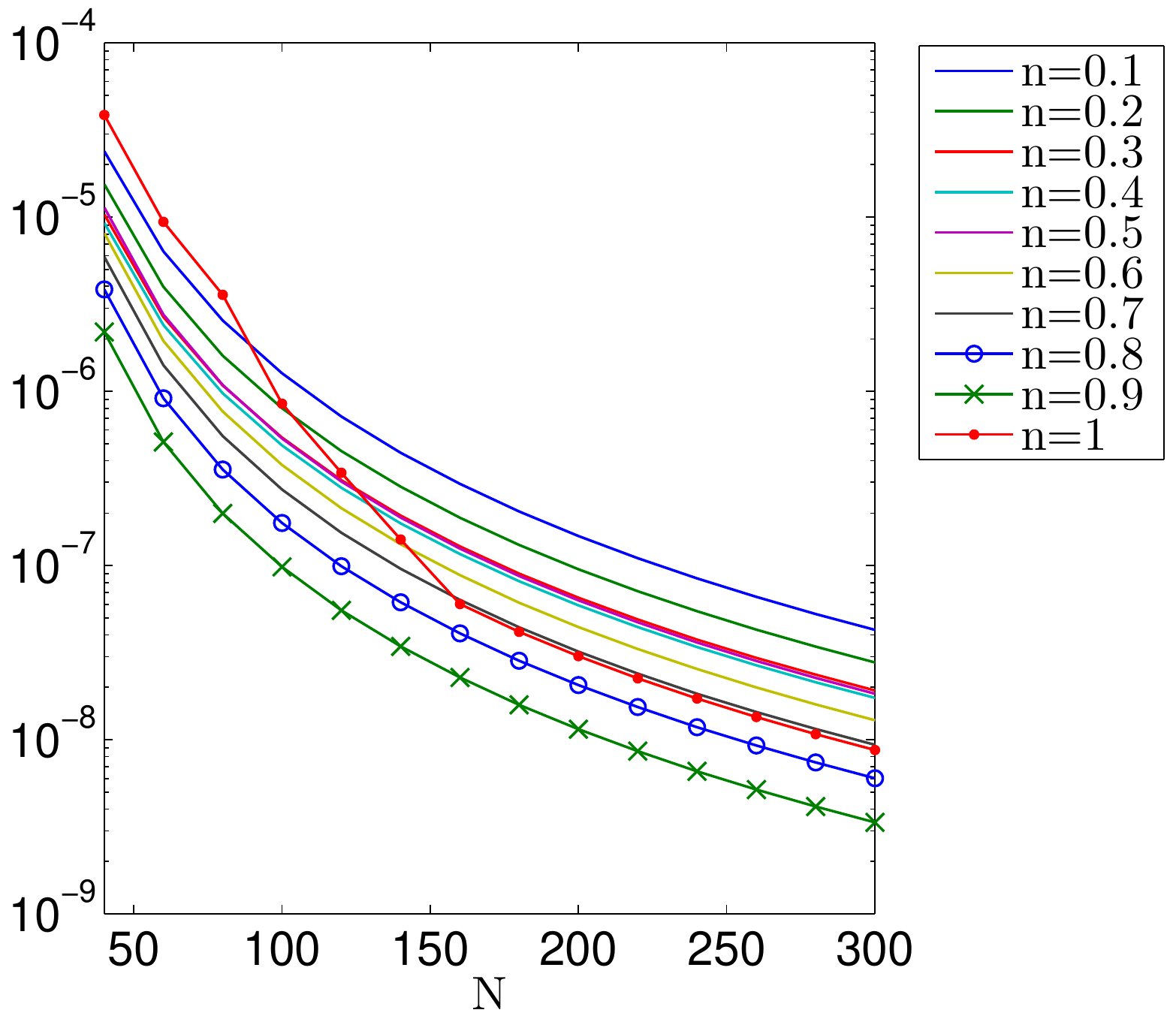}
 \put(-238,200) {{\bf (a)}}
 \put(-238,105) {\rotatebox{90}{{\bf $\delta v$}}}
 \includegraphics[scale=0.5]{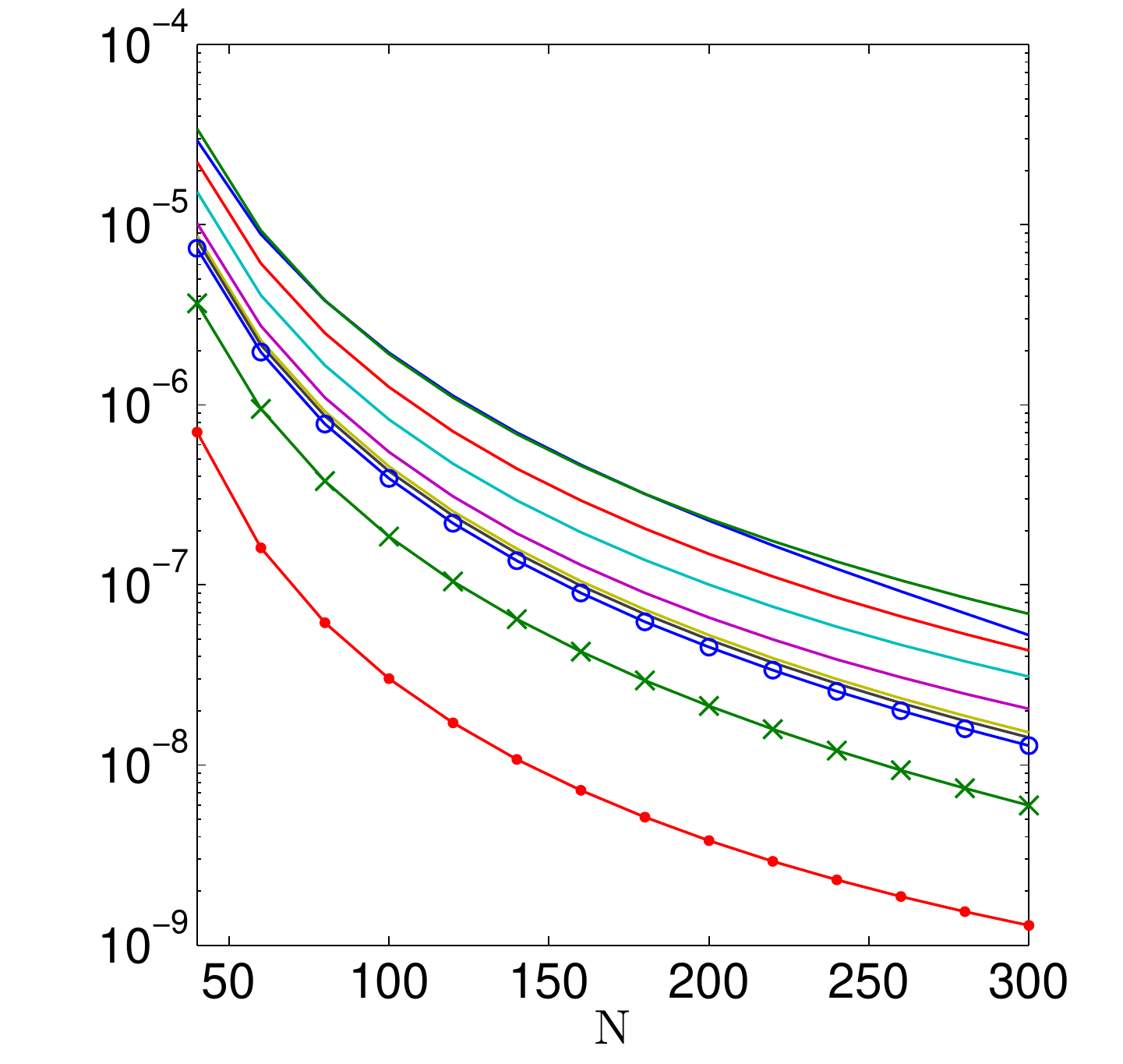}
 \put(-208,200) {{\bf (b)}}
 \caption{Relative average error of the particle velocity \eqref{BenchAcc}$_2$ obtained against the analytical benchmark over $N$ for the (a) viscosity dominated regime, (b) toughness dominated regime.}
 \label{Fig:PhiAcc}
 \end{figure}

The accuracy of computations is depicted in Fig.~\ref{Acc_Visc_01}, \ref{Fig:PhiAcc}, for varying number of nodal points $N$. A non-uniform spatial mesh was used, with meshing density increased near the ends of the interval (the same type of mesh was used for all $n$). The measures $\delta w$, $\delta v$, describing the average relative error of the crack opening and particle velocity, are taken to be:
\begin{equation}
\delta w (N) = \frac{\int_0^1 \tilde{r} \left| \hat{w}^* (\tilde{r}) - \hat{w}(\tilde{r}) \right| \, d\tilde{r} }{\int_0^1 \tilde{r} \hat{w}^* (\tilde{r}) \, d\tilde{r}} , \quad \delta v (N) = \frac{\int_0^1 \tilde{r} \left| \hat{v}^* (\tilde{r}) - \hat{v}(\tilde{r}) \right| \, d\tilde{r} }{\int_0^1 \tilde{r} \hat{v}^* (\tilde{r}) , d\tilde{r}} ,
\label{BenchAcc}
\end{equation}
where $\hat{w}^*$ and $\hat{v}^*$ denote the exact solutions for $\hat{w}$ and $\hat{v}$.

The results clearly show that the value of both error measures decreases monotonically with growing $N$. For a fixed number of nodal points $N$, $\delta w$ is lower than $\delta v$, but within the same order of magnitude. One can observe a sensitivity of the results to the value of the fluid behaviour index $n$. Here, the level of error measures can vary up to one order for a constant $N$. This trend can be alleviated by adjusting the mesh density distribution to the value of $n$ (i.e. to the varying asymptotics of solution), however such an investigation goes beyond the scope of this paper. In general, it takes fewer than $N=300$ nodal points to achieve the relative errors of the level $10^{-7}$.

In cases when the exact solution is not prescribed an alternative method of testing the solution accuracy is required. The method outlined here relies on the fact that the solution converges to the exact value at a known rate, with respect to the number of nodal points, which has been established numerically to behave as $1/N^3$. As a result the following estimation holds:

\begin{equation}
\label{sol_conv}
\int_0^1 r g_i(r)dr=A_i+\frac{B_i}{N^3}, \quad i=1,2 ,
\end{equation}
where $g_1(r)=\hat w(r)$ and $g_2(r)=\hat v(r)$. $A_i$ and $B_i$ are some constants to be found numerically.
Next, one can define the limiting value of \eqref{sol_conv} as:
\begin{equation}
\label{sol_lim}
\lim_{N \to \infty} \int_0^1 r g_i(r)dr=A_i \approx \int_0^1 r g^*_i(r)dr, \quad i=1,2 ,
\end{equation}
for $g^*_1(r)=\hat w^*(r)$, $g^*_2(r)=\hat v^*(r)$.

\begin{figure}[h!]
 \centering
 \includegraphics[scale=0.5]{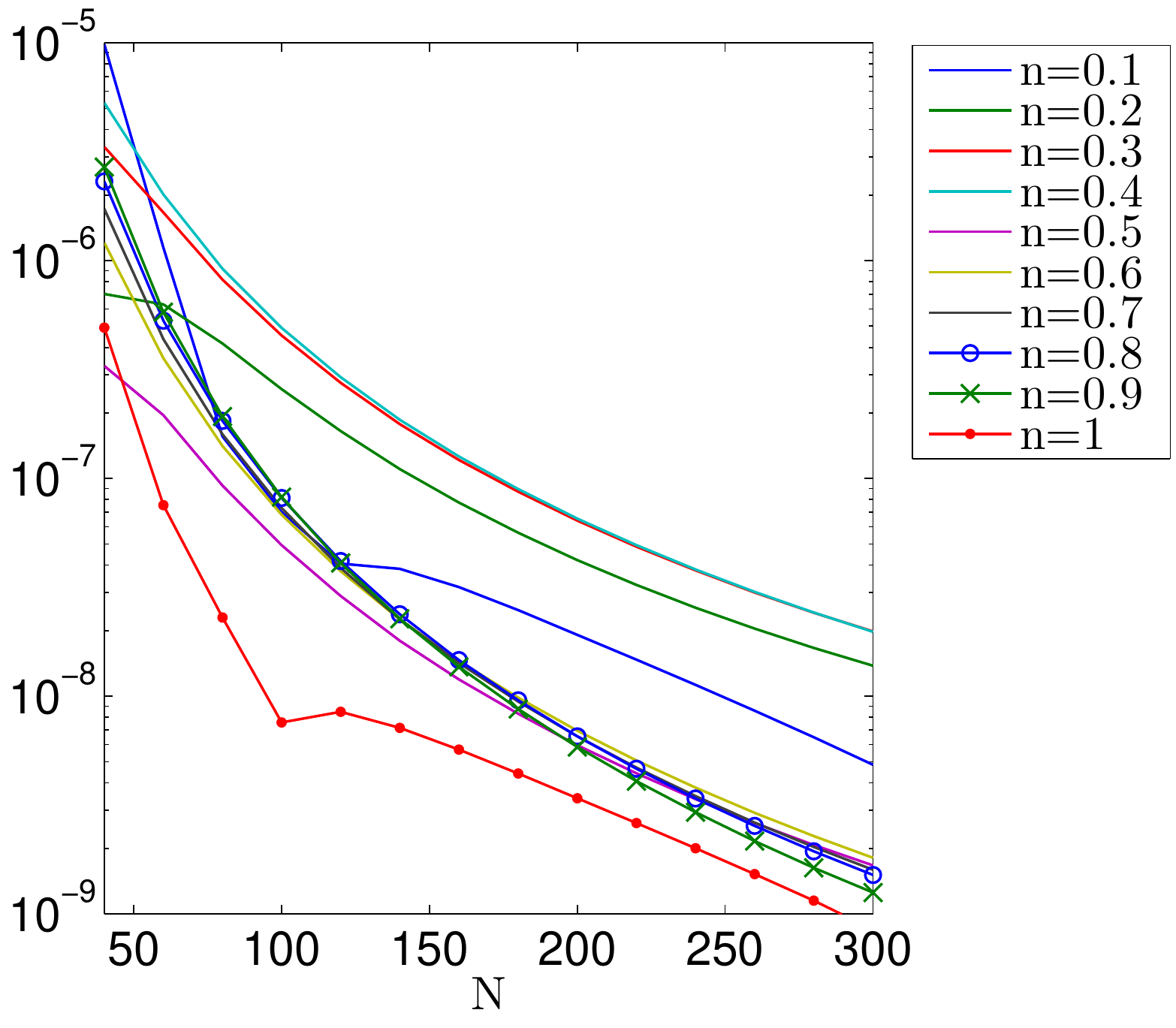}
 \put(-238,200) {{\bf (a)}}
 \put(-238,105) {\rotatebox{90}{{\bf $e_w$}}}
 \includegraphics[scale=0.5]{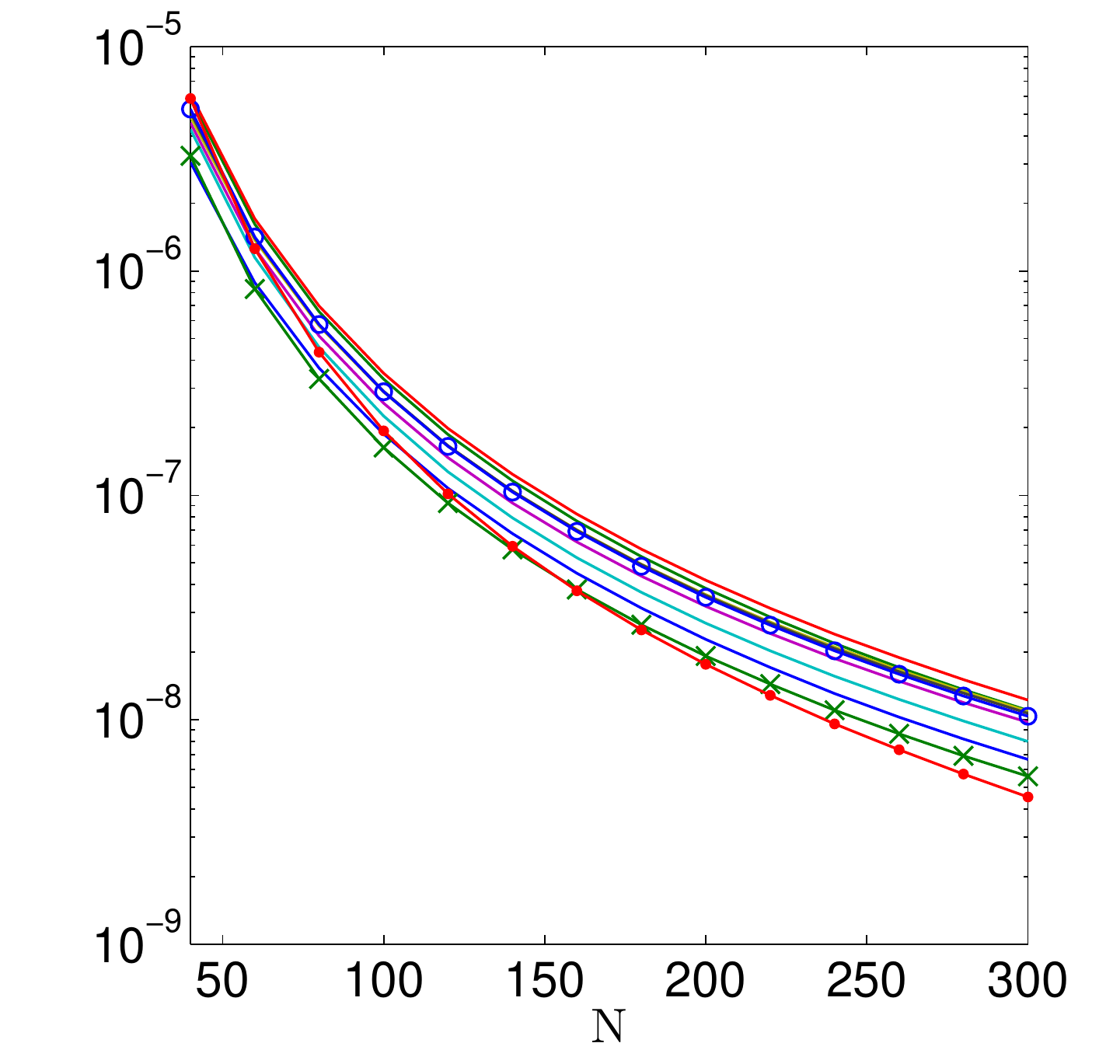}
 \put(-208,200) {{\bf (b)}}
 \caption{Rate of convergence $e_w$ \eqref{deltaG} of the numerical solution for the benchmark example:  (a) viscosity dominated regime, (b) toughness dominated regime.}
 \label{Fig:ApCon}
 \end{figure}

\begin{figure}[h!]
 \centering
 \includegraphics[scale=0.5]{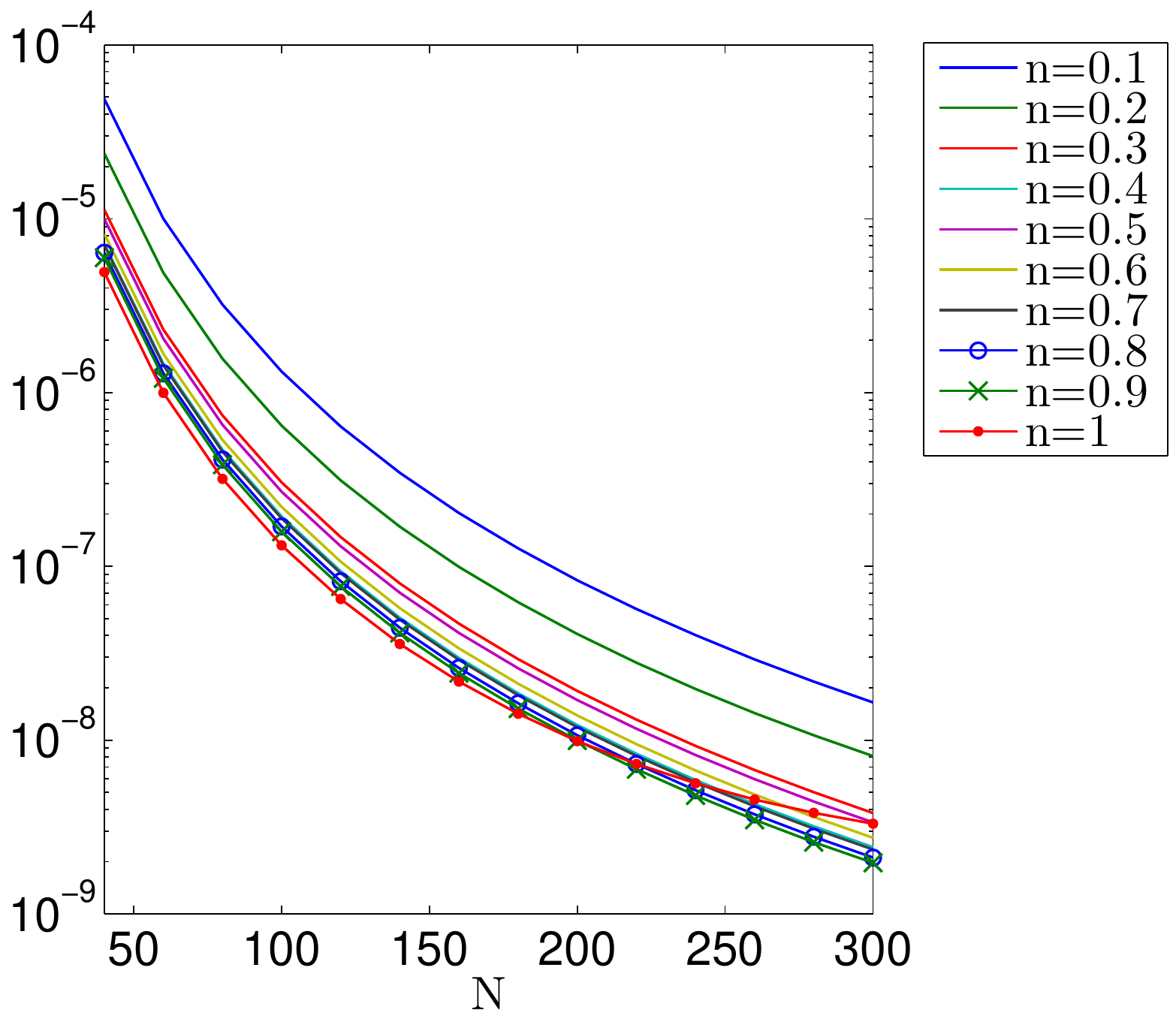}
 \put(-238,200) {{\bf (a)}}
 \put(-238,105) {\rotatebox{90}{{\bf $e_v$}}}
 \includegraphics[scale=0.5]{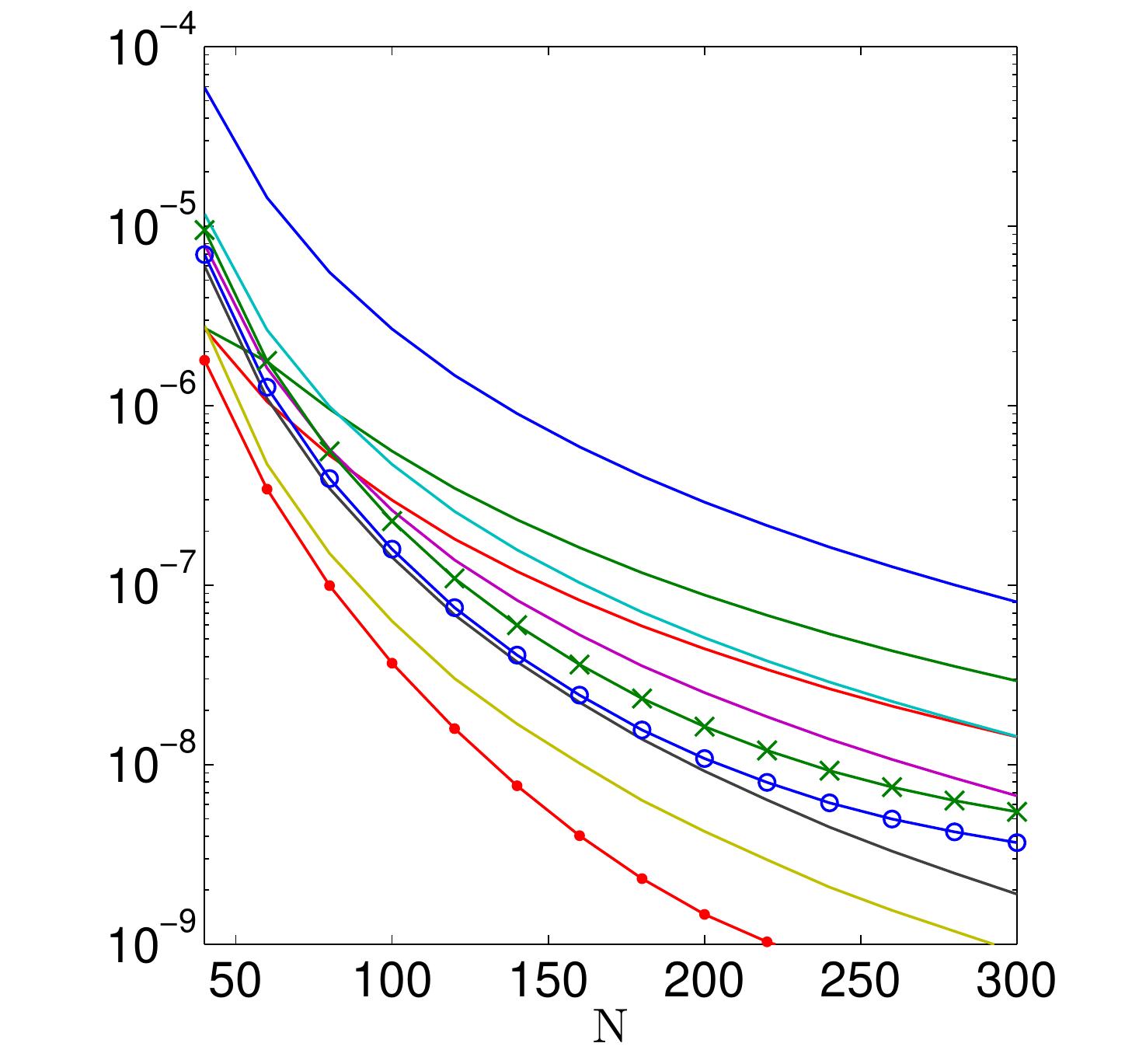}
 \put(-208,200) {{\bf (b)}}
 \caption{Rate of convergence $e_v$ \eqref{deltaG} of the numerical solution for the benchmark example: (a) viscosity dominated regime, (b) toughness dominated regime.}
 \label{Fig:PhiCon}
 \end{figure}

Knowing this, the following alternative error measures can be proposed:
\begin{equation}
e_{g_i}(N) = \frac{\left| A_i - \int_0^1 r \hat g^*_i (r) \, dr \right|}{\int_0^1 r \hat g^*_i (r) \, dr} ,\quad i=1,2.
\label{deltaG}
\end{equation}
Using this strategy, it is possible to identify the relative rate at which the solution converges: $e_w (N)$ for the aperture and $e_v (N)$ for the particle velocity. The results are shown in Fig.~\ref{Fig:ApCon}, Fig.~\ref{Fig:PhiCon}. It is notable that both $\delta w$ and $e_w$, as well as $\delta v$ and $e_v$, provide estimates of a similar order  for a fixed N. Thus, they can be considered as equivalent error measures and employed in the accuracy analysis in the cases when no exact solution is available. As such, $e_w (N)$ and $e_v (N)$ will be used in the following investigations.

\subsubsection{Impermeable solid - reference solutions}

With a suitable measure for testing the solution accuracy in place we move onto examining the solution variant most frequently studied in the literature, the case with a zero valued leak-off function and with $\hat Q_0=1$. Although there is no analytical solution to this variant of the problem, due to its relative simplicity, it is commonly used when testing numerical algorithms. For this reason it is very important that credible reference data is provided for this case, which can be easily employed to verify various computational schemes. Both the viscosity and toughness dominated regimes (for different values of the material toughness: $\hat{K}_{Ic}=\left\{1,10,100\right\}$) will be investigated. In the next subsection, accurate and simple approximations of the obtained numerical solutions will be provided.

\begin{figure}[h!]
 \centering
 \includegraphics[scale=0.5]{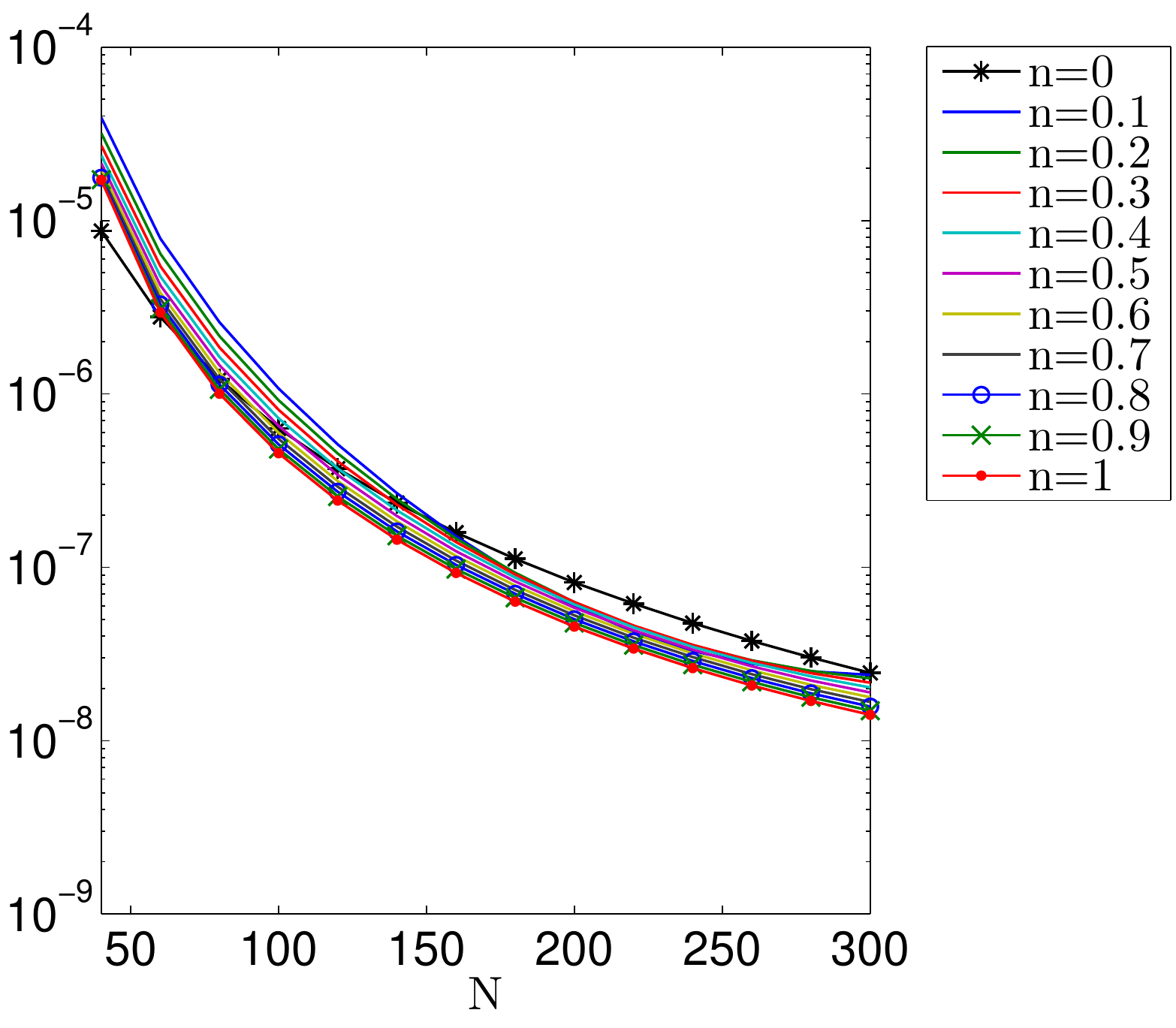}
 \put(-238,200) {{\bf (a)}}
 \put(-238,105) {\rotatebox{90}{{\bf $e_w$}}}
 \includegraphics[scale=0.5]{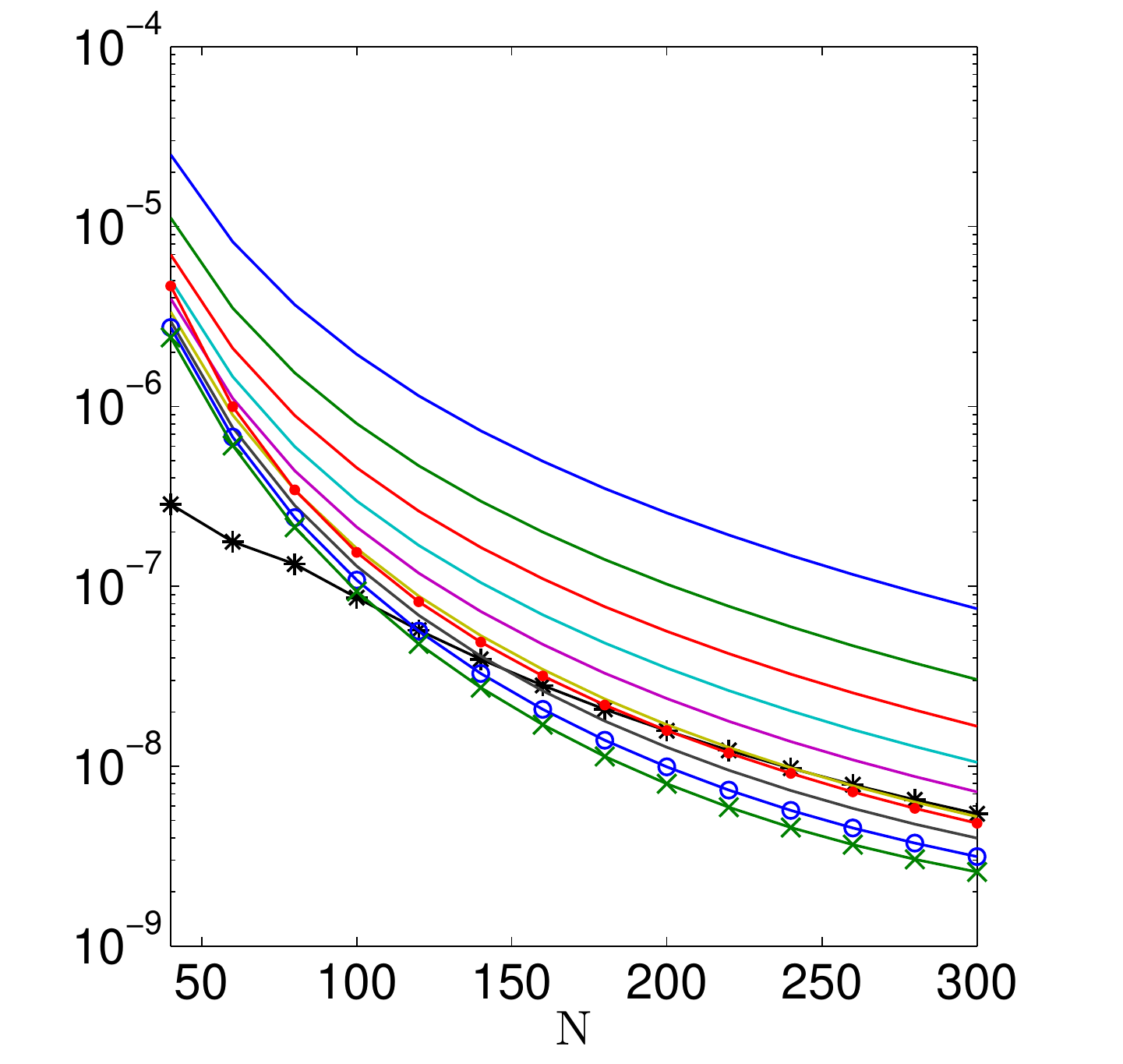}
 \put(-208,200) {{\bf (b)}}
 \caption{Rate of convergence $e_w$ \eqref{deltaG} of the numerical solution when $Q_0=1$ with no fluid leak-off for the: (a) viscosity dominated regime, (b) toughness dominated regime with $\hat{K}_{Ic}=1$.}
 \label{Fig:delG}
 \end{figure}

\begin{figure}[h!]
 \centering
 \includegraphics[scale=0.5]{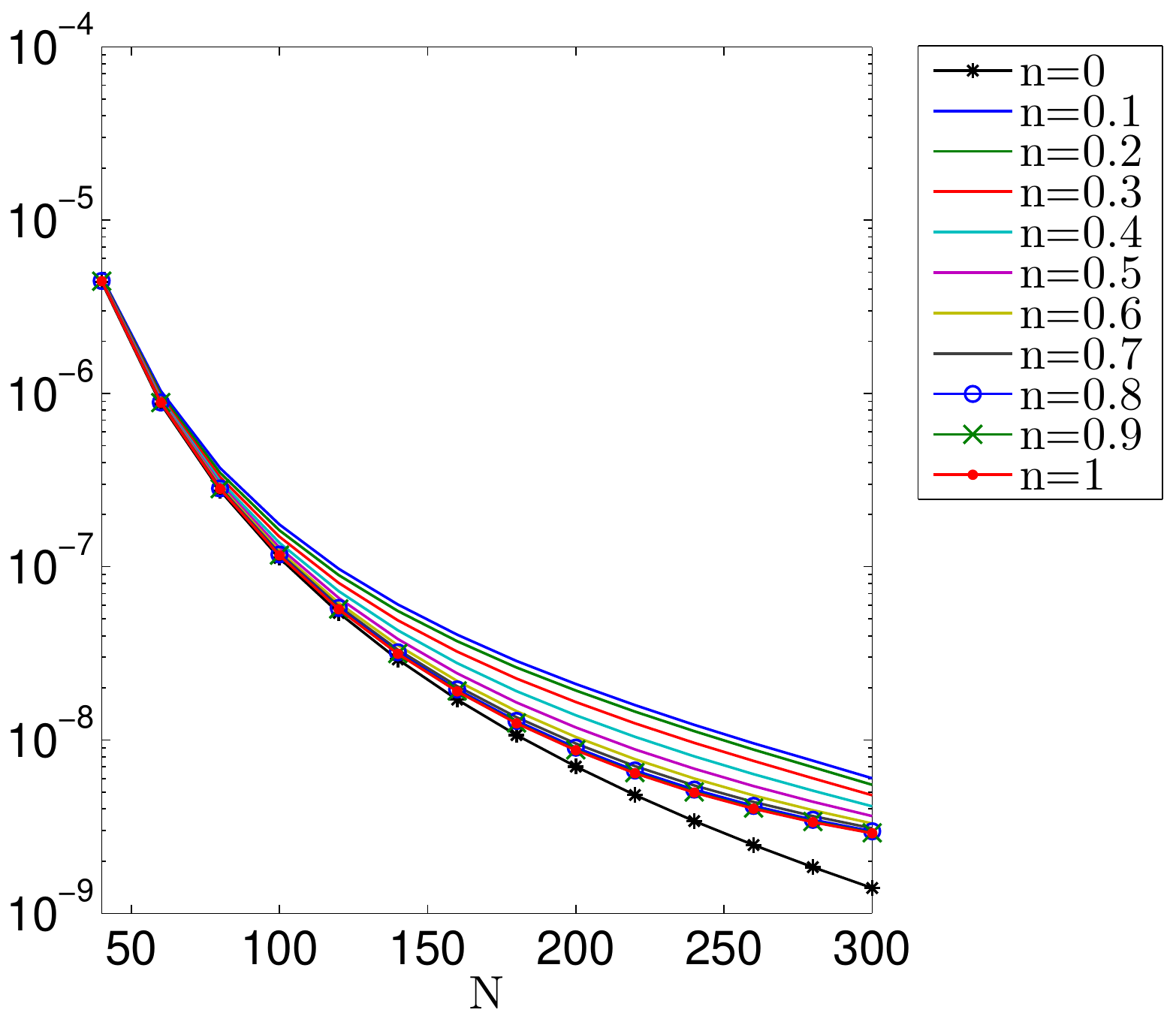}
 \put(-238,200) {{\bf (a)}}
 \put(-238,105) {\rotatebox{90}{{\bf $e_w$}}}
 \includegraphics[scale=0.5]{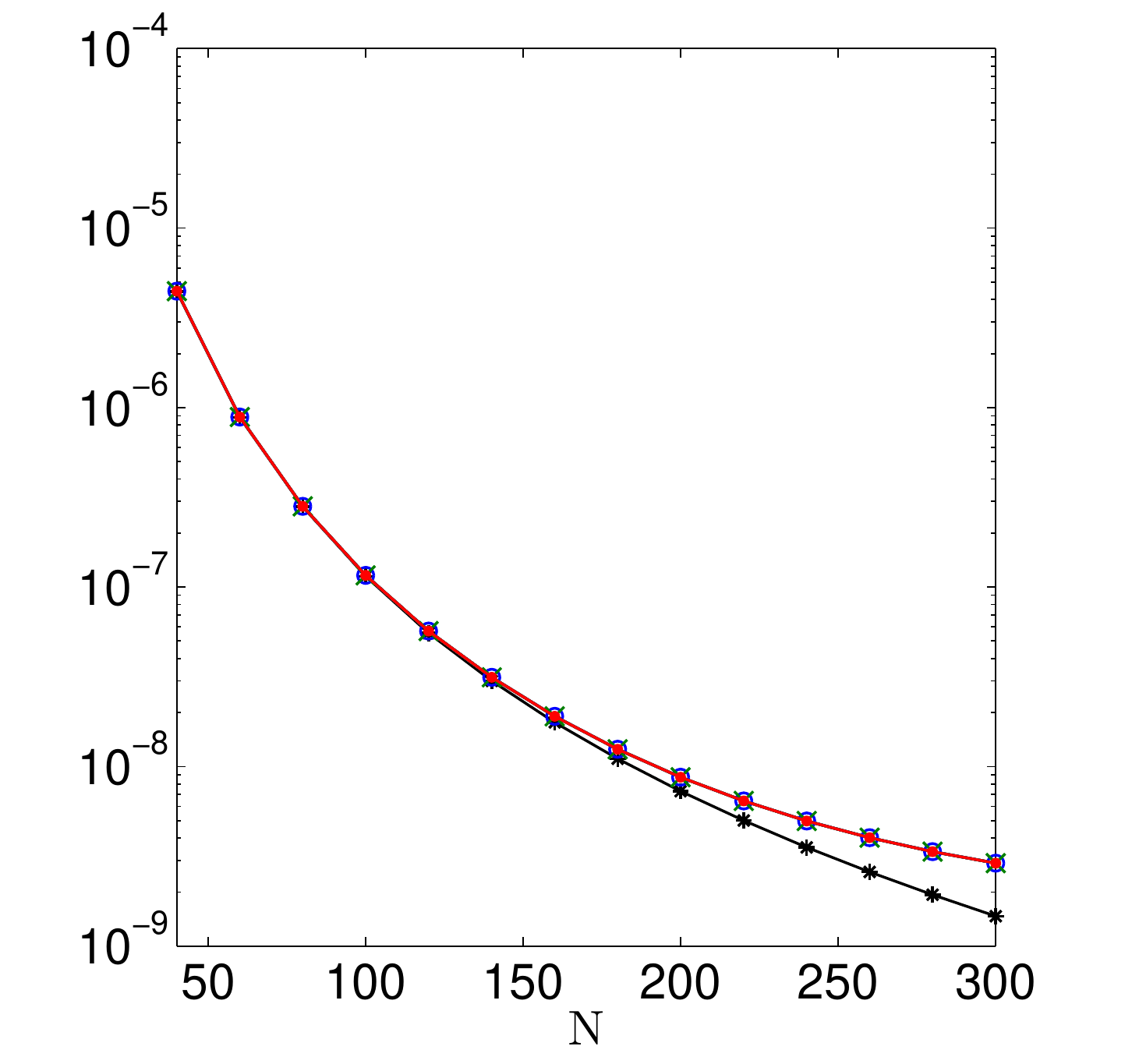}
 \put(-208,200) {{\bf (b)}}
 \caption{Rate of convergence $e_w$ \eqref{deltaG} of the numerical solution when $Q_0=1$ with no fluid leak-off for the toughness dominated regime with: (a) $\hat{K}_{Ic}=10$ and (b) $\hat{K}_{Ic}=100$.}
 \label{Fig:KI10}
 \end{figure}

\begin{figure}[h!]
 \centering
 \includegraphics[scale=0.5]{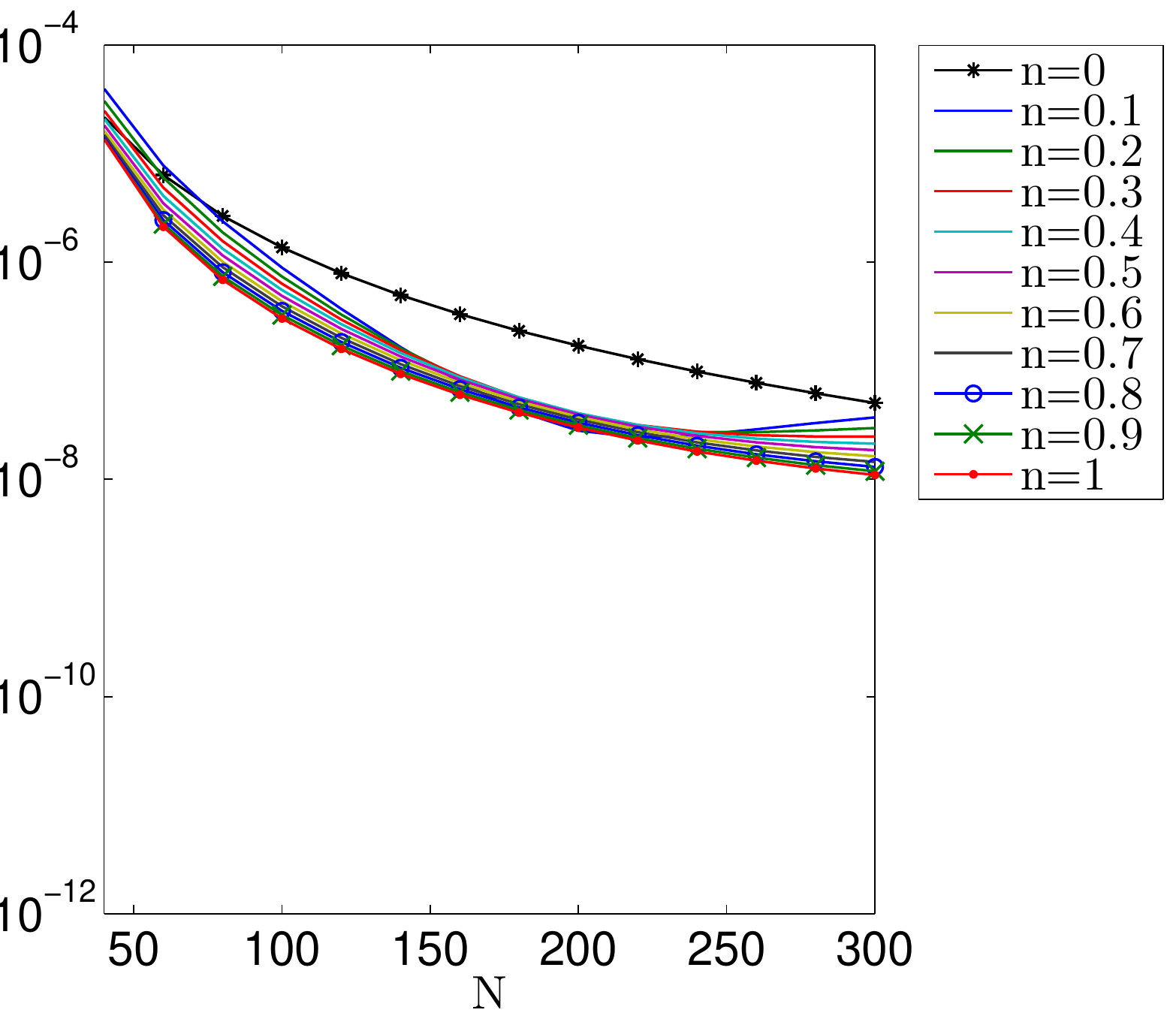}
 \put(-238,200) {{\bf (a)}}
 \put(-238,105) {\rotatebox{90}{{\bf $e_v$}}}
 \includegraphics[scale=0.5]{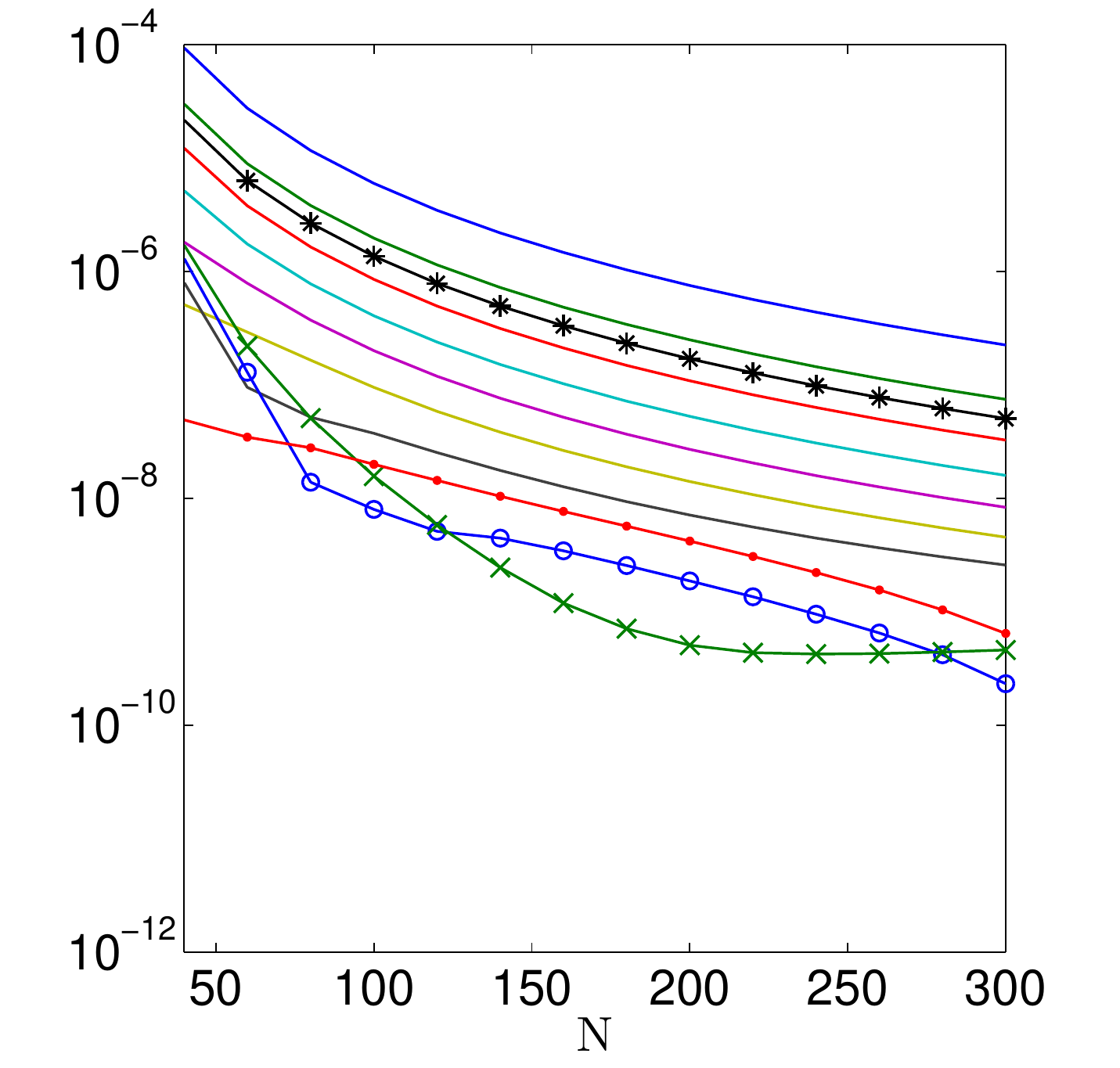}
 \put(-208,200) {{\bf (b)}}
 \caption{Rate of convergence $e_v$ \eqref{deltaG} of the numerical solution when $Q_0=1$ with no fluid leak-off for the: (a) viscosity dominated regime, (b) toughness dominated regime with $\hat{K}_{Ic}=1$.}
 \label{Fig:delH}
 \end{figure}

\begin{figure}[h!]
 \centering
 \includegraphics[scale=0.5]{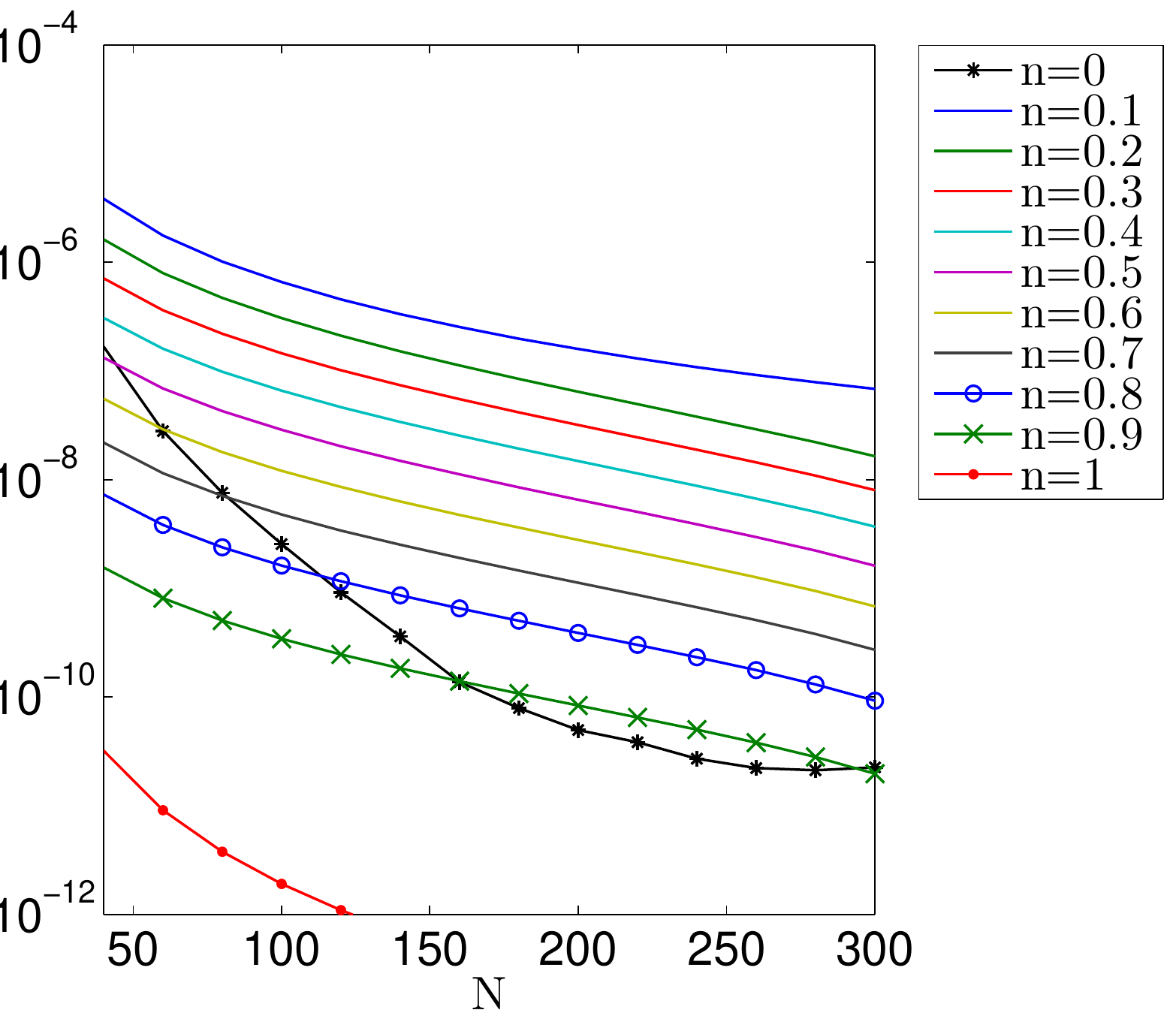}
 \put(-238,200) {{\bf (a)}}
 \put(-238,105) {\rotatebox{90}{{\bf $e_v$}}}
 \includegraphics[scale=0.5]{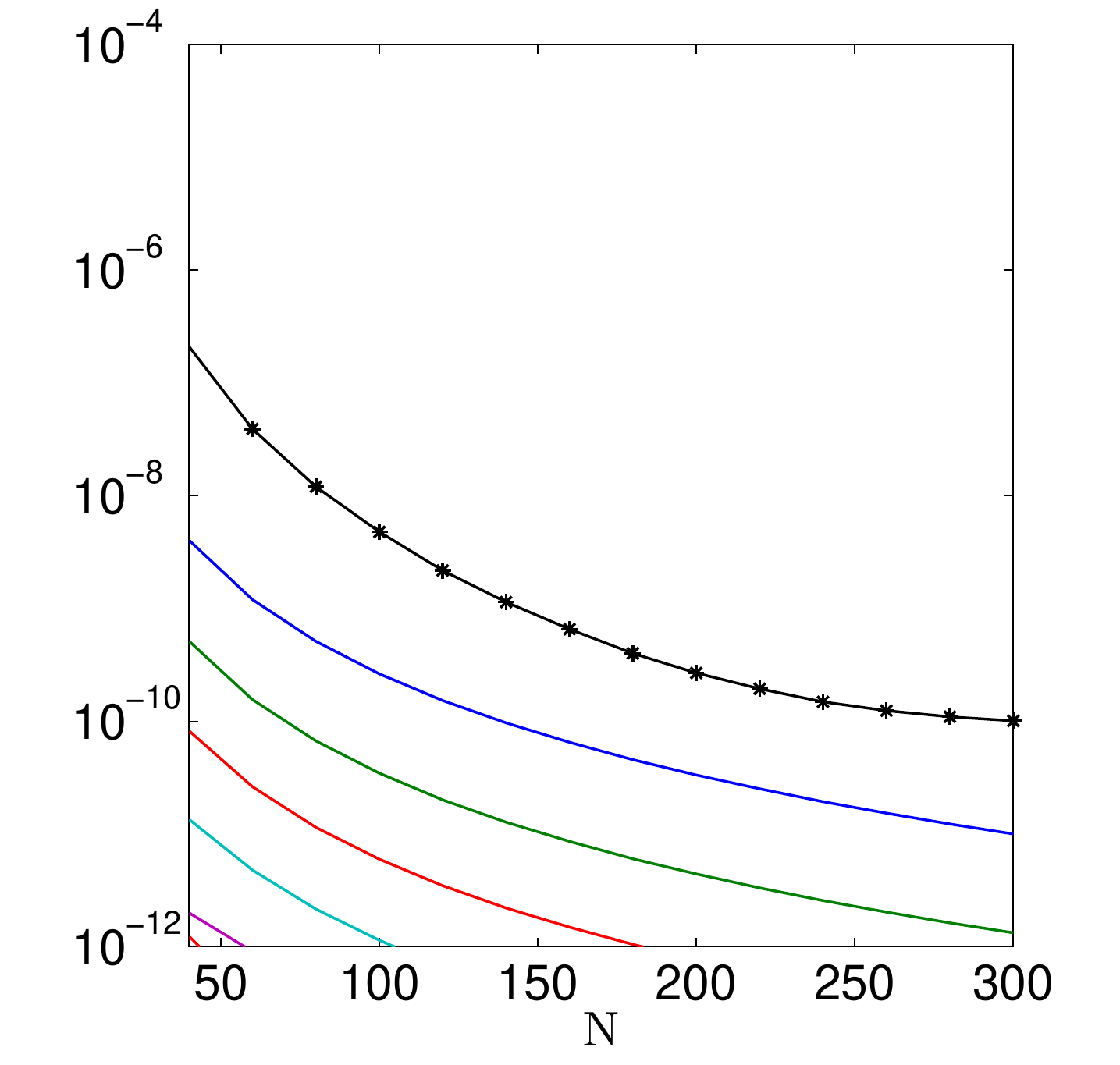}
 \put(-208,200) {{\bf (b)}}
 \caption{Rate of convergence $e_v$ \eqref{deltaG} of the numerical solution when $Q_0=1$ with no fluid leak-off for the toughness dominated regime with: (a) $\hat{K}_{Ic}=10$ and (b) $\hat{K}_{Ic}=100$.}
 \label{Fig:KI100}
 \end{figure}

The results for the crack opening and particle velocity convergence rates are shown in Figs.~\ref{Fig:delG} - \ref{Fig:KI100}.

\begin{figure}[h!]
 \centering
 \includegraphics[scale=0.5]{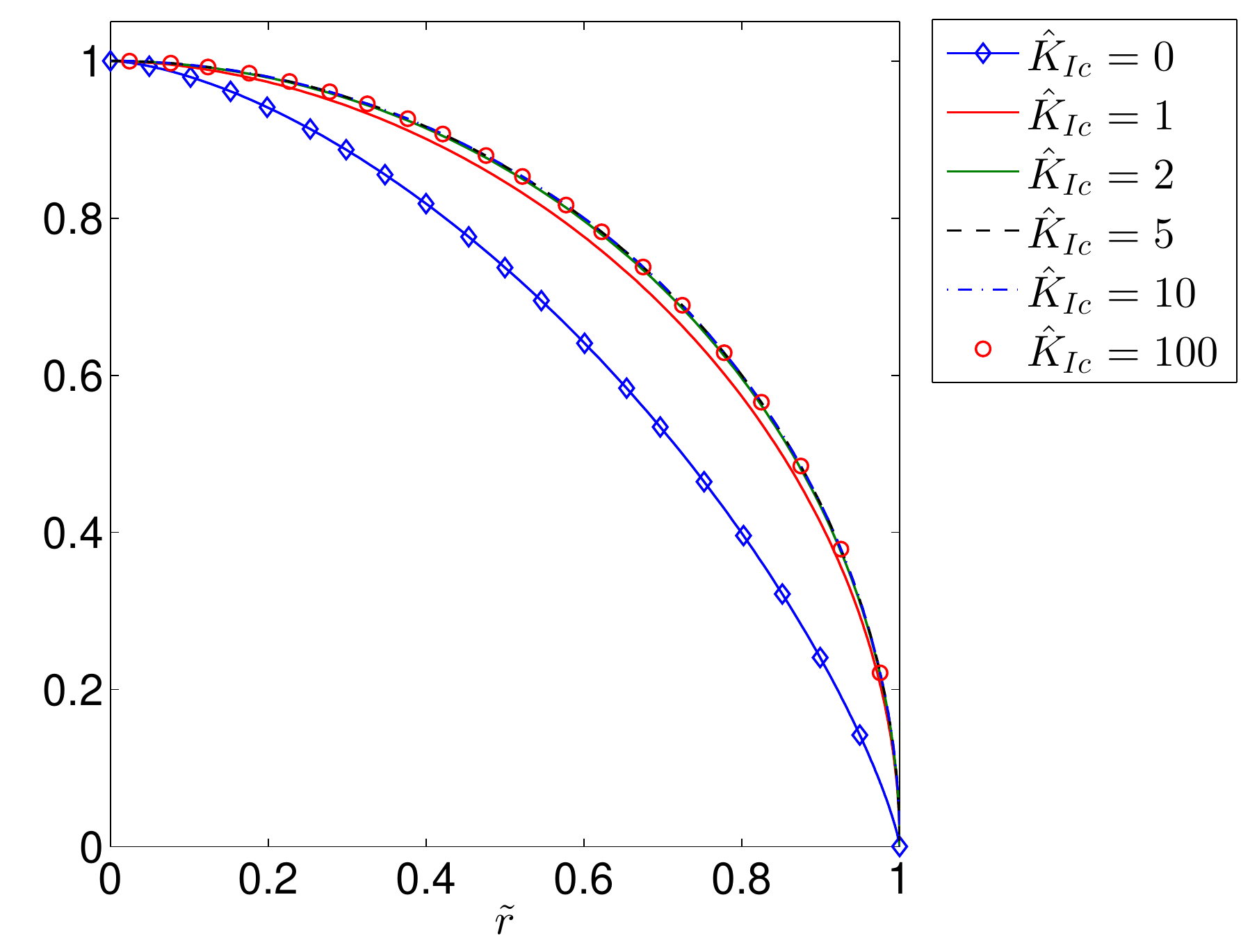} 
  \put(-270,195) {{\bf (a)}}					
  \put(-270,95) {\rotatebox{90}{{\bf{$\frac{\hat{w}(\tilde{r})}{\hat{w}(0)}$}}}}
 \includegraphics[scale=0.5]{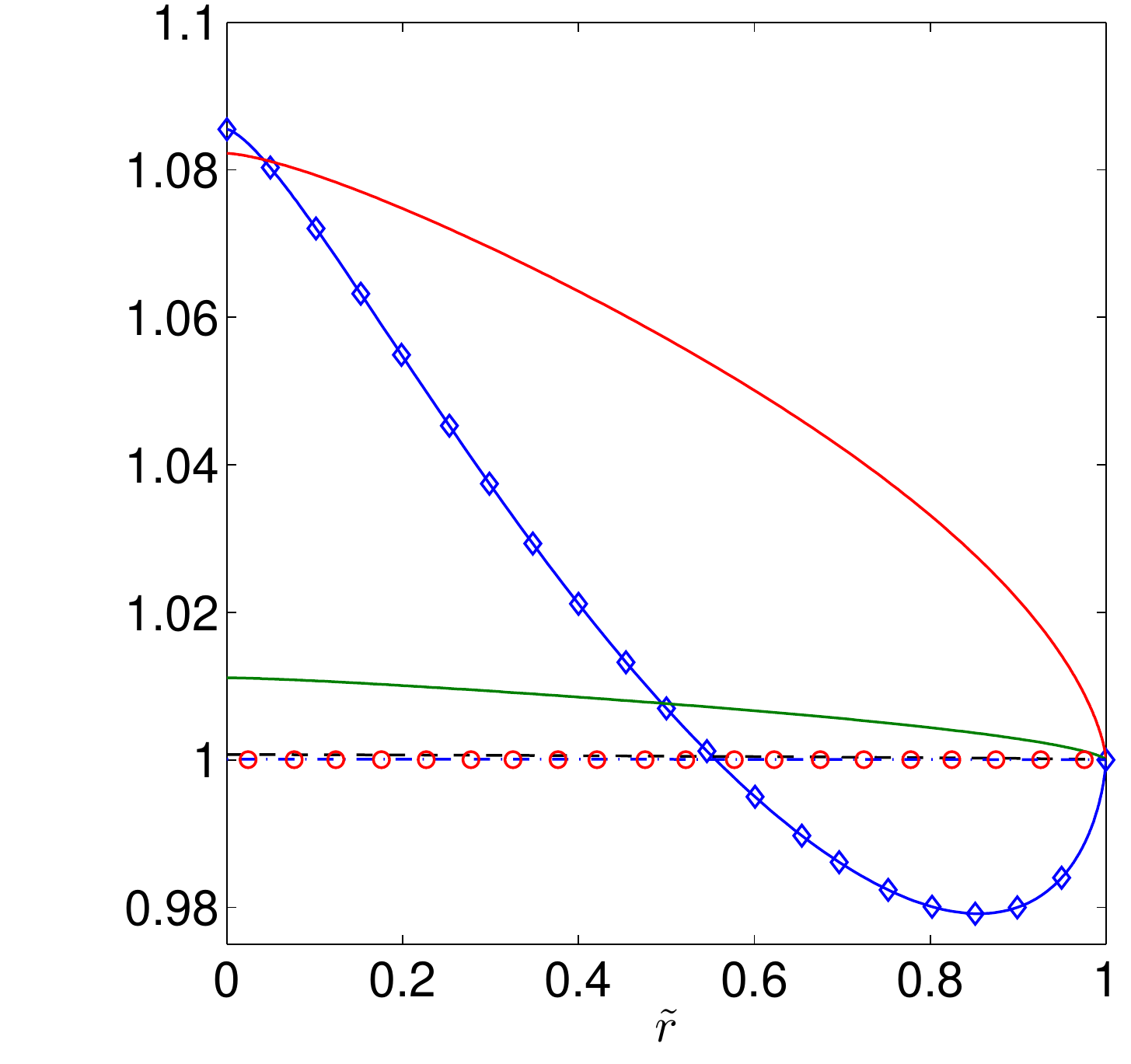} 
  \put(-208,195) {{\bf (b)}}
  \put(-208,87) {\rotatebox{90}{$\frac{\hat{w}(\tilde{r})}{\hat{w}_0 \left(1-\tilde{r}^2\right)^{\alpha_0}}$}}
 \caption{The aperture for $n=0.5$ for a different values of the fracture toughness: (a) the normalized self-similar aperture, (b) the self-similar aperture divided by the leading term of the crack tip asymptotics \eqref{apertureasymp1_otherb}. }
 \label{n050_w}
\end{figure}

\begin{figure}[h!]
 \centering
 \includegraphics[scale=0.5]{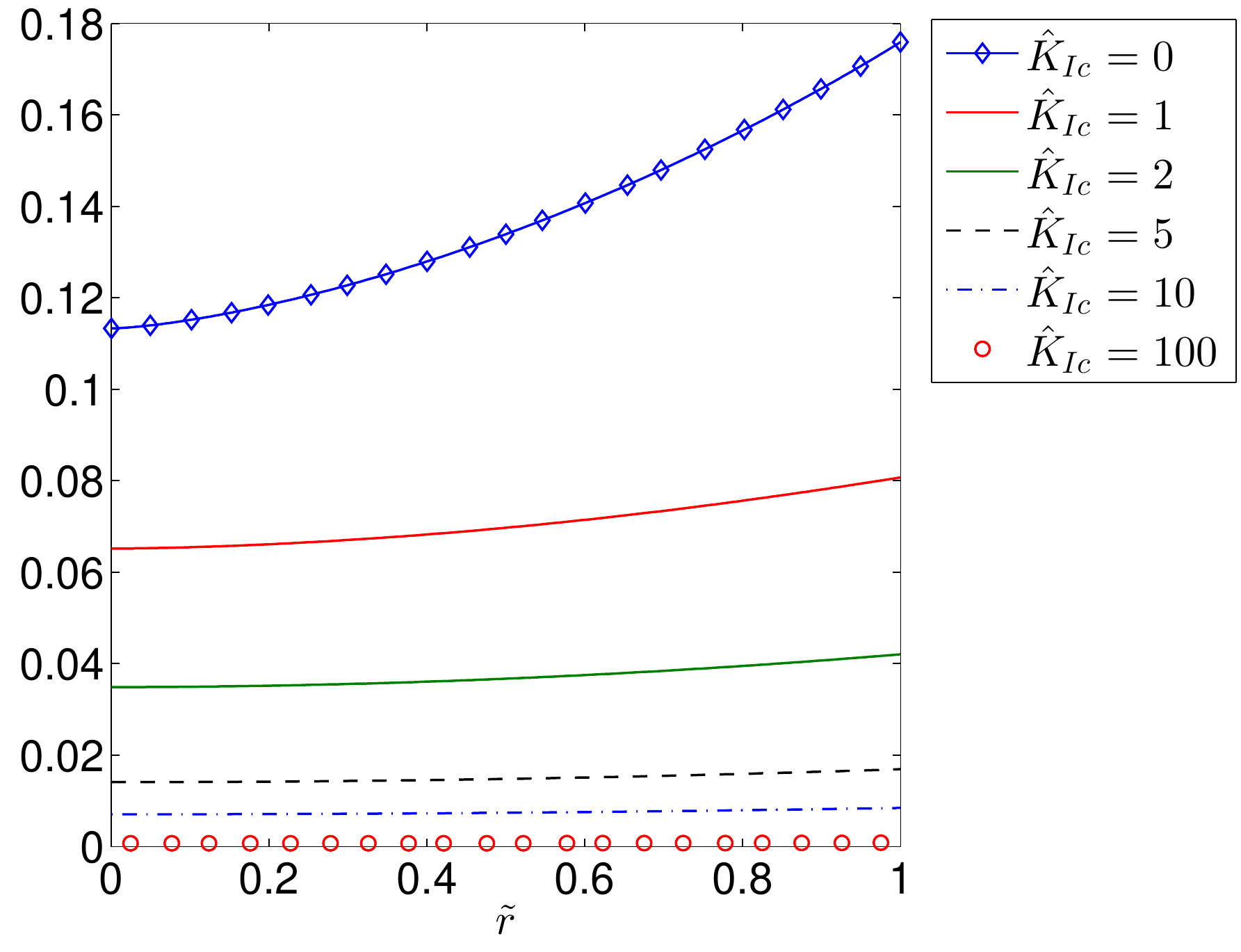} 
  \put(-270,195) {{\bf (a)}}					
  \put(-270,95)  {\rotatebox{90}{{\bf{$\tilde{r}\hat{v}(\tilde{r})$}}}}
 \includegraphics[scale=0.5]{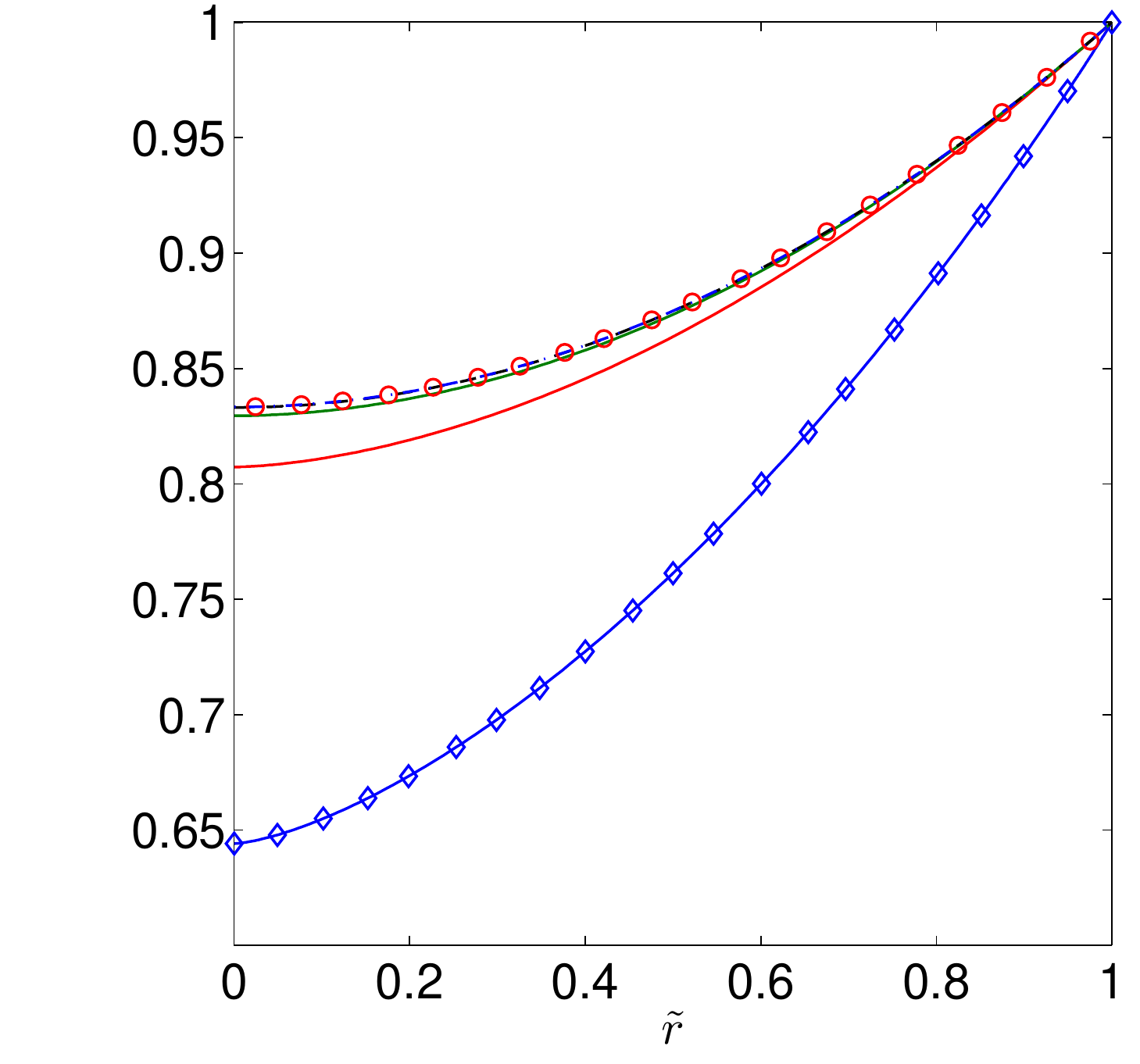} 
 \put(-208,195) {{\bf (b)}}
  \put(-208,95) {\rotatebox{90}{$\frac{\tilde{r}\hat{v}(\tilde{r})}{\hat{v}_0}$}}
 \caption{The particle velocity for $n=0.5$ for a different values of the fracture toughness: (a) the self-similar particle velocity, (b) the self-similar particle velocity divided by the leading term of the crack tip asymptotics \eqref{v_1_asymp}.}
 \label{n050_v}
\end{figure}

\begin{figure}[h!]
 \centering
 \includegraphics[scale=0.5]{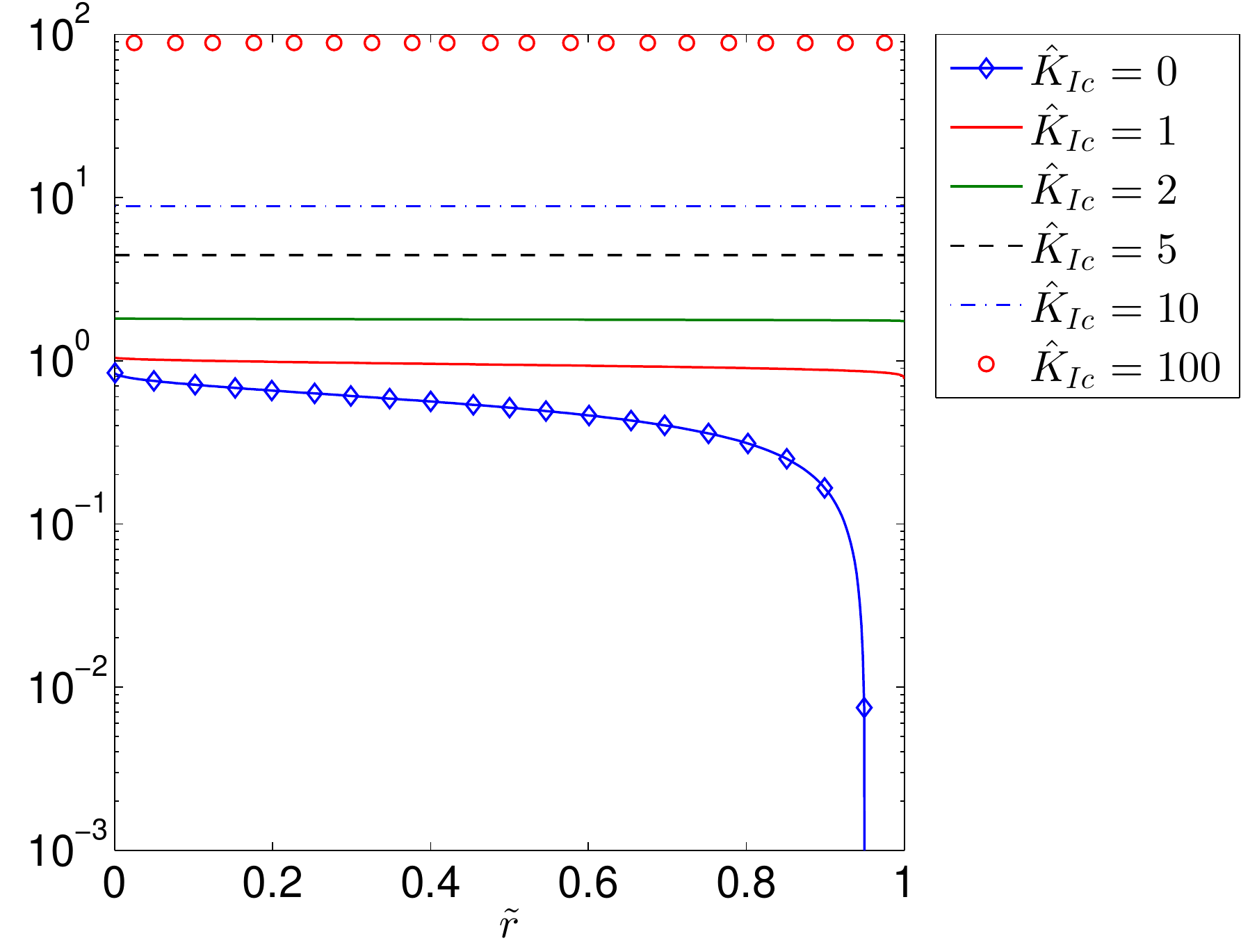} 
  \put(-270,195) {{\bf (a)}}					
  \put(-270,95)  {\rotatebox{90}{{\bf{$\hat{p}(\tilde{r})$}}}} 
 \includegraphics[scale=0.5]{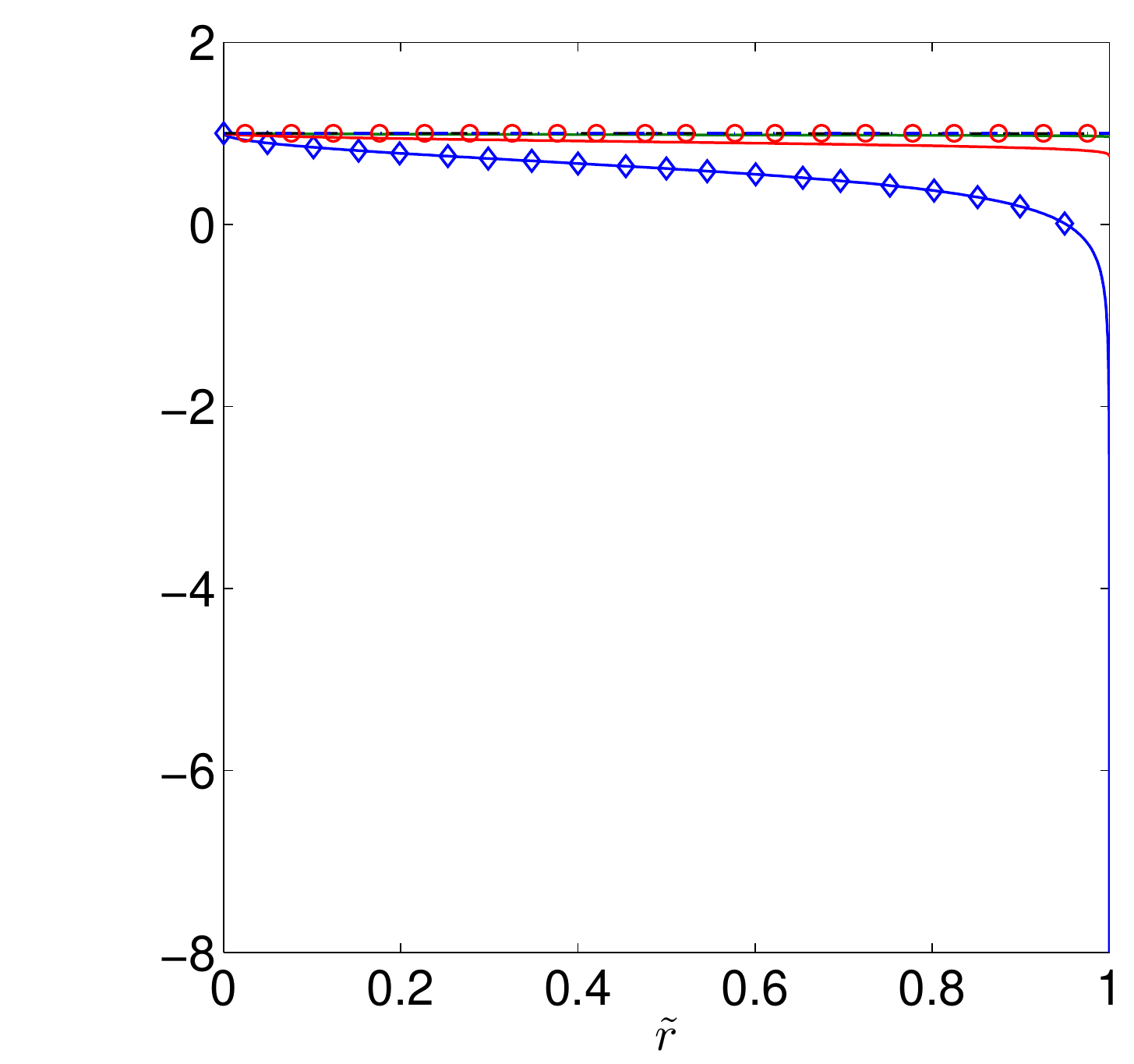} 
 \put(-198,195) {{\bf (b)}}
  \put(-198,95)  {\rotatebox{90}{$\frac{\hat{p}(\tilde{r})}{\hat{p}(0)}$}}
 \caption{The pressure function for $n=0.5$ for a different values of the fracture toughness: (a) the self-similar pressure function, (b) the self-similar pressure divided by the value of the pressure at the fracture opening.}
 \label{n050_p}
\end{figure}

\begin{figure}[h!]
 \centering
 \includegraphics[scale=0.5]{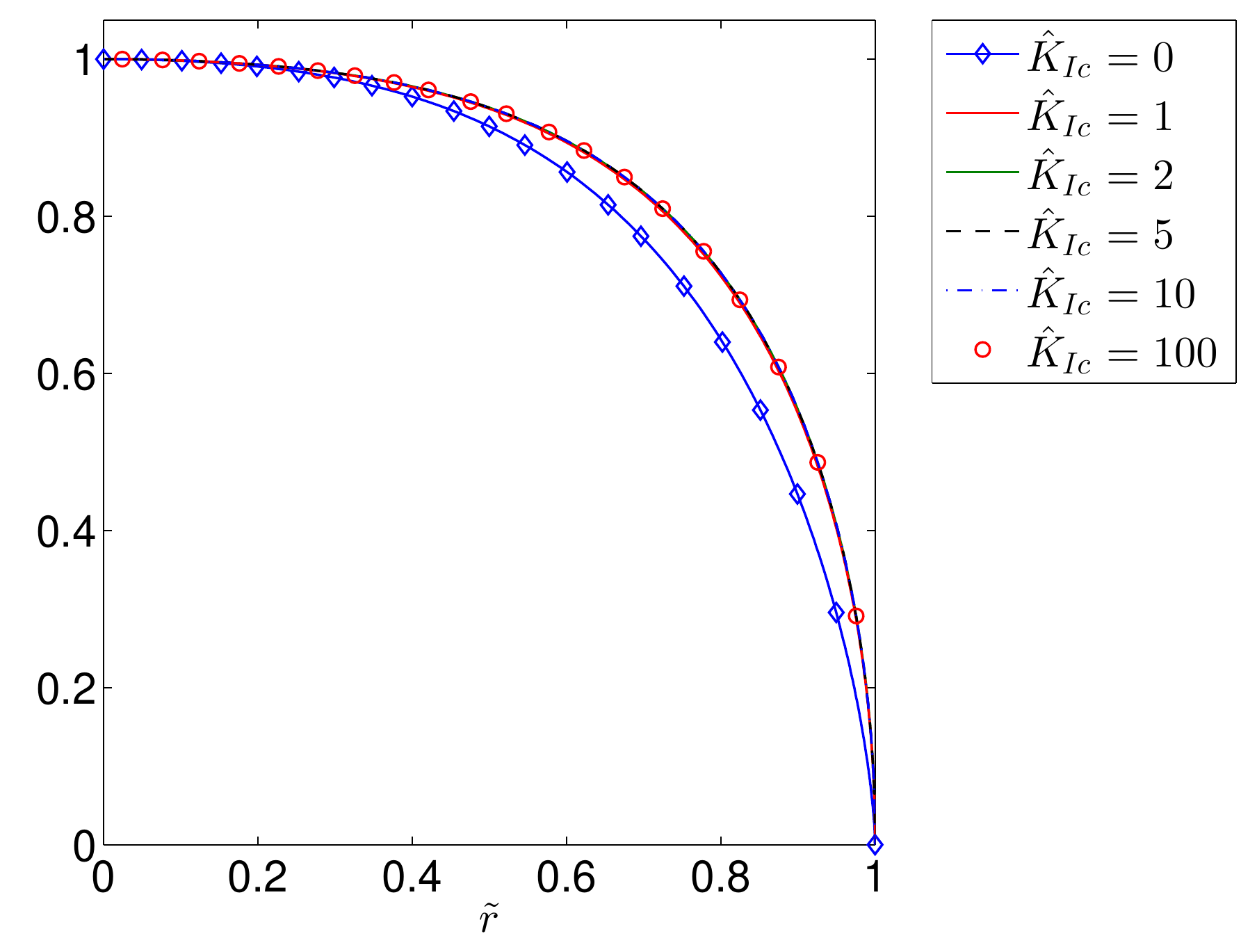}
  \put(-265,96) {\rotatebox{90}{{\bf{$2\pi rq$}}}}
\caption{The self-similar fluid flow rate for $n=0.5$ for a different values of the fracture toughness.}
 \label{n050_q}
\end{figure}

As can be seen, over the analyzed range of $N$, the computations are very accurate and converge rapidly as the mesh density is increased. In the viscosity dominated regime it can be seen that there is a lower sensitivity of $e_w$ and $e_v$ to the value of $n$, however even in the toughness dominated mode the  dependence of $e_w$ on the fluid behaviour index becomes less pronounced as $\hat{K}_{Ic}$ grows. A general trend can be observed, in that the convergence rate is magnified as the self-similar material toughness $\hat{K}_{Ic}$ increases. This is due to the fact that, for large values of $\hat{K}_{Ic}$, the solution tends to the limiting case of a uniformly pressurized immobile crack with a parabolic profile. To explain this tendency we present in Figs.~\ref{n050_w}-\ref{n050_q} some additional data for a single value of the fluid behavior index ($n=0.5$).

\begin{figure}[h!]
 \centering
 \includegraphics[scale=0.5]{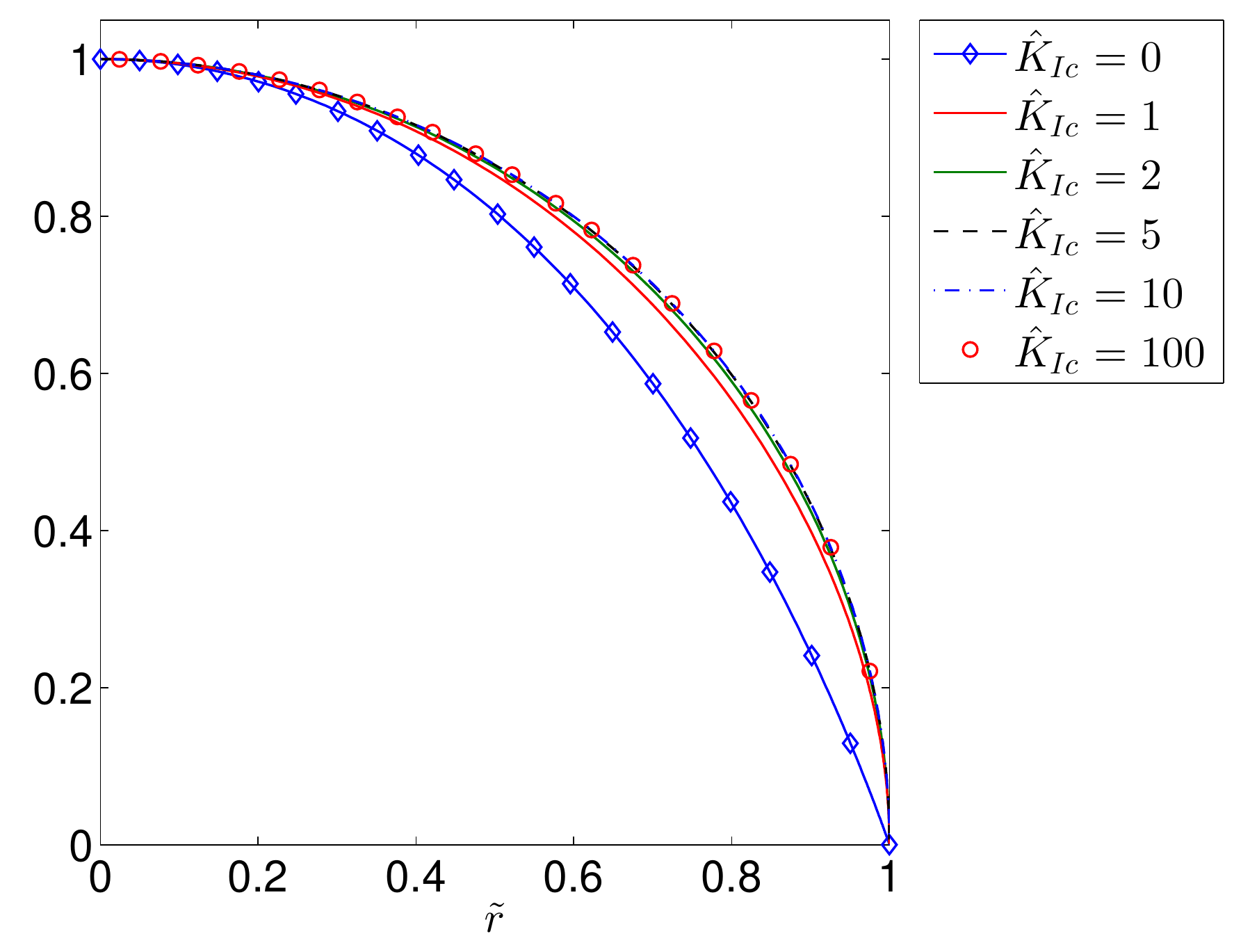}
  \put(-270,195) {{\bf (a)}}					
  \put(-270,95)  {\rotatebox{90}{{\bf{$2\pi r q$}}}}
 \includegraphics[scale=0.5]{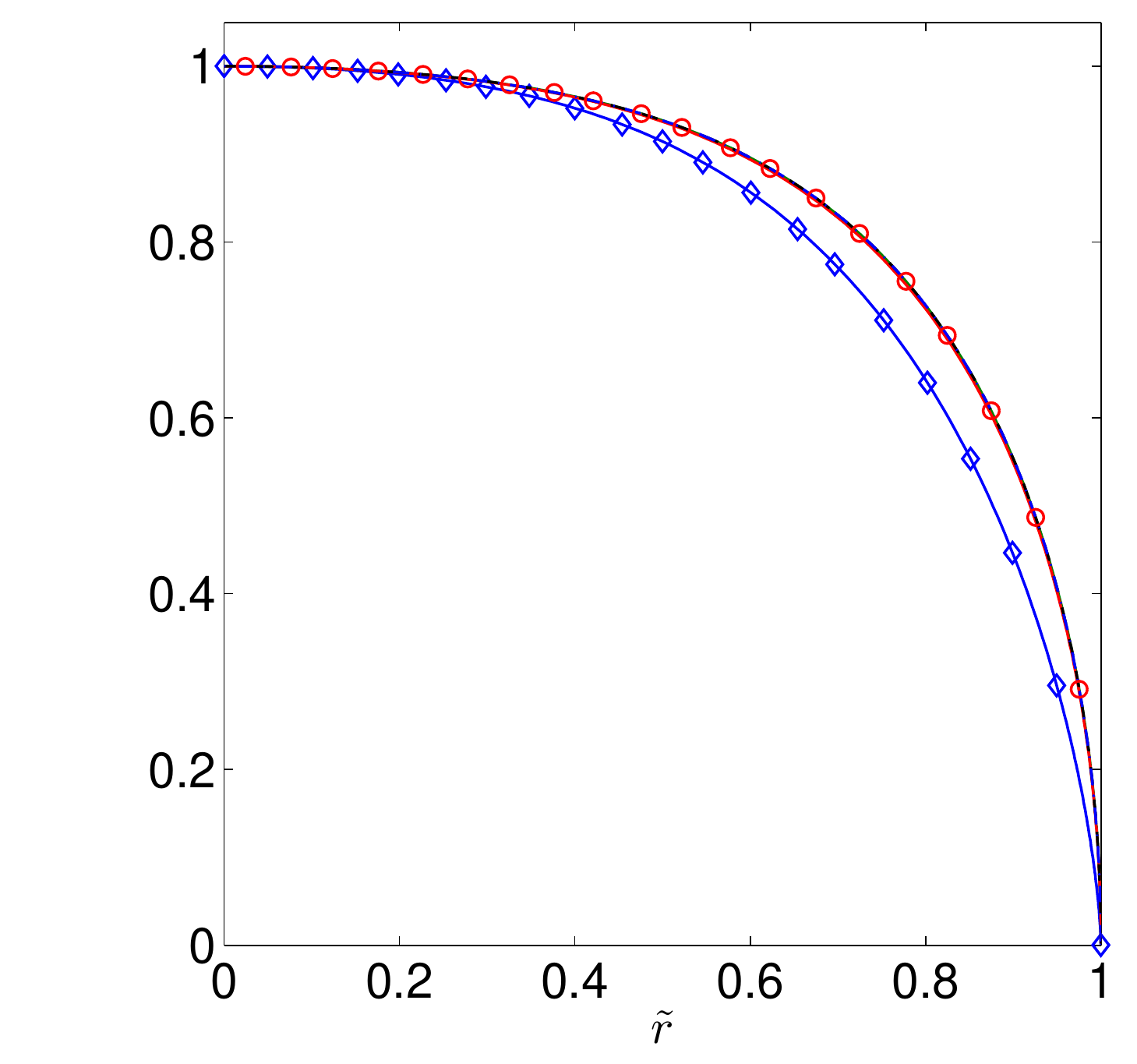}
 \put(-198,195) {{\bf (b)}}
  \put(-198,95)  {\rotatebox{90}{$2\pi r q$}}
 \caption{The self-similar fluid flow rate for a different values of the fracture toughness when the fluid behaviour index is: (a) n=0 and (b) n=1.}
 \label{n01_q}
\end{figure}

It is immediately obvious that for $\hat{K}_{Ic}>2$ the fracture aperture is almost entirely described by the leading term of its crack tip asymptotics (for $\hat{K}_{Ic}=2$ the maximal deviation between them is approximately 1 percent). For the particle velocity it can be seen that, while the effect is not as substantial as for the aperture, the crack propagation speed $\hat{v}_0$ does become a better predictor of the parameter's behaviour for larger values of the material toughness. Meanwhile, the fluid pressure increases with growing $\hat{K}_{Ic}$, eventually becoming uniformly distributed over $\tilde{r}$. As a result of the decreasing pressure gradient the velocity of the fluid flow is reduced. In Fig.~\ref{n050_q} it can be seen that the fluid flow rate rapidly converges to the limiting case with growing $\hat{K}_{Ic}$, however the rate of convergence is greater for larger values of $n$. Indeed, as can be seen in Fig.~\ref{n01_q}, for $n=1$ the curves for $\hat{K}_{Ic}=1$ and $\hat{K}_{Ic}=100$ are indistinguishable, which is not the case when $n=0$.

In fact, the behaviour of the solution as $\hat{K}_{Ic}\to\infty$ can easily be shown to take the form:
\begin{equation}
\hat{w}(\tilde{r}) \sim \frac{4}{\sqrt{\pi}}\hat{K}_I \sqrt{1-\tilde{r}^2} , \quad  \hat{p}(\tilde{r}) \sim \frac{\sqrt{\pi}}{2}\hat{K}_I ,   \quad \hat{v}_0 \sim \frac{3}{8\sqrt{\pi}\hat{K}_I(3-\rho)}, 
\label{KIInftywp}
\end{equation}
\begin{equation}
\tilde{r}\hat{v}(\tilde{r}) = \hat{v}_0 \left[ \tilde{r}^2 + \frac{3-\rho}{3} \left(1-\tilde{r}^2\right)\right] + O\left( \hat{K}_{Ic}^{-1}\right),
\end{equation}
\begin{equation}
\tilde{r}\hat{q}(\tilde{r}) = \frac{\sqrt{1-\tilde{r}^2}}{2\pi}\left[\frac{3\tilde{r}^2}{3-\rho} + \left(1-\tilde{r}^2\right)\right] + O\left( \hat{K}_{Ic}^{-1}\right), 
\label{KIInftyq}
\end{equation}
where $\rho$ is defined in Table~\ref{table:S1}. As a result the computations become far more efficient in this case and the resulting solution is calculated to a far higher level of accuracy.


Combining the results shown above in Figs.~\ref{Acc_Visc_01} - \ref{Fig:KI100}, it is clear that the computations presented here achieve a very high level of accuracy for both the aperture and particle velocity regardless of the crack propagation  regime. When using $N=300$  the accuracy of computations can almost always be assumed to be correct to a level of at least $10^{-7}$ for the fracture aperture, and $2.5\times 10^{-7}$ for the particle velocity. In this way the obtained data constitutes a very convenient and credible reference solution when testing other computational schemes.


It is worth mentioning that the efficiency of computations achieved by this algorithm means that this high level of accuracy does not come at the expense of simulation time. The final algorithm requires fewer than $20$ iterations to produce a solution. Simulation times are also very short with this scheme.



\subsubsection{Semi-analytical benchmark solutions} \label{SemiAnal}

While the numerical solutions provided above allow for the problem of a penny-shaped radial fracture to be solved rapidly, they are not necessarily in a form which can be easily utilized when testing various computational algorithms. Following the idea from \cite{Perkowska2015}, we shall also deliver simple and accurate semi-analytical approximations of the numerical solutions from the previous subsection, which can easily be used as reference data without the need for advanced computational programs. We provide below formulae mimicking the crack aperture, the particle velocity and the net fluid pressure.



All the proposed proposed relations preserve the proper asymptotic behaviour at both the fracture origin and tip. They were computed by taking solutions between $n=0.05$ and $n=0.95$, with a step-size of $n=0.05$, and creating approximation functions which predicted each parameter to a desired accuracy. These approximate solution components were then tested against results with a step-size of $n=0.025$, to ensure that the predictions were accurate over the whole range. Respective coefficients (provided in Appendix~\ref{App:SemiAnal}) used in the approximations have no set length, as the final accuracy of the solution was the deciding factor in their construction.

As a result of this approach each approximated parameter should be treated independently, which means that the guaranteed accuracy does not embrace the mutual interrelations between respective variables (e.g. the particle velocity computed according to \eqref{PoisevilleS1} from the approximate $\hat w$ and $\hat p$ is not expected to give the same accuracy as that provided by the approximation for $\hat v$). Moreover, the high level of accuracy of the approximate formulae is guaranteed over the following interval of the fluid behaviour index: $0.05<n<0.95$. The approximations for the limiting cases $n=0$ and $n=1$ are given separately in Appendix~\ref{Append:Cases}.\\



\begin{itemize}
 \item {\bf{Viscosity dominated regime ($K_{Ic}=0$)}}
\end{itemize}
For the viscosity dominated regime we propose the following approximations of the dependent variables:
\begin{equation}
\begin{aligned}
\hat{w}_{apx} (\tilde{r},n)=&{w}_0 \biggl[ (1-\tilde{r}^2)^{\alpha_0}+w_1 (1-\tilde{r}^2)^{\alpha_1} +w_2 f_2(\tilde{r}) +w_3 (1-\tilde{r}^2)^{\alpha_1+1}\tilde{r}^{2-n} +  \\
& \quad w_4 (1-\tilde{r}^2)^{\alpha_1+2}\tilde{r}^{2-n} + w_5(1-\tilde{r}^2)^{5/2}\tilde{r}^{3-n} +w_6 f_{1}(\tilde{r}) \biggr] ,
\label{apxWKI0}
\end{aligned}
\end{equation}
\begin{equation}
\tilde{r}\hat{v}_{apx}(\tilde{r},n) = v_1 + v_2 (1-\tilde{r}^2)+v_3 \tilde{r}^{2-n}+v_4(1-\tilde{r}^2)^{{\beta_2}} \tilde{r}^2 ,
\label{apxVKI0}
\end{equation}
\begin{equation}
 \begin{aligned}
\hat{p}_{apx}(\tilde{r},n) &= \hat{C}_p(n) + p_1 \tilde{r}^{1-n} + p_2 \tilde{r} \left(1-\tilde{r}^2\right)^{\alpha_0-1} + \frac{p_3}{n} + p_4 \tilde{r}\sqrt{1-\tilde{r}} \\
&\quad + \frac{p_5}{n}\left(1-\tilde{r}\right)^{\alpha_1-1} + p_6\left(1-\tilde{r}\right)^{\alpha_1}  ,
 \end{aligned}
\label{apxPKI0}
\end{equation}
\begin{equation}
\hat{v}_{0,apx} (n) = \sum_{i=0}^7 C_i n^i , \quad \hat{C}_p (n) = \frac{\sum_{i=0}^1 D_i n^i}{\sum_{k=0}^3 X_k n^k} ,
\label{v0apx}
\end{equation}
with:
\begin{equation}
f_{1}( \tilde{r} )=\sqrt{1-\tilde{r}^2}-\frac{2}{3}(1-\tilde{r}^2)^{3/2}-\tilde{r}^2\log\left|\frac{1+\sqrt{1-\tilde{r}^2}}{\tilde{r}} \right| ,
\label{fOmeg}
\end{equation}
\begin{equation}
f_2(\tilde{r})=2 \sqrt{1-\tilde{r}^2}+\tilde{r}^2 \log\left(\frac{1-\sqrt{1-\tilde{r}^2}}{1+\sqrt{1-\tilde{r}^2}}\right) .
\end{equation}
The coefficients $w_i (n)$, $v_i (n)$, $p_i (n)$, $C_i$, $D_i$, $X_k$ are given in Appendix~\ref{App:SemiAnal}, while $\alpha_0$, $\alpha_1$ and $\beta_2$ can be found in Table~\ref{table:albe}. This formulation is valid for all $0.05< n < 0.95$, with any modifications required in the limiting cases $n =0$ and $n =1$  being outlined in Appendix~\ref{Append:Cases}.

\begin{figure}[h!]
 \centering
 \includegraphics[scale=0.45]{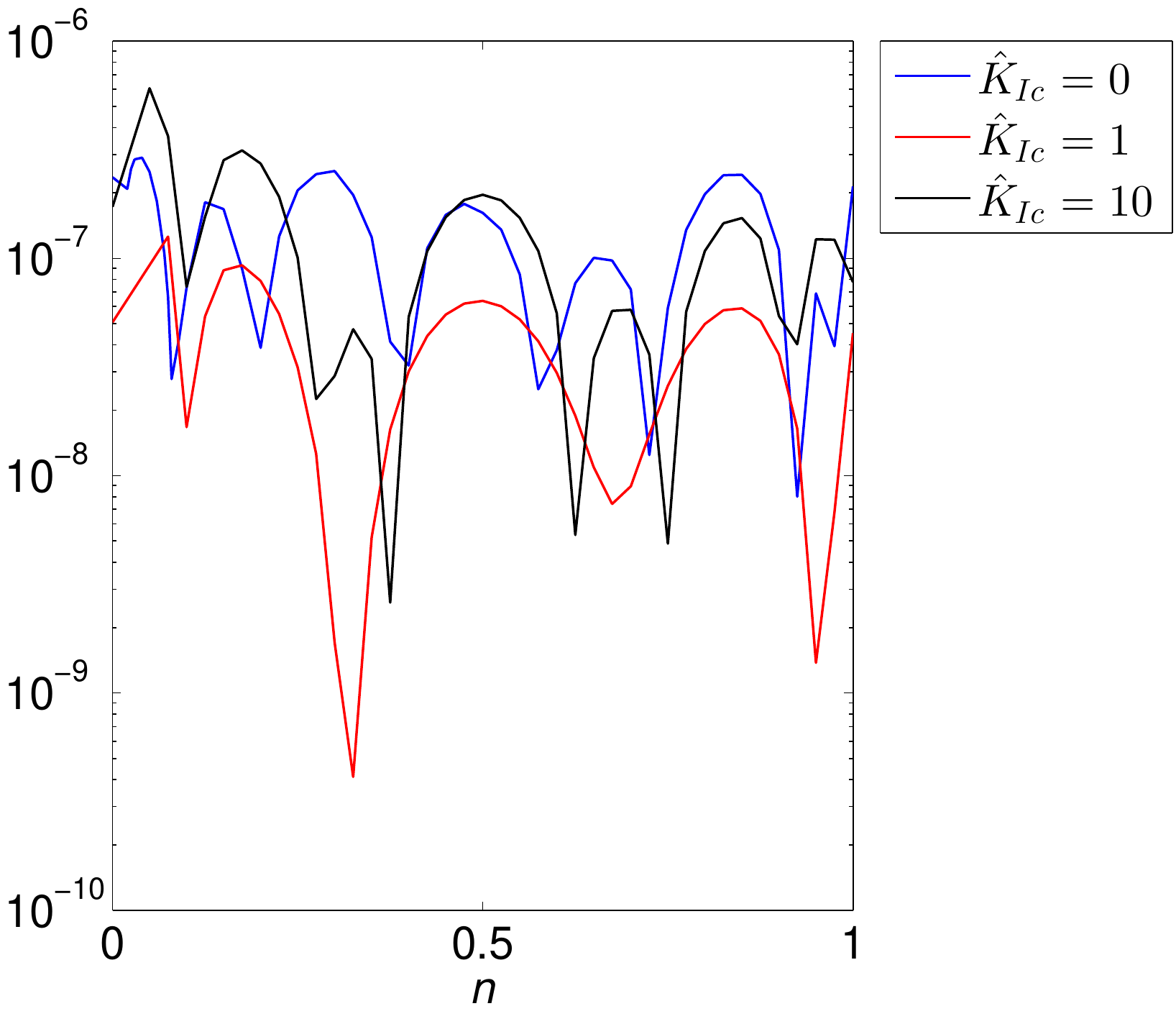}
 \put(-225,95) {\rotatebox{90}{{\bf{$\delta \hat{v}_{0,apx}$}}}}
 \caption{Relative error of approximation for the self-similar crack propagation speed $\hat{v}_0$ when evaluated using the specialized equation for $\hat{v}_{0,apx}$ \eqref{v0apx}$_1$. }
 \label{Fig:ApxErV0}
\end{figure}

Although the self-similar crack propagation speed $\hat{v}_0$ can be obtained by evaluating the general formula \eqref{apxVKI0} at the fracture front, an alternative expression \eqref{v0apx}$_1$ has been introduced. This is to ensure the highest possible level of accuracy for this important parameter, which is needed both to compute the fracture length $L(\tilde{t})$, as well as the transformations to alternative schemes in the literature (e.g. \eqref{TransLinkov}). The error of approximation of $\hat{v}_{0}$ for all considered values of the material toughness $\hat{K}_{Ic}$ is provided in Fig.~\ref{Fig:ApxErV0}.

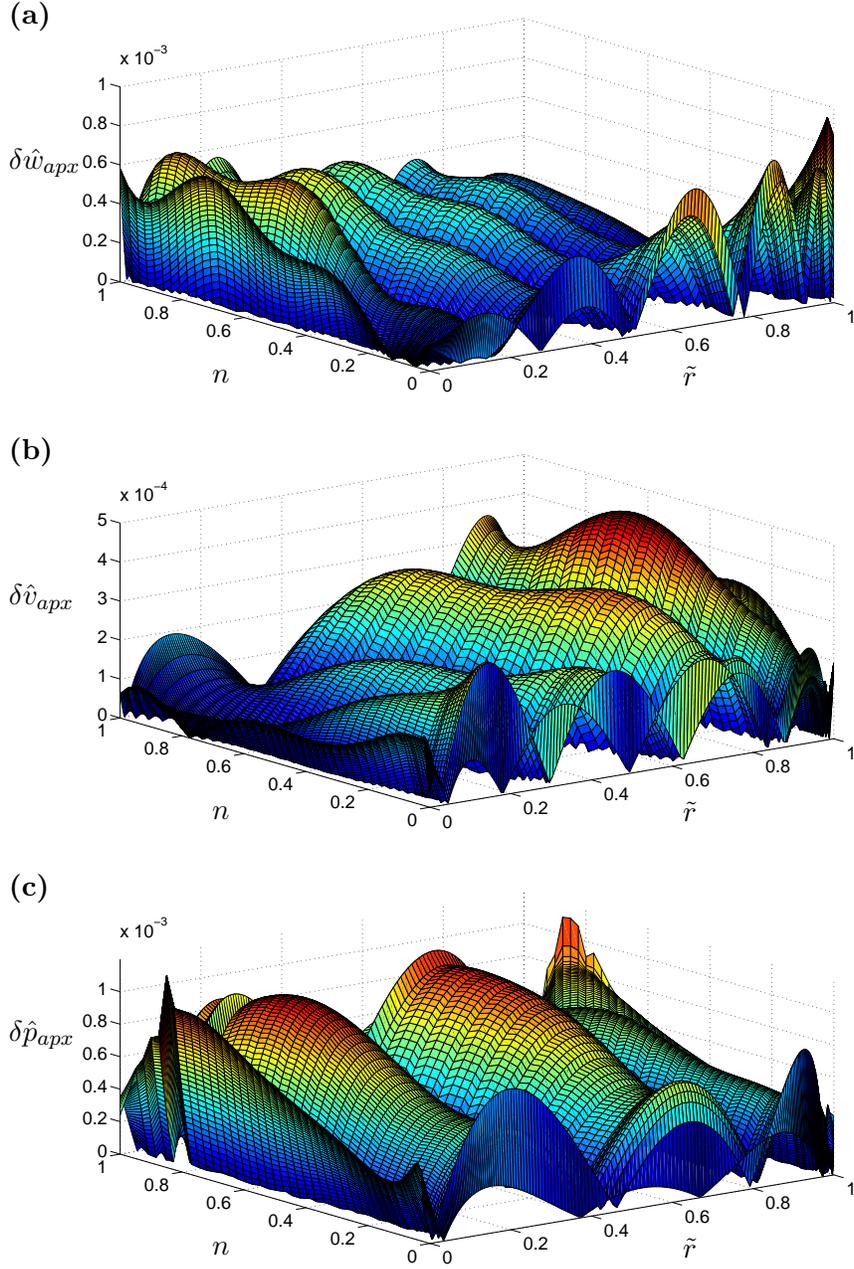
\begin{figure}[h!]
 \centering
 \includegraphics[scale=0.375]{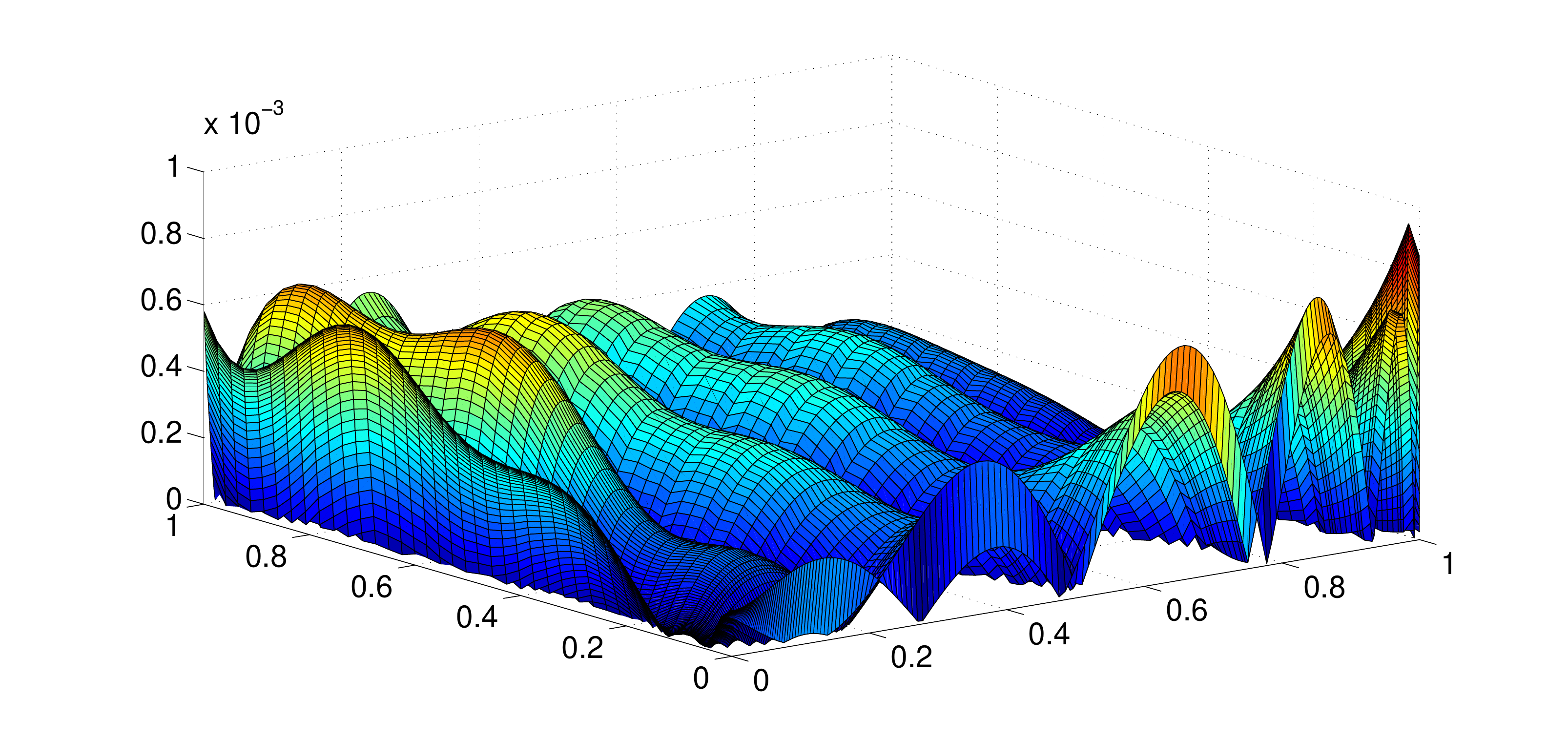}
 \put(-345,150) {{\bf{(a)}}}
 \put(-345,95) {\bf{$\delta\hat{w}_{apx}$}}
 \put(-268,14) {\bf{$n$}}
 \put(-90,13) {\bf{$\tilde{r}$}}

 \includegraphics[scale=0.375]{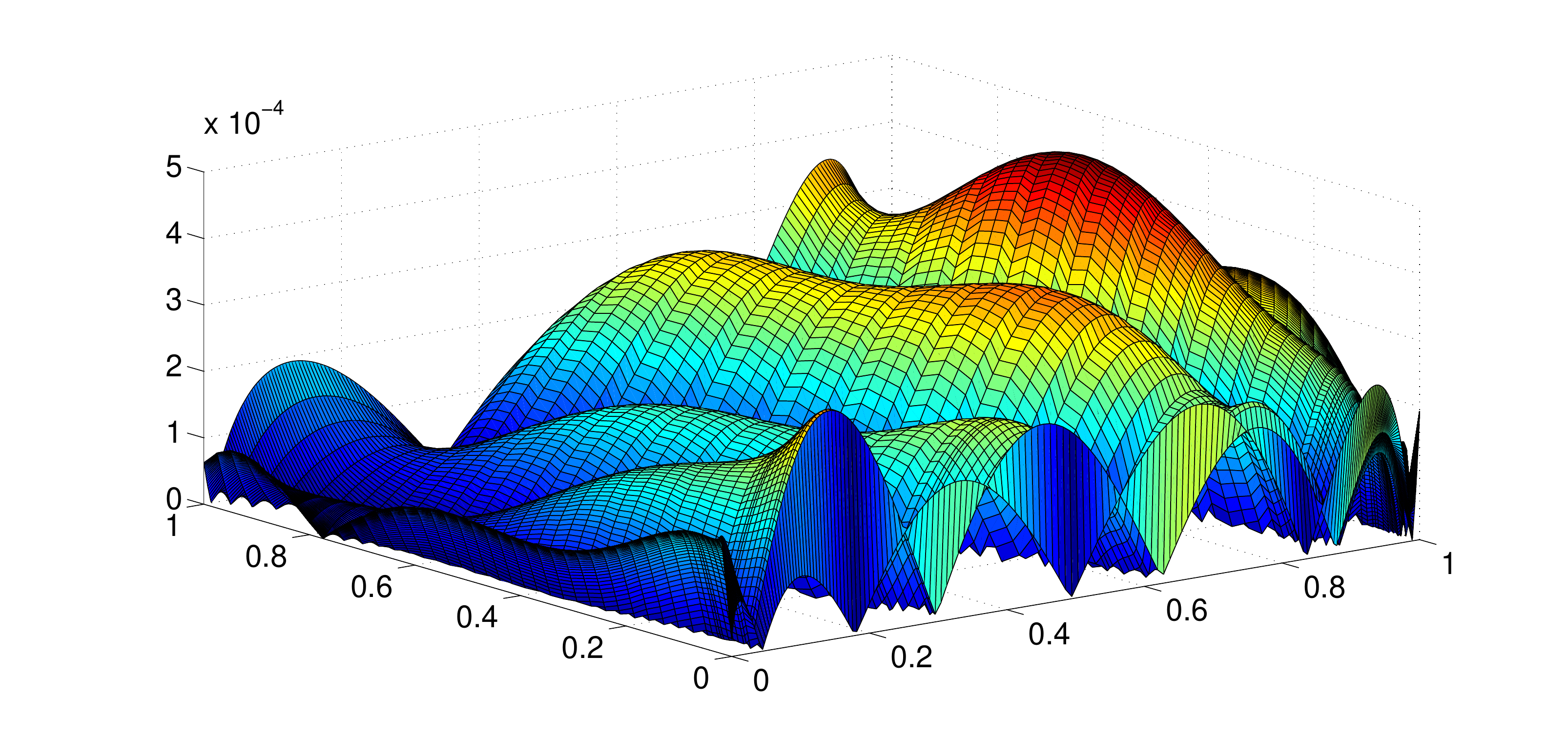}
 \put(-345,150) {{\bf{(b)}}}
 \put(-345,95) {\bf{$\delta\hat{v}_{apx}$}}
 \put(-268,14) {\bf{$n$}}
 \put(-90,13) {\bf{$\tilde{r}$}}

 \includegraphics[scale=0.375]{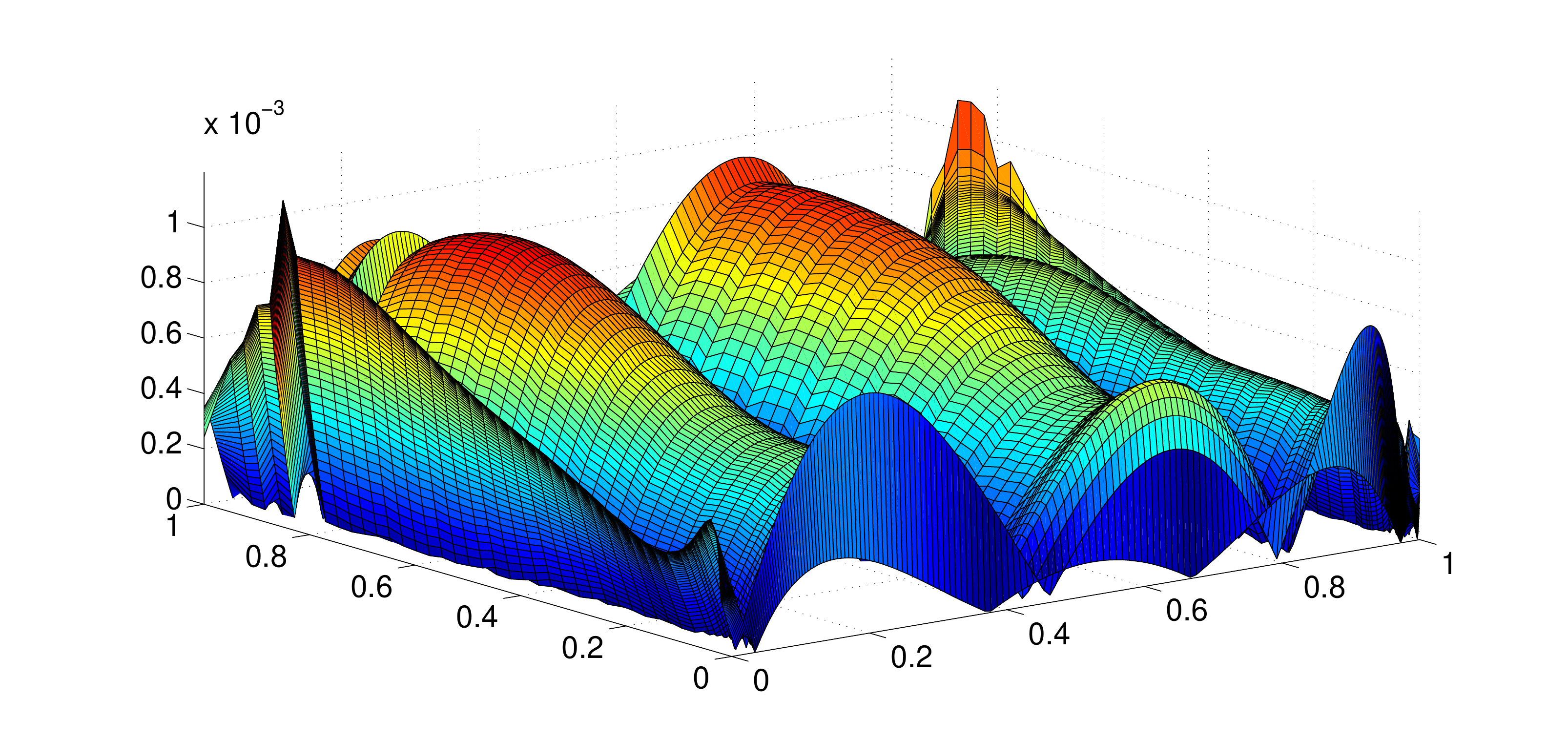}
 \put(-345,150) {{\bf{(c)}}}
 \put(-345,95) {\bf{$\delta\hat{p}_{apx}$}}
 \put(-268,14) {\bf{$n$}}
 \put(-90,13) {\bf{$\tilde{r}$}}
 \caption{Relative error of the approximations of the numerical solution for (a) the aperture \eqref{apxWKI0}, (b) the particle velocity \eqref{apxVKI0}, and the absolute error of approximation of the numerical solution for (c) the pressure \eqref{apxPKI0}, in the viscosity dominated regime ($\hat{K}_{Ic}=0$).}
 \label{Fig:ApxErKI0}
\end{figure}

Graphs demonstrating the accuracy of approximations for the aperture, particle velocity and pressure are provided in Fig.~\ref{Fig:ApxErKI0}. The respective error measures are defined as:
\begin{equation}
\delta\hat{w}_{apx}(\tilde{r},n) = \frac{ | \hat{w}_n (\tilde{r}) - \hat{w}_{apx}(\tilde{r},n)|}{\hat{w}_n (\tilde{r})} , \quad \delta\hat{v}_{apx}(\tilde{r},n) = \frac{ |  \hat{v}_n (\tilde{r}) - \hat{v}_{apx}(\tilde{r},n) |}{\hat{v}_n(\tilde{r})} ,
\end{equation}
\begin{equation}
\delta \hat{v}_{0,apx} (\tilde{n}) = \frac{| \hat{v}_{0,n} - \hat{v}_{0,apx}(n) |}{\hat{v}_{0,n}} , \quad
\delta\hat{p}_{apx}(\tilde{r},n) = | \hat{p}_n (\tilde{r}) - \hat{p}_{apx}(\tilde{r},n)| ,
\end{equation}
where $\hat{w}_n (\tilde{r})$, $\hat{v}_n (\tilde{r})$, $\hat{v}_{0,n}$ and $\hat{p}_n (\tilde{r})$ are the benchmark solutions obtained by the computational algorithm for a given value of the fluid behaviour index $n$.

It can easily be seen that the relative accuracy of the formulae for $\hat{w}_{apx}$, $\hat{v}_{apx}$, and absolute accuracy for $\hat{p}_{apx}$, are of the order $10^{-4}$ over almost the entire interval of $n$. Only for $n=0$ does the error of $\hat{w}_{apx}$ slightly exceed $10^{-3}$, while the accuracy of the pressure approximation falls below $10^{-3}$ for specific values of $n>0.8$. The accuracy of $\hat{v}_{0,apx}$, computed from \eqref{v0apx}$_1$, is reported in Fig.~\ref{Fig:ApxErV0}. It shows that the relative error is below $2\times 10^{-6}$ for any value of the fluid behaviour index.



\begin{itemize}
 \item {\bf{Toughness dominated regime ($K_{Ic}>0$)}}
\end{itemize}
In this case the form of the self-similar crack propagation speed approximation, $\hat{v}_{0,apx}$, remains as in \eqref{v0apx}$_1$. The other solution components are given in the form:
\begin{equation}
\begin{aligned}
\hat{w}_{apx} (\tilde{r},n)=&\hat{w}_0 \biggl[\sqrt{1-\tilde{r}^2}+w_1(1-\tilde{r}^2)^{\alpha_1}+w_2 (1-\tilde{r}^2)^{3/2}\log(1-\tilde{r}^2)+ \\
&\quad w_3(1-\tilde{r}^2)^{3/2}+  w_4 \tilde{r} (1-\tilde{r}^2)^{\alpha_2}+w_5 f_{1}(\tilde{r}) \biggr] ,
\label{apxWKI1}
\end{aligned}
\end{equation}
\begin{equation}
\tilde{r} \hat{v}_{apx} (\tilde{r},n) = v_1+v_2(1-\tilde{r}^2)^{\beta_1}+v_3 \tilde{r}^{2-n}+v_4(1-\tilde{r}^2) ,
\label{apxVKI1}
\end{equation}
\begin{equation}
\hat{p}_{apx}(\tilde{r},n)=p_1+p_2 f_3 (\tilde{r},n)+ p_3 (1-\tilde{r}^2)^{\alpha_1-1}+p_4 \tilde{r}^{1-n} ,
\label{apxPKI1}
\end{equation}
with:
\begin{equation}
f_3 (\tilde{r},n) =\alpha_1 \sqrt{\pi} \frac{\Gamma(\alpha_1)}{\Gamma(\alpha_1+1/2)}  {_2}F_1 \left(1, \frac{n-2}{2},\frac{1}{2}, r^2\right),
\end{equation}
where $\hat{w}_0$ is given by \eqref{w0asym1}, $f_1$ takes the form \eqref{fOmeg}, and $\alpha_1$ is in Table~\ref{table:albe}. The coefficients $w_i (n)$, $v_i (n)$, $p_i(n)$, $C_i$ are given in Appendix~\ref{App:SemiAnal} for $\hat{K}_I=\left\{ 1 , 10 \right\}$. For $n=\left\{0,1\right\}$ some parameters require alternate representations, which are outlined in Appendix~\ref{Append:Cases}.

This time the quality of approximations is better than those for the viscosity dominated regime. For $\hat{K}_{Ic}=1$ the approximation errors do not exceed $3\times 10^{-4}$ regardless of the considered variable or the value of the fluid behaviour index $n$. When analyzing the case $\hat{K}_{Ic}=10$ one can see that the accuracy of approximations improved even further, being up to two orders of magnitude better than that for $\hat{K}_{Ic}=1$.

\begin{figure}[h!]
 \centering
 \includegraphics[scale=0.375]{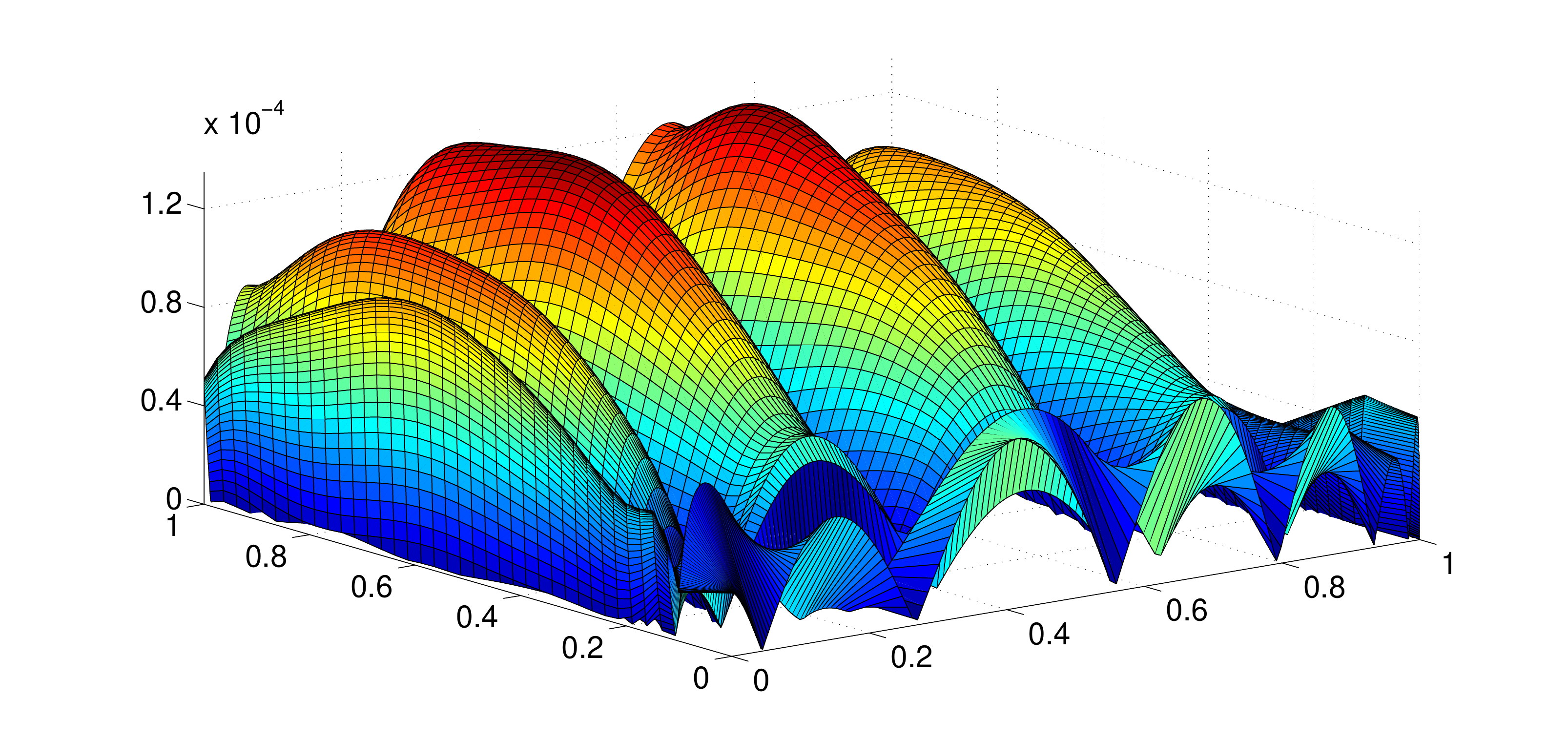}
 \put(-345,150) {{\bf{(a)}}}
 \put(-345,95) {\bf{$\delta\hat{w}_{apx}$}}
 \put(-268,14) {\bf{$n$}}
 \put(-90,13) {\bf{$\tilde{r}$}}

 \includegraphics[scale=0.375]{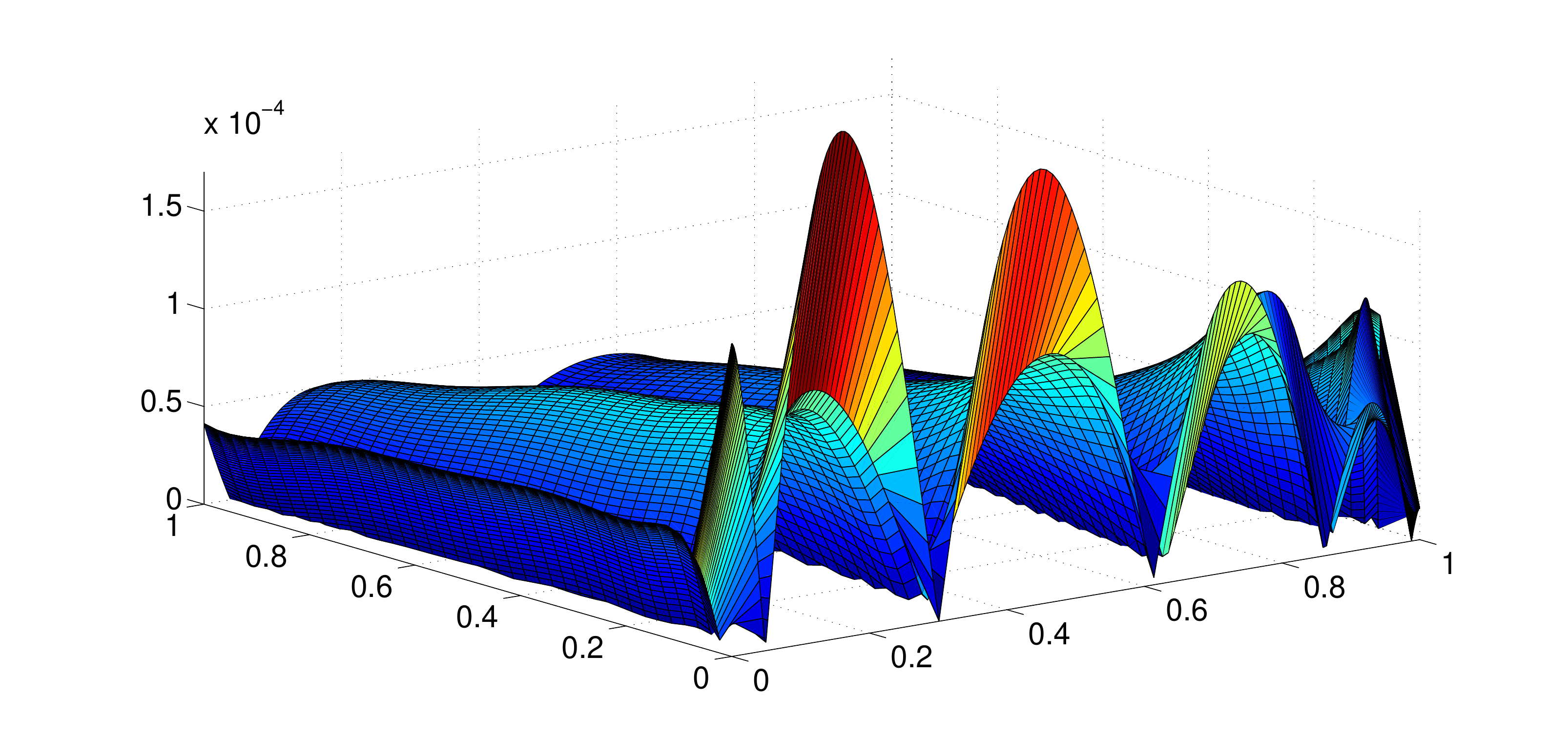}
 \put(-345,150) {{\bf{(b)}}}
 \put(-345,95) {\bf{$\delta\hat{v}_{apx}$}}
 \put(-268,14) {\bf{$n$}}
 \put(-90,13) {\bf{$\tilde{r}$}}

 \includegraphics[scale=0.375]{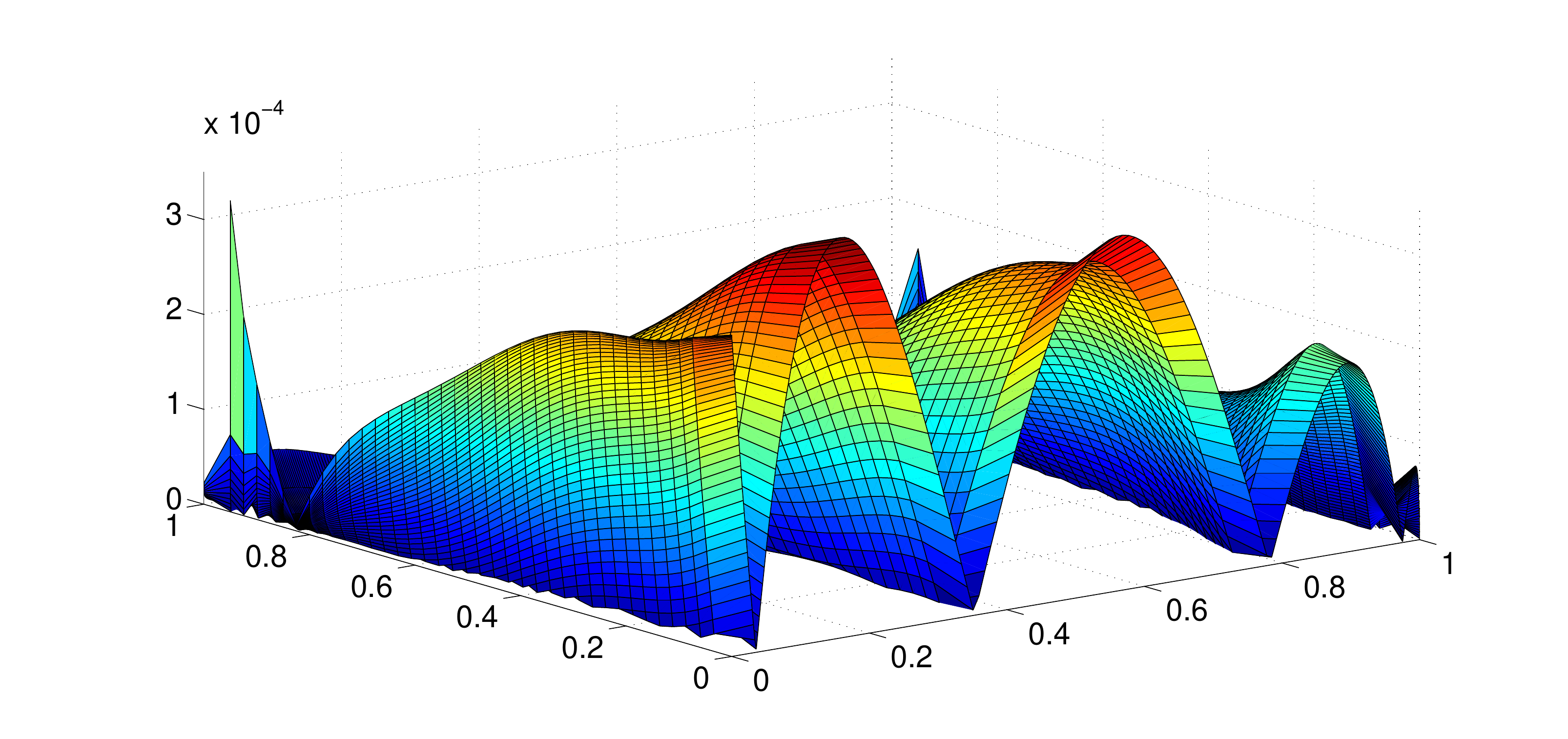}
 \put(-345,150) {{\bf{(c)}}}
 \put(-345,95) {\bf{$\delta\hat{p}_{apx}$}}
 \put(-268,14) {\bf{$n$}}
 \put(-90,13) {\bf{$\tilde{r}$}}
 \caption{Relative error of the approximations of the numerical solution for (a) the aperture \eqref{apxWKI1}, (b) the particle velocity \eqref{apxVKI1}, and the absolute error of approximation of the numerical solution for (c) the pressure \eqref{apxPKI1}, in the toughness dominated regime with $\hat{K}_I=1$.}
 \label{Fig:ApxErKI1}
\end{figure}

\begin{figure}[h!]
 \centering
  \includegraphics[scale=0.375]{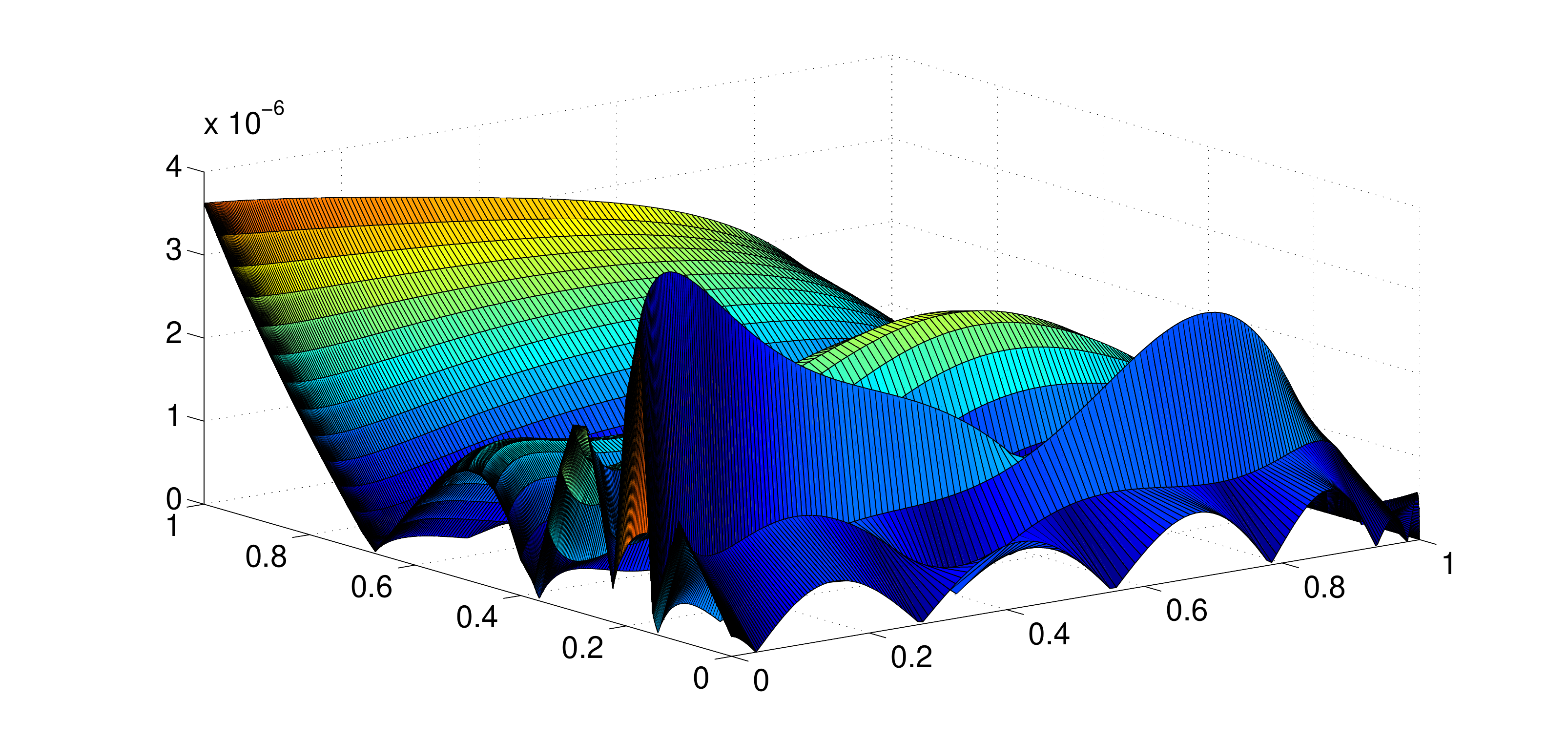}
 \put(-345,150) {{\bf{(a)}}}
 \put(-345,95) {\bf{$\delta\hat{w}_{apx}$}}
 \put(-268,14) {\bf{$n$}}
 \put(-90,13) {\bf{$\tilde{r}$}}

 \includegraphics[scale=0.375]{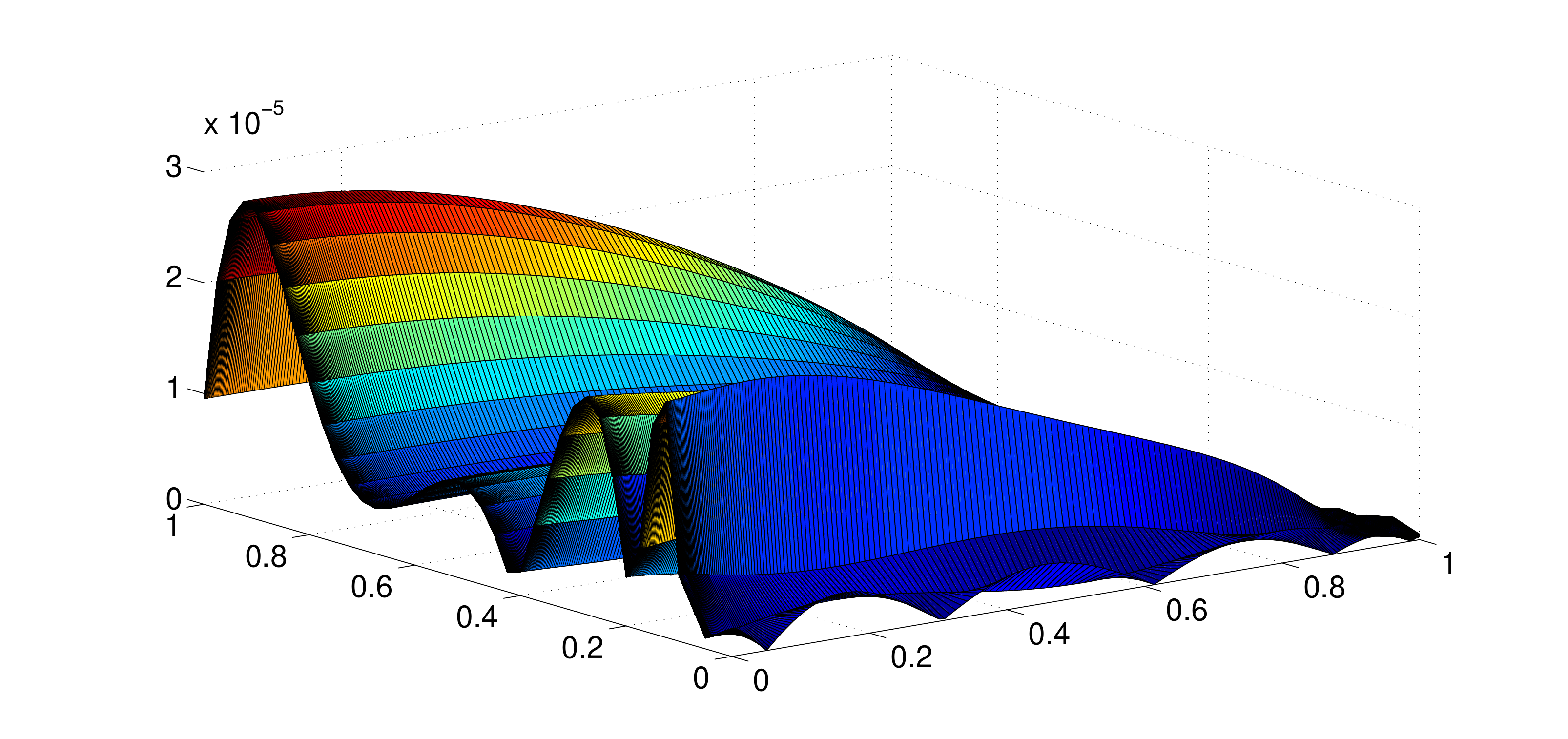}
 \put(-345,150) {{\bf{(b)}}}
 \put(-345,95) {\bf{$\delta\hat{v}_{apx}$}}
 \put(-268,14) {\bf{$n$}}
 \put(-90,13) {\bf{$\tilde{r}$}}

 \includegraphics[scale=0.375]{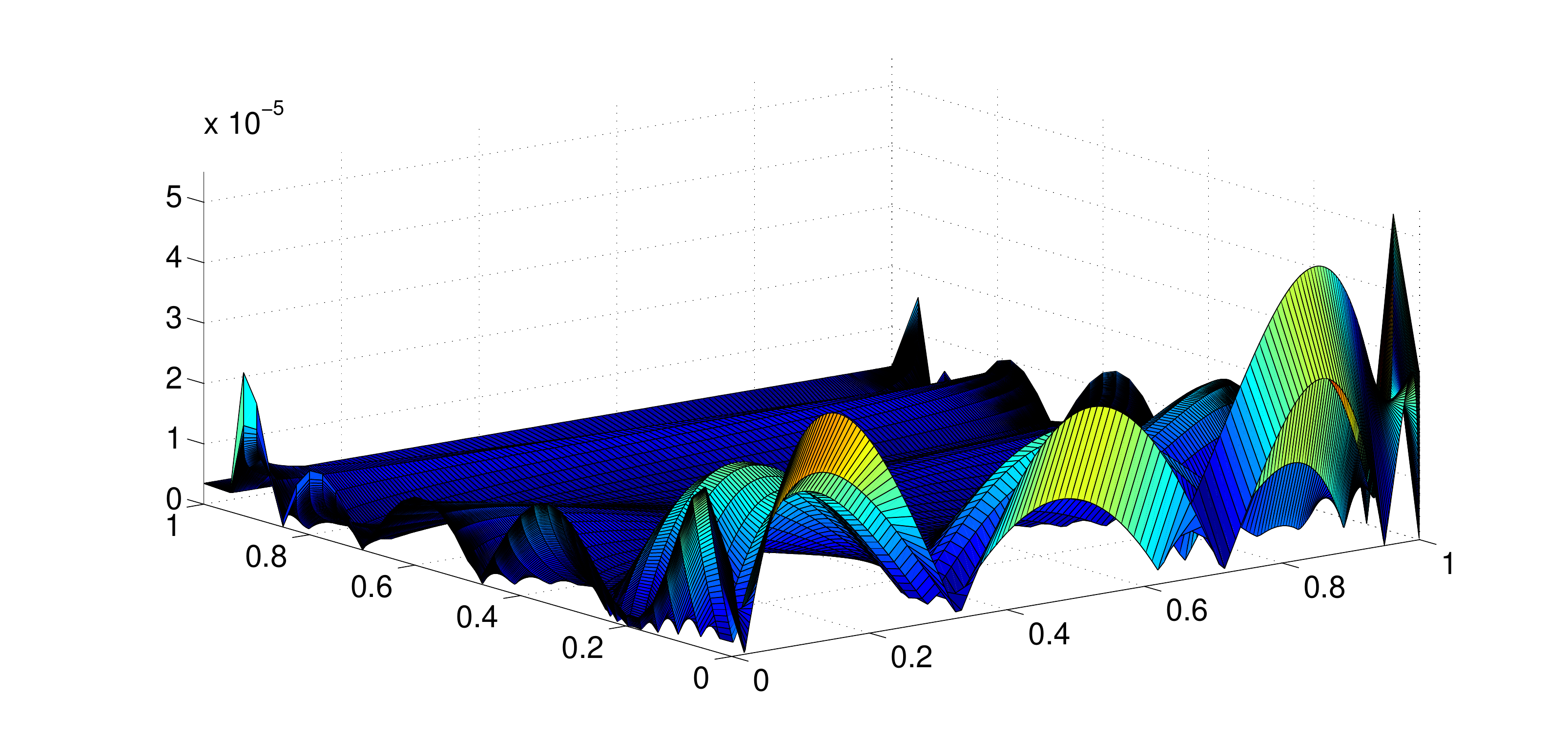}
 \put(-345,150) {{\bf{(c)}}}
 \put(-345,95) {\bf{$\delta\hat{p}_{apx}$}}
 \put(-268,14) {\bf{$n$}}
 \put(-90,13) {\bf{$\tilde{r}$}}
 \caption{Relative error of the approximations of the numerical solution for (a) the aperture \eqref{apxWKI1}, (b) the particle velocity \eqref{apxVKI1}, and the absolute error of approximation of the numerical solution for (c) the pressure \eqref{apxPKI1}, in the toughness dominated regime with $\hat{K}_I=10$.}
 \label{Fig:ApxErKI10}
\end{figure}

\subsubsection{Verification of other results from the literature} \label{Sect:Comp}

In the following, using our highly accurate numerical scheme, we will verify the results provided so far by other authors. Unfortunately, there are only a handful of papers where respective data is provided in a form which enables comparison. In most cases only graphs of the dependent variables are given. In order to make sure that the data is comparable the zero leak-off case will again be examined, taking fixed $\hat{Q}_0=1$, with transformations between the schemes outlined as necessary. Throughout this section we will use $N=300$ nodal points, which in previous sections we have shown is accurate to $7$ significant digits.

We begin by analyzing the solution delivered by \emph{Linkov} in \cite{Linkov2016} for the viscosity dominated regime ($\hat{K}_{Ic}=0$). Note that, as slightly different normalizations are used to obtain the self-similar solution, the following transformations are required to obtain a comparison between the results:
$$
\hat{w} (\tilde{r}) = \zeta^{\frac{n}{n+2}} \hat{w}^L (\tilde{r} ), \quad \hat{p} (\tilde{r}) = \zeta^{\frac{n}{n+2}} \hat{p}^L (\tilde{r} ) ,
$$
\begin{equation}
 \hat{v} (\tilde{r}) =  \zeta \hat{v}^L (\tilde{r} ) , \quad \hat{Q}_0= \frac{1}{\xi_{*,n}^3}  \zeta^{\frac{2\left(n^2+2\right)}{n+2}} \hat{Q}_0^L ,
\label{TransLinkov}
\end{equation}
$$
\hat{q}_l (\tilde{r}) =  \zeta^{\frac{n}{n+2}}  \hat{q}_l^L (\tilde{r} ) , \quad \xi_{*,n}= \left(2\pi \int_0^1  \varsigma \hat{w}^L (\varsigma ) \, d\varsigma \right)^{-\frac{1}{3}} ,
$$
where:
\begin{equation}
\zeta = \frac{3\hat{v}_0\left(n+2\right)}{2n+2} .
\end{equation}
Here $\xi_{*,n}$ is \emph{Linkov's} normalized fracture length when $Q_0=1$. It can easily be shown using the equation for fracture length from Table~\ref{table:S1} that, in order for the two formulations to coincide, the following scaling condition must be met:
\begin{equation}
\xi_{*,n}= \zeta^{\frac{2\left(n+1\right)}{3\left(n+2\right)}} .
\end{equation}
The values of the self-similar fracture opening, crack propagation speed and fracture half-length are shown in Table~\ref{table:R1}. The results obtained in \cite{Linkov2016} are included for completeness, and denoted with a superscript $L$. The notation $\hat{w}^T (0)$ represents the transformed crack opening computed according to \eqref{TransLinkov}$_1$ (this value is to be compared with $\hat{w}^L (0)$).

\begin{table}[h!]
    \centering
\begin{tabular}{||c||c|c||c|c||c|c||}
\hline
n & $\hat{v}_0$ & $\hat{w}(0)$ & $\hat{w}^T (0)$ & $\xi_{*,n}$  & $\hat{w}^L (0)$ & $\xi_{*,n}^L$  \\ [0.5ex]
\hline \hline
0 & 0.1314342 & 1.688787  &  1.688787 & 0.7332914 & 1.6889  & 0.7330 \\
\hline
0.1 & 0.1427914  & 1.602559 & 1.672277 & 0.7317711 & 1.6724 & 0.7318  \\
\hline
0.2 & 0.1527660  & 1.535686 & 1.661661 & 0.7295243 & 1.6617 & 0.7296  \\
\hline
0.3 &  0.1615208  & 1.482567 & 1.655773 & 0.7267291 & &  \\
\hline
0.4 &  0.1691971 & 1.439637 & 1.653833  & 0.7235073 & 1.6537 & 0.7236 \\
\hline
0.5  &  0.1759138  & 1.404539 & 1.655334 & 0.7199395  &  & \\
\hline
0.6 & 0.1817680 & 1.375680 & 1.659981 & 0.7160755  & 1.6599 & 0.7162  \\
\hline
0.7 & 0.1868366  & 1.351968 & 1.667648 & 0.7119399 &  & \\
\hline
0.8  &  0.1911776  & 1.332662 & 1.678369 & 0.7075363 & 1.6784 & 0.7076 \\
\hline
0.9 & 0.1948308  & 1.317280 & 1.692338 & 0.7028480 &  & \\
\hline
1  & 0.1978175  & 1.305555 & 1.709934 & 0.6978375 & 1.7092  &  0.6978\\
\hline \hline
\end{tabular}
 \caption{The values of fracture opening, crack propagation speed and half-length, given to an accuracy of seven significant figures (which defines the solution accuracy achievable for $N=300$ using the authors' solver). The final two columns, denoted with superscript $L$, show the values provided in \cite{Linkov2016}. The symbols $\hat{w}^T$ and $\xi_{*,n}$ stand for the transformed fracture opening and fracture half-length computed according to \eqref{TransLinkov}. These values are to be compared with the last two columns.}
\label{table:R1}
\end{table}

It can easily be seen that there is a high level of correspondence between the results in this paper and those provided by \emph{Linkov} for different values of the fluid behaviour index $n$. The maximum relative discrepancy is of the order $4.3\times 10^{-4}$, which considering the accuracy of our solution demonstrated in Sect.~\ref{Sect:Acc2}, describes the level of accuracy achieved by the solution from \cite{Linkov2016}. We note that, in our approach, it is sufficient to take merely $N=40$ points to have a similar accuracy (see Figs.~\ref{Acc_Visc_01}-\ref{Fig:KI10}).

Another solution to be analyzed is that from \emph{Savitski/Detournay} \cite{Savitski2002}, which provides asymptotic approximations for both the viscosity and toughness dominated regimes in the case of a Newtonian fracturing fluid. The interrelations between the self-similar crack opening and crack propagation speed given in \cite{Savitski2002} and our results are as follows:
\begin{equation}
\bar{\Omega}_{m,0}(\tilde{r}) = \left[\frac{4}{9\hat{v}_0}\right]^{\frac{1}{3}} \hat{w}(\tilde{r}) , \quad V(\tilde{r})=\frac{4}{9 \hat{v}_0} \hat{v}(\tilde{r}) .
\end{equation}
\emph{Savitski/Detournay} specify the following asymptotic approximation for the self-similar aperture:
\begin{equation}
\bar{\Omega}_{m,0} (\tilde{r}) =
2^{\frac{1}{3}} \times 3^{\frac{1}{6}} \left(1-\tilde{r}^2 \right)^{\frac{2}{3}} + O\left(\left(1-\tilde{r}^2 \right)^{\frac{5}{3}} \right) ,\quad  \tilde{r}\to 1 .
\end{equation}
Using the relevant transformations yields:
\begin{equation}
\hat{w}(\tilde{r}) = 2^{\frac{1}{3}} \times 3^{\frac{1}{6}} \left[\frac{9\hat{v}_0}{4}\right]^{\frac{1}{3}} \left(1-\tilde{r}^2 \right)^{\frac{2}{3}} + O\left(\left(1-\tilde{r}^2 \right)^{\frac{5}{3}} \right), \quad \tilde{r}\to 1.
\label{SavitskiEnd}
\end{equation}
Note that interrelation between $\hat w_0$ and $\hat v_0$ resulting from \eqref{SavitskiEnd} is exactly the same as the one defined by equations \eqref{particlevN3}-\eqref{w0v0_1} based on the speed equation. Thus, any solution in the viscosity dominated regime (for $n=1$) preserving the latter will be equivalent in terms of $\hat w_0$ and $\hat v_0$ to the data provided in \cite{Savitski2002}.

\begin{figure}[h!]
 \centering
\includegraphics[scale=0.5]{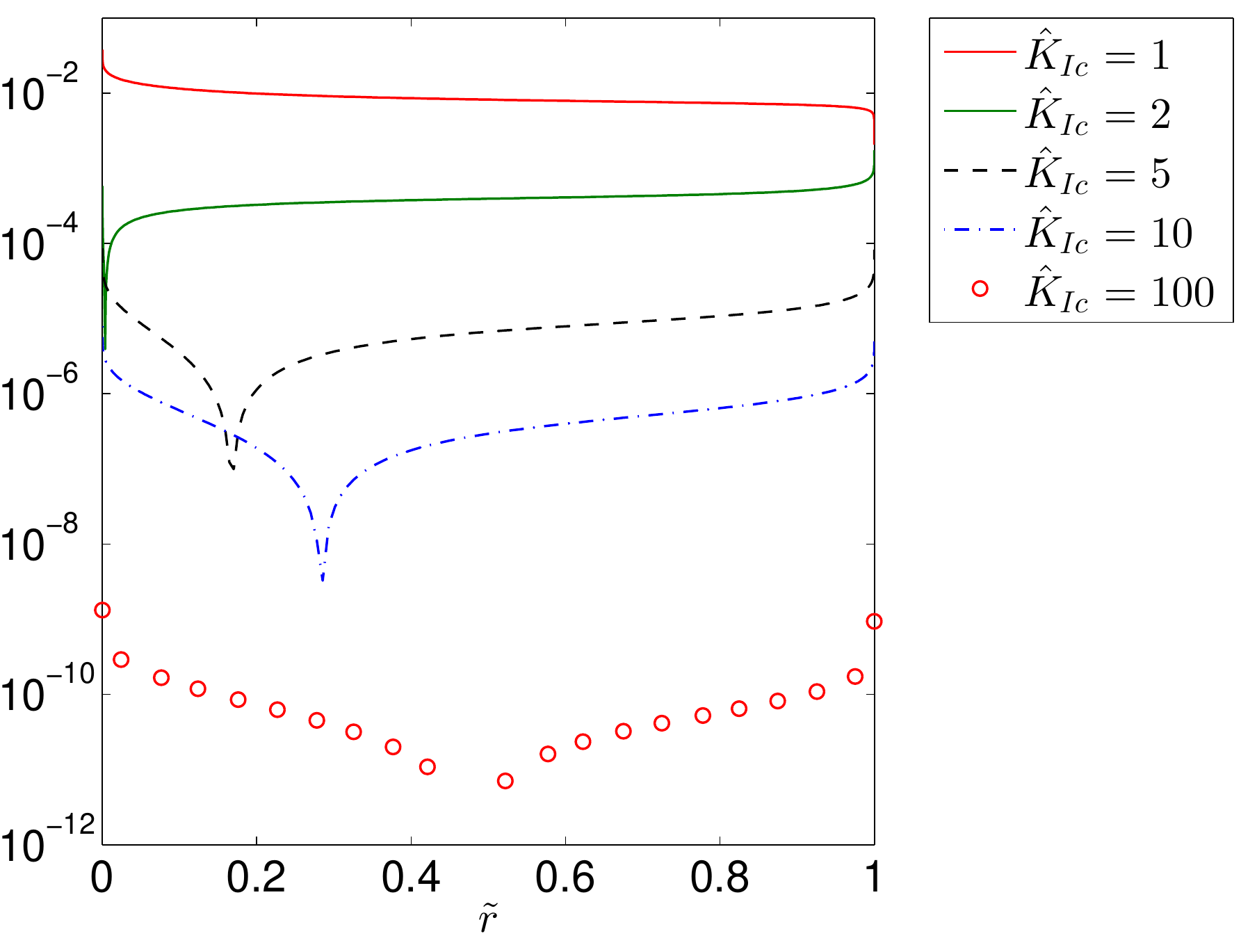}
 \put(-270,105) {\rotatebox{90}{{\bf{$\delta S$}}}}
 \caption{Comparison of the ratio between the fracture aperture and pressure for \emph{Savitski/Detournay's} solution and that presented in this paper for a few values of the fracture toughness. Here $\delta S$ shows the relative error.}
 \label{Fig:SavitskiTough}
\end{figure}

For the toughness dominated regime it is unfortunately not possible to perform the same comparison as above with the results from \cite{Savitski2002}. This is due to the fact that \emph{Savitski/Detournay's} solution is only self-similar in the limiting cases $K_I=\left\{0,\infty\right\}$, and is a time dependent function of $K_I (t)$ in the interim.
It is however possible to check the ratio between the fracture pressure and aperture with the following equality:
\begin{equation}
\frac{\hat{w}(\tilde{r})}{\hat{p}(\tilde{r})} = \frac{\Omega_{k}(\tilde{r})}{\gamma_0 \Pi_{k} (\tilde{r})} ,
\label{SavitskiTough}
\end{equation}
where $\Omega_k$ is \emph{Savitski/Detournay's} normalized aperture, $\Pi_k$ is the normalized pressure and $\gamma_0=\left(3/\pi\sqrt{2}\right)^{\frac{2}{5}}$ is the first term of the normalized asymptotic expansion of the fracture length \cite{Savitski2002}. Noting that the paper gives the limiting values for $K_{Ic}\to\infty$ as being $\Omega_{k,0}=\left(3/8\pi\right)^{\frac{1}{5}}\sqrt{1-\tilde{r}^2}$ and $\Pi_{k,0}=\pi \left( \pi / 12 \right)^{\frac{1}{5}} / 8$, it can easily be seen from \eqref{KIInftywp} that ratio \eqref{SavitskiTough} is satisfied in the limit. 
As such, we can evaluate the validity of the asymptotic fromulae from \cite{Savitski2002} by examining the relative ratio between the two sides of \eqref{SavitskiTough}, which we will label $\delta S$. The results for this metric, pertaining to the values $\hat{K}_I=\left\{1,2,5,10,100\right\}$, are provided in Fig.~\ref{Fig:SavitskiTough}.

It is evident from this comparison that there is a clear correspondence between the results of this paper and those obtained by \emph{Savitski/Detournay}. The disparity between respective data in the large toughness case, $\hat{K}_{Ic}=100$, is compatible with the error of our solution demonstrated for this model in Fig.~\ref{Fig:KI10}. This is a strong verification of the validity of the asymptotic formulae from \cite{Savitski2002}. However, the accuracy of those approximations diminishes greatly for lower values of the fracture toughness, with an error of order $10^{-1}$ when $\hat{K}_I=1$. This, in turn, provides us with an estimate of when the formula in \cite{Savitski2002} loses its practical applicability.


\section{Conclusions} \label{Sect:Conclusions}

In this paper, the problem of a penny-shaped hydraulic fracture driven by a power-law fluid has been analyzed. Following an approach similar to that in \cite{Perkowska2015,Wrobel2015} the governing equations where reformulated in terms of the aperture $w$ and the reduced particle velocity $\Phi$. Self-similar formulations have been derived for two types of time dependent function. A computational scheme based on the universal algorithm introduced in \cite{Wrobel2015} has been constructed. The accuracy of computations has been verified against a set of newly introduced analytical benchmark examples. Alternative measures of the solution accuracy have been proposed and investigated. Highly accurate numerical reference solutions for the case of an impermeable solid have been delivered. Simple and accurate approximate formulae mimicking these solutions, over whole range of the fluid behaviour index, have been given for fixed values of the material toughness. Verification of other results available in the literature has been performed.

The following conclusions can be drawn from the conducted research:
\begin{itemize}
 \item The universal algorithm for numerically simulating hydraulic fractures, introduced in \cite{Wrobel2015}, can be successfully adapted to the case of a penny-shaped fracture. It enables accurate and efficient modelling of HFs driven by the power-law fluids in both the viscosity and toughness dominated regimes.
 \item The key elements of the algorithm, which contributed to its outstanding performance, are: i) choice of proper computational variables, including the reduced particle velocity, ii) extensive utilization of the information on the solution asymptotics, combined with a fracture front tracing mechanism based on the Stefan-type condition (speed equation), iii) application of the modified form of the elasticity operator \eqref{apertureN3}, which has a non-singular kernel, that can easily be coupled with the new dependent variable - the reduced particle velocity.
 \item The newly introduced analytical benchmark solutions, with a predefined non-zero fluid leak-of, can be adjusted to mimic the HF behaviour for a class of power-law fluids in both the viscosity and toughness dominated regimes. These solutions can be directly applied to investigate the actual error of computations when testing various computational schemes.
 \item The rates of error convergence ($e_w$ and $e_v$) have been shown to be equivalent and credible error measures for analyzing the problem when no closed-form analytical solutions are available.
 \item The proposed approximate semi-analytical formulae for the case of an impermeable solid constitute a set of accurate and easily accessible reference solutions when investigating the performance of other computational algorithms.
\end{itemize}

\section*{Acknowledgement} DP and MW were funded by the European Union Seventh Framework Marie Curie Programme PARM-2 (project reference: PIAP-GA-2012-284544-PARM2). GM and MW received funding from the FP7 PEOPLE Marie Curie IRSES project TAMER under number IRSES-GA-2013-610547. GM gratefully acknowledges support from the ERC Advanced Grant Instabilities and nonlocal multiscale modeling of materials ERC-2013-ADG-340561-INSTABILITIES during his Visiting Professorship at Trento University. MP is supported by the FP7 PEOPLE Marie Curie action project CERMAT2 under number PITN-GA-2013-606878. GM acknowledges the support of grant No. 14.581.21.0027 unique identifier: RFMEFI58117X0027 by Ministry of Education and Science of the Russian Federation. All authors are grateful to the funding bodies who supported this project. DP and MW are very grateful to ISOTOP for the facilities they provided during their secondments. Both would specifically like to thank Dr Vladi Frid for his fruitful discussions when beginning the paper, and throughout their secondment periods. GM is grateful to the Royal Society for the Wolfson Research Merit Award.



\bibliography{Penny_Bib}
\bibliographystyle{unsrt}


\begin{appendix}

\section{Limiting cases: Newtonian and plastic fluids} \label{Append:Cases}

\subsection{Newtonian fluid: $n=1$}

\subsubsection{Basic formulae}

In the case of a Newtonian fluid the majority of the results remains the same as in the general case (setting $n=1$), but a few constants and functions will take alternate forms. These are detailed below.\\

The crack tip asymptotics in the viscosity dominated regime can be described by general relations \eqref{apertureasymp1_otherb}-\eqref{v_1_asymp}. However, in the toughness dominated mode one has:
\begin{equation}
 \begin{aligned}
\tilde w(\tilde r,\tilde t) &=\tilde w_0(\tilde t)  \sqrt{1-\tilde{r}^2 } + \tilde w_1(\tilde t)\left(1-\tilde r^2 \right)+ \tilde w_2(\tilde t)\left(1-\tilde r^2 \right)^{\frac{3}{2}}\log \left(1-\tilde{r}^2 \right) &\\
& \quad + O\left(\left(1-\tilde{r}^2 \right)^{\frac{3}{2}}\right) , \quad \tilde r \to 1, &
 \end{aligned}
\label{apertureasymp1_Tn1}
\end{equation}
\begin{equation}
\frac{\partial \tilde{p}}{\partial \tilde{r}} = \tilde{p}_0 (\tilde{t}) \left(1-\tilde{r}^2\right)^{-1} + \tilde{p}_1 (\tilde{t}) \left(1-\tilde{r}^2\right)^{-\frac{1}{2}}+ O\left( 1 \right) , \quad \tilde{r}\to 1 .
\end{equation}
The respective asymptotic expansions at the crack inlet, for both the viscosity and toughness dominated regimes, yield:
 \begin{equation}
\tilde{w}(\tilde{r},\tilde{t}) = \tilde{w}_0^o +\tilde{w}_1^o \tilde{r} + O\left(\tilde{r}^2 \log(\tilde{r})\right) , \quad \tilde{r}\to 0 ,
\end{equation}
 \begin{equation}
 \tilde{p}(\tilde{r},\tilde{t}) = \tilde{p}_0^o (\tilde{t}) + \tilde{p}_1^o (\tilde{t}) \log\left(\tilde{r}\right) + O\left( \tilde{r}\right) , \quad \tilde{r}\to 0 .
 \end{equation}
It should be noted that the pressure is singular at the fracture origin, which is not the case for non-Newtonian ($n<1$) fluids.

Meanwhile, the relationship between the new variable $\Omega$ and the pressure, in the time-dependent formulation, follows from the definition \eqref{new_P}:
\begin{equation}
\tilde p(\tilde r,\tilde t)= \Omega_0(\tilde t)\log (\tilde r)+C_p(\tilde t)+\int_0^{\tilde r}\Omega(\xi, \tilde{t})d\xi,
\end{equation}
where the time dependent constant $C_p(\tilde{t})$ is obtained by expanding \eqref{criterionN1} using \eqref{new_P}:
\begin{equation}
C_p (\tilde{t}) = \frac{1}{2}\sqrt{\frac{\pi}{L(\tilde{t})}} \tilde{K}_I +  \left[ 1 -\log\left(2\right)\right] \Omega_0 (\tilde{t}) - \int_0^1 \Omega(y ,\tilde{t} )\sqrt{1-y^2} \, dy .
\label{Cpt}
\end{equation}
Transforming into the self-similar formulation \eqref{SelfSimilar1}, these become:
\begin{equation}
\hat p(\tilde r)= \hat{\Omega}_0\log\left( \tilde{r}\right)+ \hat{C}_p+\int_0^{\tilde r} \hat{\Omega}(\xi) \, d\xi ,
\label{new_P2c}
\end{equation}
\begin{equation}
\hat{C}_p = \frac{\sqrt{\pi}}{2} \hat{K}_I + \left[ 1 -\log\left(2\right)\right] \hat{\Omega}_0  - \int_0^1 \hat{\Omega}(y)\sqrt{1-y^2} \, dy .
\label{Cptc}
\end{equation}
Finally, the auxiliary function ${\cal G}_n(\tilde{r})$ will now be expressed as:
\begin{equation}
{\cal G}_n (\tilde{r}) = \tilde{r}\left( \frac{\pi}{2} - \arctan \left(\frac{\tilde{r}}{\sqrt{1-\tilde{r}^2}}\right)\right) - \sqrt{1-\tilde{r}^2} \equiv \tilde{r}\arccos\left(\tilde{r}\right) - \sqrt{1-\tilde{r}^2} .
\label{aux1c}
\end{equation}


\subsubsection{Approximate semi-analytical approximation}

The semi-analytical approximations for the aperture and particle velocity remain the same as those presented in Sect.~\ref{SemiAnal}, however, the form of the pressure function must be modified. We now have:

\begin{itemize}
 \item The viscosity dominated regime ($K_{Ic}=0$):
\end{itemize}
Here the form of the aperture \eqref{apxWKI0} and particle velocity \eqref{apxVKI0} approximations remain the same as in the general case, but the approximation of the pressure takes the form:
 	\begin{equation}
	 \begin{aligned}
	\hat{p}_{apx}(\tilde{r},n) &= \hat{C}_p(n) + p_1 \log(\tilde{r}) + p_2 \tilde{r} \left(1-\tilde{r}^2\right)^{-\frac{1}{3}} + p_3 + p_4 \tilde{r}\sqrt{1-\tilde{r}} \\
	&\quad + p_5\left(1-\tilde{r}\right)^{\frac{2}{3}} + p_6\left(1-\tilde{r}\right)^{\frac{5}{3}},
	 \end{aligned}
	\label{n1apxPv}
	\end{equation}
with  $\hat{C}_p(n)$ remaining as in \eqref{v0apx}$_2$.
\begin{itemize}
 \item The toughness dominated regime ($K_{Ic}>0$):
\end{itemize}
Here the form of the aperture \eqref{apxWKI1} and particle velocity \eqref{apxVKI1} approximations remain the same as in the general case, but the approximation of the pressure is now:
	\begin{equation}
	\hat{p}_{apx}(\tilde{r},n) = p_1+p_2 \log(1-\tilde{r}^2)+p_3\log(\tilde{r}) +p_4 \tilde{r} \sqrt{1-\tilde{r}} .
	\label{n1apxP}
	\end{equation}

\subsection{Perfectly plastic fluid: $n=0$}

\subsubsection{Basic formulae}

In the case of a perfectly plastic fluid, alongside changes to the system asymptotics and reformulated equations, the degeneration of the Poiseuille equation means that it cannot be used to define the particle velocity $\tilde{v}$, or the reduced particle velocity $\Phi$. As a result fundamental changes to the scheme are required. These are outlined below. \\

The crack tip asymptotics in the viscosity dominated regime remains in the same form as was outlined in \eqref{apertureasymp1_otherb}-\eqref{v_1_asymp}. In the toughness dominated mode however it now yields:
\begin{equation}
 \begin{aligned}
\tilde w(\tilde r,\tilde t) &=\tilde w_0(\tilde t) \sqrt{1-\tilde{r}^2 }+ \tilde w_1(\tilde t)\left(1-\tilde r^2 \right)^{\frac{3}{2}}\log \left(1-\tilde{r}^2 \right)  + \tilde w_2(\tilde t)\left(1-\tilde r^2 \right)^{\frac{3}{2}} &\\
& \quad +  O\left(\left(1-\tilde{r}^2 \right)^{\frac{5}{2}}\right) , \quad \tilde r \to 1, &
 \end{aligned}
\label{apertureasymp1_Tn0}
\end{equation}
\begin{equation}
\frac{\partial \tilde{p}}{\partial \tilde{r}} = \tilde{p}_0 (\tilde{t}) \left(1-\tilde{r}^2\right)^{-\frac{1}{2}} +  O\left( 1 \right) , \quad \tilde{r}\to 1 .
\end{equation}
The fracture opening and the fluid pressure can be estimated at the crack inlet as:
  \begin{equation}
\tilde{w}(\tilde{r},\tilde{t}) = \tilde{w}_0^o (\tilde{t})+ O\left(\tilde{r}^2 \log(\tilde{r})\right) , \quad \tilde{r}\to 0 ,
\end{equation}
 \begin{equation}
 \tilde{p}(\tilde{r},\tilde{t}) = \tilde{p}_0^o (\tilde{t}) + \tilde{p}_1^o (\tilde{t}) \tilde{r} + O\left(\tilde{r}^2 \right) , \quad \tilde{r}\to 0 .
 \end{equation}


Meanwhile, the relationship between the modified pressure derivative and the pressure follows from the definition \eqref{new_P}:
\begin{equation}
\tilde{p}(\tilde{r},\tilde{t}) = \tilde{r} \Omega_0 (\tilde{t} )  + C_p (\tilde{t}) + \int_0^{\tilde{r}} \Omega ( \xi , \tilde{t}) \, d\xi ,
\end{equation}
where the constant $\hat{C}_p$ takes the form \eqref{criterionN1}:
\begin{equation}
C_p = \frac{1}{2}\sqrt{\frac{\pi}{L(\tilde{t})}} \tilde{K}_I - \frac{\pi}{4}\Omega_0 (\tilde{t}) - \int_0^{1} \hat{\Omega}(y, \tilde{t})\sqrt{1-y^2} \, dy .
\end{equation}
Note, from the form of the above, that the pressure is not be singular at the injection point in this case. Transforming into the self-similar formulation \eqref{SelfSimilar1} these become:
\begin{equation}
\hat{p}(\tilde{r}) = \tilde{r}\hat{\Omega}_0  + \hat{C}_p + \int_0^{\tilde{r}} \Omega(\xi ) \, d\xi ,
\end{equation}
\begin{equation}
\hat{C}_p = \frac{\sqrt{\pi}}{2}\hat{K}_I - \frac{\pi}{4}\hat{\Omega}_0 - \int_0^{1} \hat{\Omega}(y)\sqrt{1-y^2} \, dy .
\end{equation}
It can be shown that the relationship between $\Omega$ and the fracture aperture \eqref{newaperture5} still holds, with the function ${\cal G}_n (\tilde{r})$ being given by:
\begin{equation}
 \begin{aligned}
{\cal G}_n (\tilde{r}) &= -\frac{\pi}{8} \left[ \sqrt{1-\tilde{r}^2} + \tilde{r}^2 \log \left(\frac{\tilde{r}}{1+\sqrt{1-\tilde{r}^2}}\right) \right] &\\
&\equiv -\frac{\pi}{8} \left[ \sqrt{1-\tilde{r}^2} - \tilde{r}^2 \arctanh\left(\sqrt{1-\tilde{r}^2} \right) \right] . &
\end{aligned}
\end{equation}
In practice however, the degeneration of the Poiseuille equation means that a new scheme for solving the governing equations must be devised. The first step towards this is to note that the fracture aperture can be expressed as a non-linear integral equation:
\begin{equation}
\hat{w} (\tilde{r}) = -\frac{8}{\pi}\int_0^1 \frac{1}{\hat{w}(y)} {\cal K}(y,\tilde{r}) , dy + \frac{4}{\sqrt{\pi}} \hat{K}_I \sqrt{1-\tilde{r}^2} ,
\label{AperturePlastic}
\end{equation}
while the crack-propagation speed is calculated from the fluid balance equation \eqref{fluidbalanceS1} as follows:
\begin{equation}
\hat{v}_0 = \frac{\hat{Q}_0}{2\pi \left[ \left(3-\rho\right)\int_0^1 \tilde{r}\hat{w}(\tilde{r}) \, d\tilde{r} + \frac{1-\rho}{\gamma}\int_0^1 \tilde{r} \hat{q}_l \, d\tilde{r}\right]} .
\label{v0Plastic}
\end{equation}
The reduced particle velocity $\hat{\Phi}$ can be determined by integrating \eqref{fluidmassS1}:
\begin{equation}
\hat{\Phi}(\tilde{r}) = \frac{\hat{v}_0}{\hat{w}(\tilde{r})} \int_{\tilde{r}}^1 \xi \left[ \left(3-\rho\right)\hat{w}(\xi)+\left(1-\rho\right)\frac{\hat{q}_l(\xi)}{\gamma} \right] \, d\xi .
\label{PVPlastic}
\end{equation}

\subsubsection{Approximate semi-analytical approximation}

Finally, as a result of changes to the system behaviour and asymptotics, the semi-analytical approximations presented in Sect.~\ref{SemiAnal} take the following form when $n=0$:

\begin{itemize}
 \item The viscosity dominated regime ($K_{Ic}=0$):
\end{itemize}
Here the form of the aperture approximation \eqref{apxWKI0} remains the same as in the general case. However the approximations of the particle velocity and pressure are now:
\begin{equation}
\tilde{r}\hat{v}_{apx} (\tilde{r},n)=\left( v_1 \tilde{r} + v_2 \right) / \left( \tilde{r}^3 + v_3 \tilde{r}^2 + v_4 \tilde{r} + v_5 \right) ,
\label{n0ApxV}
\end{equation}
\begin{equation}
	 \begin{aligned}
	\hat{p}_{apx}(\tilde{r},n) &= \hat{C}_p(n) + p_1 \tilde{r} + p_2 \tilde{r}\log\left(1-\tilde{r}\right) + p_3 + p_4 \tilde{r}\sqrt{1-\tilde{r}} \\
	&\quad + p_5\left(1-\tilde{r}^2\right)\log\left(1-\tilde{r}^2\right) + p_6\left(1-\tilde{r}\right) + p_7 \left(1-\tilde{r}\right)^2 ,
	 \end{aligned}
	\label{n0apxPv}
	\end{equation}
with  $\hat{C}_p(n)$ remaining as in \eqref{v0apx}$_2$.
\begin{itemize}
 \item The toughness dominated regime ($K_{Ic}>0$):
\end{itemize}
Here the pressure approximation \eqref{apxPKI1} remains the same as in the general case. However, the aperture and particle velocity approximations become:
\begin{equation}
 \begin{aligned}
\hat{w}_{apx}(\tilde{r},n)=&\hat{w}_0 (\sqrt{1-\tilde{r}^2}+w_1(1-\tilde{r}^2)^{3/2}+w_2 (1-\tilde{r}^2)^{3/2}\log(1-\tilde{r}^2)+ \\
& w_3 (1-\tilde{r}^2)^3\log(1-\tilde{r}^2)+ w_4 (1-\tilde{r}^2)^{5/2}\tilde{r}^2+w_5 f_1 ,
 \label{n0ApxW}
 \end{aligned}
\end{equation}
\begin{equation}
\tilde{r}\hat{v}_{apx}(\tilde{r},n)=v_1+v_2(1-\tilde{r}^2)^2 \log(1-\tilde{r}^2)+v_3 (1-\tilde{r}^2)^2+v_4(1-\tilde{r}^2)^2 \tilde{r}^2 \log(\tilde{r})  ,
\label{n0ApxV_1}
\end{equation}
with $f_1$ being given in \eqref{fOmeg} and $\hat{w}_0$  in \eqref{w0asym1}.

\section{Analytical benchmarks} \label{App:AnalBench}

In the following we will present a way to construct a set of analytical benchmark solutions that satisfy the system of governing equations \eqref{particlevS1}-\eqref{BCS1} for the self-similar problem. Those solutions can be easily extended through the relations \eqref{particlevN1}-\eqref{fluidbalanceN1}, \eqref{criterionN1}, \eqref{BCN1}, \eqref{fluidmassK1} and \eqref{newaperture5} to the time dependent forms. In this way one can formulate a set of analytical benchmark examples for both, the self-similar and the time dependent versions of the problem.

The basic concept employed to derive the self-similar solutions is the same as that in \cite{Wrobel2015} for the KGD model. We assume that the crack aperture can be expressed as a weighted sum of properly chosen base functions:
\begin{equation}
\label{ad_m_w}
\hat{w}(\tilde{r}) = \sum_{i=0}^M \lambda_i h_i (\tilde{r}).
\end{equation}
The functions $h_i$ are selected in a way that enables one to: i) comply with the asymptotic representation \eqref{apertureasymp1_otherb}, ii) satisfy the respective boundary conditions \eqref{BCS1}, iii) compute analytically the elasticity operator \eqref{apertureS1}. The multipliers $\lambda_i$ are to be chosen properly to ensure the physically justified behaviour and desired properties of the solution.

Provided that iii) is satisfied, the fluid pressure function can be computed in a closed form from \eqref{apertureS1} to give:
\begin{equation}
\label{ad_m_p}
\hat{p}(\tilde{r}) = \sum_{i=0}^M \lambda_i \pi_i (\tilde{r}),
\end{equation}
where each function $\pi_i$ corresponds to respective function $h_i$.

The self-similar stress intensity factor follows immediately from the asymptotic bahviour of functions $h_i$ and complies with \eqref{criterionS1}. Next, the self-similar crack propagation speed, $\hat v_0$ can be determined according to \eqref{particlevS1}, while the particle velocity is computed from (62) to produce:
\begin{equation}
\label{ad_m_v}
\hat v (\tilde r)=\left\{ -\left[\sum_{i=0}^M \lambda_i h_i (\tilde{r})\right]^{n+1} \cdot \sum_{i=0}^M \lambda_i \pi'_i (\tilde{r})\right\}^{1/n}.
\end{equation}

Consequently, the reduced particle velocity is defined by employing \eqref{ad_m_p} in \eqref{SSdef}$_1$. The influx magnitude, $\hat{Q}_0$, is computed from  \eqref{BCS1}, while the modified pressure derivative can be obtained from the definition \eqref{SSdef}$_2$, \eqref{SSOmega0}. Finally, the benchmark leak-off function is determined by a transformation of \eqref{fluidmassS1} as:
\begin{equation}
\label{ad_m_ql}
\hat q_l(\tilde r)=\frac{\gamma}{1-\rho}\left[(\rho-3)\hat w(\tilde r)-\frac{1}{r\hat v_0}\left(\hat w(\tilde r)\hat \phi(\tilde r)\right)'\right],
\end{equation}
where the quantities on the right hand side are taken according to \eqref{ad_m_w}-\eqref{ad_m_v}.

In this way, by using different values of the coefficients $\lambda_i$ and different functions $h_i(\tilde r)$, $\pi_i(\tilde r)$ one can construct a number of self-similar problems for various fluid behaviour indices and crack propagation regimes, for which there exist  known purely analytical solutions in the form \eqref{ad_m_w}-\eqref{ad_m_v}. The values of pumping rate, $\hat{Q}_0$, and the self-similar material toughness, $\hat K_{Ic}$, can be tuned by the choice of magnitudes of respective coefficients $\lambda_i$.

The examples of base functions $h_i(\tilde{r})$, $\pi_i(\tilde{r})$ are collected in Table \ref{table:Anal1}.

\begin{table}[h!]
 \centering
\begin{tabular}{||c|c|c||}
\hline
$i$ & $\pi_i(\tilde{r})$ & $h_i(\tilde{r})$ \\ [0.5ex]
\hline \hline
&&\\
$1$ & $1$ & $\frac{8}{\pi} \sqrt{1-\tilde{r}^2}$ \\[2mm]
\hline
&&\\
$2$ &$\tilde{r}$ & $\sqrt{1-\tilde{r}^2} + \tilde{r}^2 \log\left(\frac{1+\sqrt{1-\tilde{r}^2}}{\tilde{r}}\right)$ \\[2mm]
\hline
&&\\
$3$ & $\tilde{r}^{1-n}$ & $ \frac{2\Gamma\left(\frac{3}{2}-\frac{n}{2}\right)\Gamma\left(\frac{n}{2}-1\right)}{\Gamma\left(2-\frac{n}{2}\right)\Gamma\left(\frac{n}{2}-\frac{1}{2}\right)}\left[ \tilde{r}^{2-n} - \frac{\Gamma\left(\frac{n}{2}-\frac{1}{2}\right)}{\sqrt{\pi}\Gamma\left(\frac{n}{2}\right)}{_2F_1}\left(\frac{1}{2} , \frac{n}{2} - 1 ; \frac{n}{2} ; \tilde{r}^2 \right) \right]$ \\[4mm]
\hline
&&\\
$4$ & $\tilde{r}^{2-n}$ & $\frac{2\Gamma\left(2-\frac{n}{2}\right) \Gamma\left(\frac{n-3}{2}\right)}{\Gamma\left(\frac{5-n}{2}\right) \Gamma\left(\frac{n}{2}-1\right)}\left[ \tilde{r}^{3-n}  - \frac{\Gamma\left(\frac{n}{2}-1\right)}{\sqrt{\pi}\Gamma\left(\frac{n-1}{2}\right)}  {_2F_1}\left(\frac{1}{2}, \frac{n-3}{2} ; \frac{n-1}{2} ; \tilde{r}^2 \right) \right]$ \\[4mm]
\hline
&&\\
$5$ & $\log (\tilde{r} )$ & $ \frac{8}{\pi}\left[ \tilde{r}\arccos\left(\tilde{r}\right) + \left(\log(2)-2\right)\sqrt{1-\tilde{r}^2} \right]$ \\[3mm]
\hline
&&\\
$6$ & $\tilde{r} {_2F_1} \left(\frac{1}{2}-\alpha , 1 ; \frac{1}{2} ; \tilde{r}^2 \right)$ & $\frac{2\sqrt{\pi}\left(1-\tilde{r}^2\right)^{\alpha}}{1+2\alpha}\biggl[ \frac{\Gamma\left(\alpha+\frac{1}{2}\right)}{\Gamma\left(1+\alpha\right)}{_2F_1}\left(\frac{1}{2},\frac{1}{2}+\alpha ; 1+\alpha ; 1-\tilde{r}^2 \right)$ \\
& & $+ \frac{4\Gamma\left(\frac{3}{2}+\alpha\right)}{\left(1+2\alpha\right)\Gamma\left(\alpha\right)}{_2F_1}\left( -\frac{1}{2} , \frac{1}{2}+\alpha ; 1 + \alpha ; 1 - \tilde{r}^2 \right) \biggr]$ \\
& & $- \frac{4}{1+2\alpha}\log\left(\frac{1+\sqrt{1-\tilde{r}^2}}{\tilde{r}} \right) $ \\[4mm]
\hline
&&\\
$7$ & $\arctanh (\tilde{r})$ & $4\left[ \EllipticE{1-\tilde{r}^2} - \EllipticK{1-\tilde{r}^2} + \log\left(\frac{1+\sqrt{1-\tilde{r}^2}}{\tilde{r}}\right)\right]$ \\[3mm]
\hline
\end{tabular}
 \caption{Table showing the components of the benchmark solutions. Here ${_2F_1}$ is the Gaussian hypergeometric function, while functions $K$, $E$ represent the complete elliptic integral of the first and second kinds respectively. }
\label{table:Anal1}
\end{table}

To provide a very simple example of a numerical benchmarks which can be created using the aforementioned methodology, we consider the following composite functions:
\\

\begin{equation}
h_A (\tilde{r},\alpha) = h_6 (\tilde{r},\alpha) + \frac{\pi}{1+2\alpha} h_1 - \frac{2}{1+2\alpha} h_2(\tilde{r}) ,
 \label{Bench:comp3}
\end{equation}
\begin{equation}
h_B (\tilde{r},n) = -h_3 (\tilde{r},n) + \frac{n\sqrt{\pi}\Gamma\left(\frac{3-n}{2}\right)}{2\Gamma\left(2-\frac{n}{2}\right)}h_1 + \frac{2\left(1-n\right)\Gamma\left(\frac{3-n}{2}\right)}{\sqrt{\pi}\Gamma\left(2-\frac{n}{2}\right)} h_2 (\tilde{r}),
 \label{Bench:comp2}
\end{equation}
with the corresponding pressure terms:
\begin{equation}
\pi_A (\tilde{r},\alpha) = \pi_6 (\tilde{r},\alpha) + \frac{\pi}{1+2\alpha} \pi_1 - \frac{2}{1+2\alpha} \pi_2(\tilde{r}) ,
 \label{Bench:comp1}
\end{equation}
\begin{equation}
\pi_B (\tilde{r},n) = -\pi_3 (\tilde{r},n) + \frac{n\sqrt{\pi}\Gamma\left(\frac{3-n}{2}\right)}{2\Gamma\left(2-\frac{n}{2}\right)} \pi_1 + \frac{2\left(1-n\right)\Gamma\left(\frac{3-n}{2}\right)}{\sqrt{\pi}\Gamma\left(2-\frac{n}{2}\right)} \pi_2 (\tilde{r})
 .
 \label{Bench:comp4}
\end{equation}
Then the asymptotic behaviour of the respective functions at the fracture tip is:
\begin{equation}
h_A (\tilde{r}, \alpha) = \frac{2\sqrt{\pi}\Gamma\left(\alpha+\frac{1}{2}\right)}{\Gamma\left(\alpha+1\right)} \left(1-\tilde{r}^2 \right)^{\alpha} + O\left(\left(1-\tilde{r}^2 \right)^{\min \left( \frac{5}{2} , \alpha+1 \right)} \right) , \quad  \tilde{r}\to 1 ,
\label{Bench_prop1}
\end{equation}
\begin{equation}
\frac{d\pi_A (\tilde{r}, \alpha)}{d\tilde{r}} = \frac{\sqrt{\pi}\left(1-2\alpha\right)\Gamma\left(2-\alpha\right)}{\Gamma\left(\frac{3}{2}-\alpha\right)}\left(1-\tilde{r}^2\right)^{\alpha-2} + O\left(\left(1-\tilde{r}^2\right)^{\alpha-1}\right) , \quad  \tilde{r}\to 1 ,
\label{Bench_prop1b}
\end{equation}
\begin{equation}
h_B (\tilde{r},n) = O\left(\left(1-\tilde{r}^2\right)^{\frac{5}{2}} \right) , \quad \tilde{r} \to 1 ,
\label{Bench_prop3}
\end{equation}
\begin{equation}
\frac{d \pi_B(\tilde{r} , n)}{d \tilde{r}} = \left(1-n\right)\left[\frac{2\Gamma\left(\frac{3-n}{2}\right)}{\sqrt{\pi}\Gamma\left(2-\frac{n}{2}\right)}-1 \right] + O \left( 1-\tilde{r}^2 \right) , \quad \tilde{r}\to 1 ,
\end{equation}
It can easily be seen from the above equations that the functions $h_A$ and $\pi_A$ will provide the proper first term of the crack tip asymptotics for the aperture \eqref{apertureasymp1_otherb} and pressure derivative \eqref{dp_asym_1}, \eqref{dp_asym_2}, provided that $\alpha$ is taken in accordance with Table \ref{table:albe}. Further terms may also be constructed, although subsequent (known) asymptotic terms of $h_A$ and $\pi_A$ must be accounted for. Additionally the behaviour of $h_B$, $\pi_B$ at the crack tip ensures that it will not interfere with the final asymptotics of the benchmark at the fracture front in a notable way.

Meanwhile, at the crack inlet, we have:
\begin{equation}
h_A (\tilde{r},\alpha) = \frac{2}{1+2\alpha} \left[ 3 + \frac{4\alpha}{1+2\alpha} - H\left(\alpha-\frac{1}{2}\right)\right] + O(\tilde{r}^2 \log(\tilde{r})) , \quad \tilde{r}\to 0,
\end{equation}
\begin{equation}
\frac{d \pi_A(\tilde{r} , \alpha)}{d \tilde{r}} = \frac{2\alpha-1}{1+2\alpha} + O ( \tilde{r}^2 ) , \quad \tilde{r}\to 0 ,
\end{equation}
\begin{equation}
h_B (\tilde{r},n) = -\frac{2n\sqrt{\pi}\left(1-n\right)\sec\left(\frac{n\pi}{2}\right)}{\left(2-n\right)\Gamma\left(2-\frac{n}{2}\right)\Gamma\left(\frac{n-1}{2}\right)}  + O(\tilde{r}^{2-n}) ,\quad \tilde{r}\to 0 ,
\end{equation}
\begin{equation}
 \frac{d \pi_B (\tilde{r},n)}{d \tilde{r}} = -\left(1-n\right) \tilde{r}^{-n} + O(1) , \quad \tilde{r}\to 0
\end{equation}
where $H$ is the harmonic number function and $\alpha$ can be taken in accordance with Table \ref{table:albe}. From this it can be easily seen that the required asymptotic representations of the aperture \eqref{wAsym02} and pressure derivative \eqref{sourceN2} will be satisfied by $h_B$ and $\pi_B$, while the fracture opening asymptotics of $h_A$ and $\pi_A$ will not prevent the benchmark from displaying the correct behaviour. As with the crack tip, here further asymptotic terms can be accounted for using additional functions.

In this way, by linear combination of functions \eqref{Bench:comp3}--\eqref{Bench:comp4} and other functions from Table \ref{table:Anal1} one can build a benchmark example for the viscosity dominated regime of crack propagation for a number of shear-thinning fluids, provided that $\alpha=\alpha_0$. Moreover, by incorporation of function $h_0$ from Table \ref{table:Anal1} we obtain a solution which mimics the toughness dominated mode.

The above strategy have been successfully employed to create  a set of analytical benchmark examples for the the varying crack propagation regimes and fluid behaviour indices.

\section{Coefficients of the approximate solutions} \label{App:SemiAnal}

For any value of the fluid behaviour index $n$ and self-similar material toughness $\hat{K}_I$, the self-similar crack propagation speed $\hat{v}_0$ is given in the form \eqref{v0apx}$_1$. The values of respective coefficients $C_i$ are provided in Table~\ref{v0_apxC} for $\hat{K}_I=\left\{0,1,10\right\}$.


\begin{table}[h!]
 \centering
\begin{tabular}{||c||c|c|c|c||}
\hline
$\hat{K}_I$ & $C_0$ & $C_1$ & $C_2$ & $C_3$   \\
\hline \hline
&&&&\\
0 & 0.1314342 & 0.1210766 & -0.0781383 & 0.031537   \\[2mm]
\hline
&&&&\\
1 & 0.06125898 & 0.050859704 & -0.029318586 & 0.012385442   \\[2mm]
\hline
&&&&\\
10 & 7.04065$\times 10^{-3}$ & 3.602954$\times 10^{-3}$ & -2.00895$\times 10^{-3}$ & 1.373533$\times 10^{-3}$  \\[2mm]
\hline \hline
\end{tabular}

\vspace{4mm}

\begin{tabular}{||c||c|c|c|c||}
\hline
$\hat{K}_I$ &  $C_4$ & $C_5$ & $C_6$ & $C_7$ \\
\hline \hline
&&&&\\
0 &  -5.293135$\times 10^{-3}$ & -6.62796$\times 10^{-3}$ & 5.350374$\times 10^{-3}$ & -1.521311 $\times 10^{-3}$ \\[2mm]
\hline
&&&&\\
1 & -2.920989$\times 10^{-3}$ & -2.8172727$\times 10^{-4}$ & 4.8397784$\times 10^{-4}$ & -1.2631848$\times 10^{-4}$ \\[2mm]
\hline
&&&&\\
10  & -1.0841455$\times 10^{-3}$ & 7.441777$\times 10^{-4}$ & -3.330152$\times 10^{-4}$ & 6.79385$\times 10^{-5}$ \\[2mm]
\hline \hline
\end{tabular}
\caption{Values of the coefficients $C_i$ used to approximate $\hat{v}_0$ \eqref{v0apx} for different values of the fracture toughness.}
\label{v0_apxC}
\end{table}

Meanwhile, the coefficients of the constant $\hat{C}_p(n)$, which takes the form \eqref{v0apx}$_2$ in the viscosity dominated case ($\hat{K}_{Ic}=0$), are provided in Table~\ref{Cp_tab1}.

\begin{table}[h!]
 \centering
 \begin{tabular}{||c|c|c|c|c|c||}
\hline
$D_0$ & $D_1$ & $X_0$ & $X_1$ & $X_2$ & $X_3$ \\
\hline \hline
&&&&&\\
3.5484 & -3.1946 & 3.711 & -1.3516 & -3.3625 & 1 \\[2mm]
\hline \hline
\end{tabular}
\caption{Values of the coefficients $D_i$, $X_k$ used to approximate the constant $\hat{C}_p(n)$ in equation \eqref{v0apx}$_2$.}
\label{Cp_tab1}
\end{table}

The remaining coefficients for the fracture aperture, pressure and particle velocity approximations are outlined for different values of the fracture toughness below.

\subsection{Viscosity dominated regime ($\hat{K}_{Ic}=0$)}


In the general case $0<n<1$ the coefficients of approximation for the aperture \eqref{apxWKI0}, particle velocity \eqref{apxVKI0} and pressure \eqref{apxPKI0} are given as:
\begin{equation}
{\cal Z} (n) = \frac{\sum_{k=0}^5 r_k n^k}{\left(1-n\right)^\kappa \sum_{k=0}^5 s_k n^k} ,
\label{genFormApx0}
\end{equation}
with the values of $r_k$, $s_k$ and $\kappa$ for the case $0<n<1$ being listed in Tables~\ref{Tab:apx0} and \ref{Tab:apx0b}.

\begin{table}[h!]
 \centering
\begin{tabular}{||c||c|c|c|c|c|c||}
\hline \hline
${\cal Z}(n)$ & $r_0$ & $r_1$ & $r_2$ & $r_3$ & $r_4$ & $r_5$  \\
\hline \hline
&&&&&&\\
${w}_0$ & 1.087913 & 0.629465 & 0.1884191 & -0.0954601 & 0.0539965 & 0 \\[2mm]
\hline
&&&&&&\\
$w_1$ & 0.0731578 & -0.0940037 & -0.2924713 & 0.712854 & -0.220774  & 0 \\[2mm]
\hline
&&&&&&\\
$w_2$ & -0.0813068 & 0.1374238 & -0.0672009 & -0.0557795 & 0  & 0 \\[2mm]
\hline
&&&&&&\\
$w_3$ & 0.1130671 & -0.458432 & -0.549883 & 0 & 0  & 0 \\[2mm]
\hline
&&&&&&\\
$w_4$ & -0.3394015 & 1.968425 & -0.324536 & 0 & 0  & 0 \\[2mm]
\hline
&&&&&&\\
$w_5$ & -0.4207775 & 2.729404 & 0 & 0 & 0  & 0 \\[2mm]
\hline
&&&&&&\\
$w_6$ & 0.374811 & -0.595337 & 0.4492 &  0.0240865 & 0  & 0 \\[2mm]
\hline \hline
&&&&&&\\
$v_1$ & - 0.0618879 & 0.238355 & 0.488614 & -0.089777 & 0  & 0 \\[2mm]
\hline
&&&&&&\\
$v_2$ & 0.106085 & -0.0105322 & -0.43386 & -0.0150819 & 0  & 0 \\[2mm]
\hline
&&&&&&\\
$v_3$ &  0.0260021 & 0.0203881 & -0.0379568 & 0.0258418 & -6.69655$\times 10^{-3}$  & 0 \\[2mm]
\hline
&&&&&&\\
$v_4$ & -0.0127769 & -0.0152235 & 0.0201527 & 0 & 0  & 0 \\[2mm]
\hline \hline
&&&&&&\\
$p_1$ & -1.383 & 0.6689 & 0 & 0 & 0 & 0 \\[2mm]
\hline
&&&&&&\\
$p_2$ & -18.738 & -7.314 & 7.802 & 0 & 0 & 0 \\[2mm]
\hline
&&&&&&\\
$p_3$ & 9.470147 & -26.2166 & 23.92346 & -7.16925 & 0 & 0 \\[2mm]
\hline
&&&&&&\\
$p_4$ & 0.1491 & -0.09304 & -0.13218 & 0.16745 & -0.12976 & 0.07958 \\[2mm]
\hline
&&&&&&\\
$p_5$ & -0.0754673 & -0.463258 & 1.755936 & -1.882529 & 0.732565 & -0.05901 \\[2mm]
\hline
&&&&&&\\
$p_6$ & -27.292 & -94.974 & 111.858 & 0 & 0 & 0 \\[2mm]
\hline \hline
\end{tabular}
 \caption{The values of coefficients $r_i$ used in approximation \eqref{genFormApx0} in the general case $0<n< 1$ with $\hat{K}_I=0$.}
 \label{Tab:apx0}
\end{table}
\begin{table}[h!]
 \centering
\begin{tabular}{||c||c|c|c|c|c|c|c||}
\hline \hline
${\cal Z}(n)$ & $s_0$ & $s_1$ & $s_2$ & $s_3$ & $s_4$ & $s_5$ & $\kappa$ \\
\hline \hline
&&&&&&&\\
${w}_0$ & 0.613792 & 1 & 0 & 0 & 0  & 0 & 0 \\[2mm]
\hline
&&&&&&&\\
$w_1$ & 1.30785 & -1.57716 & 0.820255 & 1 & 0  & 0 & 0 \\[2mm]
\hline
&&&&&&&\\
$w_2$ & 0.504215 & -0.2551376 & -0.436244 & 1 & 0  & 0 & 0\\[2mm]
\hline
&&&&&&&\\
$w_3$ & 0.2952694 & 0.319092 & -0.2805504 & 0.0738782 & 1  & 0 & 0 \\[2mm]
\hline
&&&&&&&\\
$w_4$ & 1.022663 & 0.28412 & -1.162825 & 1.77880 & 1  & 0 & 0 \\[2mm]
\hline
&&&&&&&\\
$w_5$ & 2.02325 & -0.427459 & -1.46776 & 3.51378 & 1  & 0 & 0 \\[2mm]
\hline
&&&&&&&\\
$w_6$ & 0.57009 & -1.09863 & 1 & 0 & 0  & 0 & 0 \\[2mm]
\hline \hline
&&&&&&&\\
$v_1$ & -3.3351 $\times 10^{-6}$ & 2.37989 & 1 & 0 & 0  & 0 & 0 \\[2mm]
\hline
&&&&&&&\\
$v_2$ & -5.84509$\times 10^{-6}$ & 4.0795 & 2.2347 & 1 & 0   & 0 & 0\\[2mm]
\hline
&&&&&&&\\
$v_3$ & -2.50863$\times 10^{-6}$ & 1 & 0 & 0 & 0  & 0 & 0 \\[2mm]
\hline
&&&&&&&\\
$v_4$ & 1.75635 & 0.685504 & 1 & 0 & 0  & 0 & 0 \\[2mm]
\hline \hline
&&&&&&&\\
$p_1$ & 2.3357 & 4.248 & 2.4022 & 1 & 0 & 0 & 1 \\[2mm]
\hline
&&&&&&&\\
$p_2$ & 0 & 33.212 & 47.87 & 1 & 0 & 0 & 0 \\[2mm]
\hline
&&&&&&&\\
$p_3$ & 16.78564 & -30.8988 & 8.118 & 9.44912 & -4.39755 & 1 & 0 \\[2mm]
\hline
&&&&&&&\\
$p_4$ & 1 & 0 & 0 & 0 & 0 & 0 & 0 \\[2mm]
\hline
&&&&&&&\\
$p_5$ & 0.133757 & 0.959086 & -2.15943 & 1 & 0 & 0 & 0 \\[2mm]
\hline
&&&&&&&\\
$p_6$ & 495.12 & 686.2 & -734.6 & 1 & 0 & 0 & 0 \\[2mm]
\hline \hline
\end{tabular}
 \caption{The value of constant coefficients $s_i$ and $\kappa$ used in approximation \eqref{genFormApx0} in the general case $0<n< 1$ with $\hat{K}_I=0$.}
 \label{Tab:apx0b}
\end{table}

In the case of a Newtonian fluid $n=1$ the coefficients used to approximate the aperture \eqref{apxWKI0} and the particle velocity \eqref{apxVKI0} remain the same as in the general case. The coefficients of the pressure approximation \eqref{n1apxPv} are now given by:
\begin{equation}
p_1=-0.0715 , \quad p_2=-0.22233 , \quad p_3=114.7455 ,
\end{equation}
$$
p_4 = 0.0413 , \quad p_5=-0.12312 , \quad p_6=-0.0237 .
$$
For the perfectly plastic fluid $n=0$ the coefficients used to approximate the aperture \eqref{apxWKI0}, particle velocity \eqref{n0ApxV} and pressure \eqref{n0apxPv} are as follows:
$$
{w}_0 = 1.773 , \quad w_1 = 0.06 , \quad w_2 = -0.1638 , \quad w_3 =0.335 ,
$$
$$
w_4=-0.289 , \quad w_5 = -0.179 , \quad w_6=0.6607 ,
$$
\begin{equation}
v_1 = 4.8656 , \quad v_2 = 1.703 , \quad v_3 = -20.484 , \quad v_4 = 51.4 , \quad v_5 = 18.07 ,
\end{equation}
$$
p_1=-0.5921 , \quad p_{2}=0.28201 , \quad p_3=0.264, \quad p_4=0.127 ,
$$
$$
p_5=0.099 , \quad p_6 =-0.6436 , \quad p_7=0.3806 .
$$

\subsection{Toughness dominated regime with $\hat{K}_I=1$}


In the general case $0<n<1$ the approximation coefficients for the aperture \eqref{apxWKI1}, particle velocity \eqref{apxVKI1} and pressure \eqref{apxPKI1} are given in the form:
\begin{equation}
{\cal Z} (n) = \frac{\sum_{k=0}^7 r_k n^k}{\sum_{k=0}^5 s_k n^k} ,
\label{genFormApx1}
\end{equation}
with the values of $r_k$, $s_k$ for the case $0<n< 1$ being listed in Tables~\ref{Tab:apx1} and \ref{Tab:apx1b}.

\begin{sidewaystable}[h!]
    \centering
\begin{tabular}{||c||c|c|c|c|c|c|c||}
\hline \hline
${\cal Z} (n)$ & $r_0$ & $r_1$ & $r_2$ & $r_3$ & $r_4$ & $r_5$ & $r_6$  \\
\hline \hline
&&&&&&&\\
$w_1$ & 0.01306626 & 0.0775474 & -0.05721985 & -0.0640863 & 0.082593 & -0.01774346 & 0 \\[2mm]
\hline
&&&&&&&\\
$w_2$ & 0.0878826 & -0.1161146 & 0.232977 & -0.2632192 & 0 & 0 & 0  \\[2mm]
\hline
&&&&&&&\\
$w_3$ & -0.02452076 & -0.0963696 & 0.465517 & -0.768842 & 0.916261 & -0.76577 & 0.376085  \\[2mm]
\hline
&&&&&&&\\
$w_4$ & 1.044817 $\times 10^{-3}$ & -0.0625069 & 0.0359431 & 1.138043$\times 10^{-3}$ & -4.5913$\times 10^{-3}$ & 0 & 0  \\[2mm]
\hline
&&&&&&&\\
$w_5$ & 0.1573167 & -0.383264 & 0.240315 & -0.1076469 & 0.1434914 & -0.126323 & 0.03972025   \\[2mm]
\hline \hline
&&&&&&&\\
$v_1$ & -5.33568$\times 10^{-3}$ & 0.0490222 & 0.1536204 & 0.1200713 & 2.598823$\times 10^{-3}$ & 0 & 0   \\[2mm]
\hline
&&&&&&&\\
$v_2$ & 1.116405$\times 10^{-3}$ & -1.752448$\times 10^{-3}$ & 9.66676$\times 10^{-4}$ & -1.993698$\times 10^{-4}$ & 0 & 0 & 0 \\[2mm]
\hline
&&&&&&&\\
$v_3$ & 5.16309$\times 10^{-3}$ & -3.942155$\times 10^{-3}$ & 2.435618$\times 10^{-3}$ & 0 & 0 & 0 & 0  \\[2mm]
\hline
&&&&&&&\\
$v_4$ & 8.42106$\times 10^{-3}$ & -4.082466$\times 10^{-3}$ & -0.0754209 & 0.01562056 & -0.01000874 & 2.352954$\times 10^{-3}$ & 0 \\[2mm]
\hline \hline
&&&&&&&\\
$p_1$ & 13.254 & -13.317 & 0 & 0 & 0 & 0 & 0 \\[2mm]
\hline
&&&&&&&\\
$p_2$ & -0.1896 & 0.368634 & -0.17891 & 0 & 0 & 0 & 0  \\[2mm]
\hline
&&&&&&&\\
$p_3$ & 2.85655 & -3.3178 & 0.96667 & 0 & 0 & 0 & 0  \\[2mm]
\hline
&&&&&&&\\
$p_4$ & -1.41826 & 1.60526 & -0.500383 & 0 & 0 & 0 & 0  \\[2mm]
\hline \hline
\end{tabular}
 \caption{The values of coefficients $r_1,\hdots,r_6$ used in approximation \eqref{genFormApx1} in the general case $0<n< 1$ with $\hat{K}_I=1$.}
 \label{Tab:apx1}
\end{sidewaystable}

\begin{sidewaystable}[h!]
 \centering
\begin{tabular}{||c||c||c|c|c|c|c|c||}
\hline \hline
${\cal Z}(n)$ & $r_7$ & $s_0$ & $s_1$ & $s_2$ & $s_3$ & $s_4$ & $s_5$ \\
\hline \hline
&&&&&&&\\
$w_1$ &  0 & 0 & -7.11987 $\times 10^{-5}$ & 0.5363626 & 1 & 0 & 0 \\[2mm]
\hline
&&&&&&&\\
$w_2$ & 0 & -1.036047$\times 10^{-3}$ & 7.22007 & -2.613708 & 12.86834 & -5.560094 & 1 \\[2mm]
\hline
&&&&&&&\\
$w_3$ & -0.0811716 & 0 & 0 & 1 & 0 & 0 & 0  \\[2mm]
\hline
&&&&&&&\\
$w_4$ & 0 & 1 & 0 & 0 & 0 & 0 & 0 \\[2mm]
\hline
&&&&&&&\\
$w_5$ & 0 &  1 & 0 & 0 & 0 & 0 & 0 \\[2mm]
\hline \hline
&&&&&&&\\
$v_1$ & 0 & 1.128027$\times 10^{-5}$ & 0.891702 & 1.62677 & 1 & 0 & 0 \\[2mm]
\hline
&&&&&&&\\
$v_2$ & 0 & -1.066292$\times 10^{-5}$ & 0.865467 & 1.724926 & 1.837862 & 1 & 0 \\[2mm]
\hline
&&&&&&&\\
$v_3$ & 0 & 3.128464$\times 10^{-5}$ & 0.861807 & 0.02594735 & 1.430577 & -0.703677 & 1 \\[2mm]
\hline
&&&&&&&\\
$v_4$ & 0 & -9.85603$\times 10^{-6}$ & 1.79581 & 1 & 0 & 0 & 0 \\[2mm]
\hline \hline
&&&&&&&\\
$p_1$ & 0 & 14.6064 & -11.804 & -7.0187 & 7.3799 & -4.1367 & 1 \\[2mm]
\hline
&&&&&&&\\
$p_2$ & 0 & 5.4791 & 4.8663 & -4.7430 & -1.34197 & -4.34704 & 1 \\[2mm]
\hline
&&&&&&&\\
$p_3$ & 0 & 6.47756 & 5.5563 & -5.9133 & -2.0187 & -5.0902 & 1 \\[2mm]
\hline
&&&&&&&\\
$p_4$ & 0 & 4.14283 & 5.2029 & -1.85248 & -2.80587 & -5.6837 & 1 \\[2mm]
\hline \hline
\end{tabular}
 \caption{The value of constant coefficients $r_7$ and $s_{k}$ used in approximation \eqref{genFormApx1} in the general case $0<n< 1$ with $\hat{K}_I=1$.}
 \label{Tab:apx1b}
\end{sidewaystable}

For the Newtonian fluid $n=1$ the coefficients for the approximate pressure function \eqref{n1apxP} are now given by:
\begin{equation}
p_1=0.90064, \quad p_2= 9.053 \times 10^{-3}, \quad p_3=-0.0126243 , \quad p_4=-7.3525 \times 10^{-4} ,
\end{equation}
while those for the aperture \eqref{apxWKI1} and particle velocity \eqref{apxVKI1} remain the same as in the general case.

In the case of a perfectly plastic fluid $n=0$ the coefficients used to approximate the aperture \eqref{n0ApxW} and particle velocity \eqref{n0ApxV} are as follows:
$$
w_1 =0.20403 , \quad w_2=-0.073008 , \quad w_3=-0.65676 , \quad w_4= -0.6802 , \quad w_5=0.14507 ,
$$
\begin{equation}
v_1=0.061258 , \quad v_2= 9.584 \times 10^{-4} , \quad v_3=-4.939 \times 10^{-3} , \quad v_4=-4.12\times 10^{-3} ,
\end{equation}
while those for the pressure function \eqref{apxPKI1} remain the same as in the general case.

\subsection{Toughness dominated regime with $\hat{K}_I = 10$}


In the general case $0<n<1$ the approximation coefficients for the aperture \eqref{apxWKI1}, particle velocity \eqref{apxVKI1} and pressure \eqref{apxPKI1} are given in the form:
\begin{equation}
{\cal Z} (n) = \frac{\sum_{k=0}^6 r_k n^k}{\sum_{k=0}^5 s_k n^k} ,
\label{genFormApx2}
\end{equation}
with the values of $r_k$, $s_k$ for the case $0<n< 1$ being listed in Tables~\ref{Tab:apx10} and \ref{Tab:apx10b}.

\begin{sidewaystable}[h!]
    \centering
\begin{tabular}{||c||c|c|c|c|c|c|c||}
\hline \hline
${\cal Z} (n)$ & $r_0$ & $r_1$ & $r_2$ & $r_3$ & $r_4$ & $r_5$ & $r_6$  \\
\hline \hline
&&&&&&&\\
$w_1$ & 2.091489$\times10^{-5}$ & -2.17337$\times10^{-5}$ & 0 & 0 & 0 & 0 & 0 \\[2mm]
\hline
&&&&&&&\\
$w_2$ & 1.576727$\times10^{-5}$ & -2.197917$\times10^{-5}$ & 0 & 0 & 0 & 0 & 0 \\[2mm]
\hline
&&&&&&&\\
$w_3$ & -1.305999$\times10^{-5}$ & 1.981974$\times10^{-5}$ & 0 & 0 & 0 & 0 & 0  \\[2mm]
\hline
&&&&&&&\\
$w_4$ & 9.1626$\times10^{-6}$ & -7.9008$\times10^{-4}$ & 3.71779$\times10^{-3}$ & -7.84178$\times10^{-3}$ & 8.92224$\times10^{-3}$ & -5.33619$\times10^{-3}$ & 1.315743$\times10^{-3}$  \\[2mm]
\hline
&&&&&&&\\
$w_5$ & 1078013$\times10^{-4}$ & -8.97955$\times10^{-4}$ & 1.822423$\times10^{-4}$ & 0 & 0 & 0 & 0  \\[2mm]
\hline \hline
&&&&&&&\\
$v_1$ & -9.10529$\times10^{-6}$ & 6.74877$\times10^{-3}$ & 0.02089243 & 0.01419788 & -2.20858$\times10^{-5}$ & 0 & 0  \\[2mm]
\hline
&&&&&&&\\
$v_2$ & 3.9550128$\times 10^{-8}$ & 3.298671 $\times 10^{-8}$ & -2.0487 $\times 10^{-7}$ & 0 & 0 & 0 & 0   \\[2mm]
\hline
&&&&&&&\\
$v_3$ & 8.61754$\times10^{-7}$ & -1.455384$\times10^{-6}$ & 6.6048$\times10^{-7}$ & 0 & 0 & 0 & 0 \\[2mm]
\hline
&&&&&&&\\
$v_4$ & 1.769907$\times10^{-4}$ & -1.483712$\times10^{-3}$ & -0.0455816 & 0 & 0 & 0 & 0 \\[2mm]
\hline \hline
&&&&&&&\\
$p_1$ & 0.884925 & 3.86403 & 8.86889 & -3.30243$\times10^{-3}$ & 0 & 0 & 0 \\[2mm]
\hline
&&&&&&&\\
$p_2$ & -1.323758$\times10^{-4}$ & 2.79849$\times10^{-4}$ & -1.506993$\times10^{-4}$ & 0 & 0 & 0 & 0 \\[2mm]
\hline
&&&&&&&\\
$p_3$ & 0.02033428 & -0.0862176 & 0.1551598 & -0.1360064 & 0.0477962 & 0 & 0 \\[2mm]
\hline
&&&&&&&\\
$p_4$ & -0.0165688 & 0.0751266 & -0.142603 & 0.1293825 & -0.0461426 & 0 & 0 \\[2mm]
\hline \hline
\end{tabular}
 \caption{The value of constant coefficients $r_k$ used in approximation \eqref{genFormApx2} in the general case $0<n< 1$ with $\hat{K}_I=10$.}
 \label{Tab:apx10}
\end{sidewaystable}


\begin{sidewaystable}[h!]
    \centering
\begin{tabular}{||c||c|c|c|c|c|c||}
\hline \hline
${\cal Z}(n)$ & $s_0$ & $s_1$ & $s_2$ & $s_3$ & $s_4$ & $s_5$  \\
\hline \hline
&&&&&&\\
$w_1$ &  4.622623$\times10^{-5}$ & -1.900085$\times10^{-3}$ & 0.1066525 & -0.1166793 & 1 & 0 \\[2mm]
\hline
&&&&&&\\
$w_2$ & 1.722383$\times10^{-3}$ & 0.06665135 & 1 & 0 & 0 & 0  \\[2mm]
\hline
&&&&&&\\
$w_3$ & 6.52219$\times10^{-5}$ & -2.700796$\times10^{-3}$ & 0.0875316 & -0.2049092 & 1 & 0  \\[2mm]
\hline
&&&&&&\\
$w_4$ & 1 & 0 & 0 & 0 & 0 & 0 \\[2mm]
\hline
&&&&&&\\
$w_5$ & 0.0569477 & 0.1781103 & 1 & 0 & 0 & 0  \\[2mm]
\hline \hline
&&&&&&\\
$v_1$ & 4.40059$\times10^{-4}$ & 0.950784 & 2.494946 & 1 & 0 & 0  \\[2mm]
\hline
&&&&&&\\
$v_2$ & 2.3357$\times10^{-3}$ & -0.0486563 & 1 & 0 & 0 & 0  \\[2mm]
\hline
&&&&&&\\
$v_3$ & 3.87371$\times10^{-5}$ & 0.0666403 & 0.369115 & 0.686029 & 1.67268 & 1  \\[2mm]
\hline
&&&&&&\\
$v_4$ & 0.034835 & 14.4869 & 3.31537 & 4.40656 & -3.29718 & 1  \\[2mm]
\hline \hline
&&&&&&\\
$p_1$ & 0.0997469 & 0.436514 & 1 & 0 & 0 & 0  \\[2mm]
\hline
&&&&&&\\
$p_2$ & 0.03608244 & 0.1856344 & 0.3893436 & 1 & 0 & 0  \\[2mm]
\hline
&&&&&&\\
$p_3$ & 0.4682724 & 1 & 0 & 0 & 0 & 0 \\[2mm]
\hline
&&&&&&\\
$p_4$ & 0.382409 & 1 & 0 & 0 & 0 & 0  \\[2mm]
\hline \hline
\end{tabular}
 \caption{The value of constant coefficients $s_k$ used in approximation \eqref{genFormApx2} in the general case $0<n< 1$ with $\hat{K}_I=10$.}
 \label{Tab:apx10b}
\end{sidewaystable}

For a Newtonian fluid $n=1$ the coefficients for the approximate pressure function \eqref{n1apxP} are now given by:
\begin{equation}
p_1=8.86228, \quad p_2= 9.23151 \times 10^{-6} , \quad p_3=-1.384716 \times 10^{-5}, \quad p_4=-8.6771 \times 10^{-11},
\end{equation}
while those for the aperture \eqref{apxWKI1} and particle velocity \eqref{apxVKI1} remain the same as in the general case.

Finally, in the case of a perfectly plastic fluid $n=0$ the coefficients used to approximate the aperture \eqref{n0ApxW} and particle velocity \eqref{n0ApxV} are as follows:
$$
w_1=2.2352 \times 10^{-3}, \quad w_2=-7.2 \times 10^{-4} , \quad w_3=-6.893\times 10^{-3} ,
$$
\begin{equation}
 w_4=-7.137 \times 10^{-3} , \quad w_5=1.843 \times 10^{-3} ,
\end{equation}
$$
v_1=7.040647 \times 10^{-3} , \quad v_2= 2.8 \times 10^{-7} , \quad v_3= -8.3\times 10^{-6} , \quad v_4= -8.4 \times 10^{-6} ,
$$
while those for the pressure function \eqref{apxPKI1} remain the same as in the general case.


\end{appendix}

\end{document}